\documentclass[reprint,showpacs,superscriptaddress,english,twocolumn,notitlepage,longbibliography,nofootinbib]{revtex4-1}
\usepackage{times}
\usepackage{graphicx}
\usepackage{float}
\usepackage{latexsym,amsmath,amssymb,bm,euscript}
\usepackage{color}
\usepackage{subfigure}
\usepackage{epstopdf}
\usepackage[colorlinks=true,linkcolor=blue,citecolor=blue,urlcolor=blue]{hyperref}
\usepackage{hyperref}
\usepackage{ulem}
\usepackage[dvipsnames]{xcolor}
\usepackage{dcolumn}     
\usepackage{bbm}

\newcommand{\be}[0]{\begin{equation}}
\newcommand{\ee}[0]{\end{equation}}
\def\ba#1\ea{\begin{align*}#1\end{align*}}  
\def\baa#1\eaa{\begin{align}#1\end{align}}  
\newcommand{\up}[0]{\uparrow}
\newcommand{\dn}[0]{\downarrow}
\newcommand{\bmat}[0]{\begin{bmatrix}}
\newcommand{\emat}[0]{\end{bmatrix}}

\newcommand{\beq}{\begin{equation}}
\newcommand{\eneq}{\end{equation}}

\def\qq{\mathbf{q}}
\def\kk{\mathbf{k}}

\def\RR{\mathbf{R}}

\def\rr{\mathbf{r}}
\def\GG{\mathbf{G}}
\def\QQ{\mathbf{Q}}

\def\kk{\mathbf{k}}
\def\qq{\mathbf{q}}

\def\GG{\mathbf{G}}
\def\QQ{\mathbf{Q}}
\def\RR{\mathbf{R}}

\def\ee{\epsilon}

\def\up{\uparrow}

\def\ee{\epsilon}

\def\qq{\mathbf{q}}
\def\kk{\mathbf{k}}

\def\RR{\mathbf{R}}

\def\rr{\mathbf{r}}
\def\GG{\mathbf{G}}
\def\QQ{\mathbf{Q}}

\def\hH{{ \hat{H} }}

\begin{document}

\title{Symmetric Kondo Lattice States in Doped Strained Twisted Bilayer Graphene}

\author{H. Hu}
\affiliation{Donostia International Physics Center, P. Manuel de Lardizabal 4, 20018 Donostia-San Sebastian, Spain}

\author{G.~Rai}
\affiliation{I. Institute of Theoretical Physics, University of Hamburg, Notkestrasse 9, 22607 Hamburg, Germany}

\author{L.~Crippa}
\affiliation{Institut f\"ur Theoretische Physik und Astrophysik and W\"urzburg-Dresden Cluster of Excellence ct.qmat, Universit\"at W\"urzburg, 97074 W\"urzburg, Germany}

\author{J.~Herzog-Arbeitman}
\affiliation{Department of Physics, Princeton University, Princeton, New Jersey 08544, USA}

\author{D.~C\u{a}lug\u{a}ru}
\affiliation{Department of Physics, Princeton University, Princeton, New Jersey 08544, USA}

\author{T.~Wehling}
\affiliation{I. Institute of Theoretical Physics, University of Hamburg, Notkestrasse 9, 22607 Hamburg, Germany}
\affiliation{The Hamburg Centre for Ultrafast Imaging, 22761 Hamburg, Germany}

\author{G.~Sangiovanni}
\affiliation{Institut f\"ur Theoretische Physik und Astrophysik and W\"urzburg-Dresden Cluster of Excellence ct.qmat, Universit\"at W\"urzburg, 97074 W\"urzburg, Germany}

\author{R.~Valent\'\i}
\affiliation{Institut f\"ur Theoretische Physik, Goethe Universit\"at Frankfurt, Max-von-Laue-Strasse 1, 60438 Frankfurt am Main, Germany}

\author{ A. M. Tsvelik}
\affiliation{Division of Condensed Matter Physics and Materials Science, Brookhaven National Laboratory, Upton, NY 11973-5000, USA}

\author{B. A. Bernevig}
\email{bernevig@princeton.edu} 
\affiliation{Department of Physics, Princeton University, Princeton, New Jersey 08544, USA}
\affiliation{Donostia International Physics Center, P. Manuel de Lardizabal 4, 20018 Donostia-San Sebastian, Spain}
\affiliation{IKERBASQUE, Basque Foundation for Science, Bilbao, Spain}

\begin{abstract} 
We use the topological heavy fermion (THF) model~\cite{HF_MATBLG} and its Kondo Lattice (KL) formulation~\cite{Spin_MATBLG} to study the possibility of a symmetric Kondo state in  twisted bilayer graphene. Via a large-$N$ approximation, we find a symmetric Kondo state in the KL model at fillings  $\nu=0,\pm 1,\pm 2$ where a KL model can be constructed~\cite{Spin_MATBLG}. 
 In the symmetric Kondo state, all symmetries are preserved and the local moments are Kondo screened by the conduction electrons.
 At the mean-field level of the THF model at $\nu=0,\pm 1, \pm 2, \pm 3$ we also find a similar symmetric state that is adiabatically connected to the symmetric Kondo state~\cite{coleman2015introduction}. We study the stability of the symmetric state by comparing its energy with the ordered (symmetry-breaking) states found in Ref.~\cite{HF_MATBLG} and find the ordered states to have lower energy 
 at $\nu=0,\pm 1,\pm 2$. However, moving away from integer fillings by doping holes to the light bands, our mean-field calculations find the energy difference between the ordered state and the symmetric state to be reduced, which suggests the loss of ordering and a tendency towards Kondo screening.
 We expect that including the Gutzwiller projection in our mean-field state will further reduce the energy of the symmetric state. 
 In order to include many-body effects beyond the mean-field approximation, we also performed dynamical mean-field theory (DMFT) calculations on the THF model in the non-ordered phase.
 The spin susceptibility follows a Curie behavior at $\nu=0, \pm 1,\pm 2$ down to $\sim 2\text{K}$ where the onset of screening of the local moment becomes visible. This hints to very low Kondo temperatures at these fillings, in agreement with the outcome of our mean-field calculations. At non-integer filling $\nu=\pm 0.5,\pm 0.8,\pm 1.2$ DMFT shows deviations from a $1/T$-susceptibility at much higher temperatures, suggesting a more effective screening of local moments with doping.
Finally, we study the effect of a $C_{3z}$-rotational-symmetry-breaking strain via mean-field approaches and find that a symmetric phase (that only breaks $C_{3z}$ symmetry) can be stabilized at sufficiently large strain at $\nu=0,\pm 1, \pm 2$. Our results suggest that a symmetric Kondo phase is strongly suppressed at integer fillings, but could be stabilized either at non-integer fillings or by applying strain.
\end{abstract}


\maketitle

{\it Introduction---} 
The  experiments on 
magic-angle $(\theta=1.05^{\circ})$ twisted bilayer graphene (MATBLG)~\cite{bistritzer2011moire,Balents2020,Andrei2021} have established the existence of a variety of interesting phases~\cite{CAO20,LU19,STE20,XIE21a,KER19,JIA19,WON20,ZON20,CHO21a,PAR21c,LU21,Saito2021,ROZ21,PhysRevLett.128.217701,Seifert2020,Otteneder2020,Lisi2021,PhysRevResearch.3.013153,Hesp2021,PhysRevMaterials.6.024003,Jaoui2022,Grover2022}, including correlated insulating phases~\cite{Cao2018,CAO20a,POL19,LIU21e, XIE19, CHO19, NUC20,SAI21,DAS21,WU21a, PhysRevLett.127.197701} and superconductivity~\cite{Cao2018_sc,CAO21,YAN19,diez2021magnetic,DiBattista2022}.
Their discovery has been followed by considerable theoretical efforts
~\cite{EFI18,VAF20,PAD18,PAD20,GUI18,DOD18,HEJ21,KHA21,PO18a,KON20,CHR20,KEN18,HUA20a,GUO18,CHA21,KAN21,WU19,BAL20,FER20,WIL20,CEA19, Yu2022,PhysRevLett.125.236804,PhysRevLett.129.076401,MATBLG_hf_DS} 
aimed at understanding their origin. 
An extended Hubbard model has been constructed to analyze the interacting physics~\cite{KAN19,KAN18,KOS18,OCH18,VAF21,OCH18,XU18,XU18b,VEN18,YUA18,DA19,DA21,CHI20a,KAN21,SEO19}, however, 
due to the non-trivial topology of the flat bands
~\cite{LIU19,ZOU18,SON19,PO19,LIA20a,HEJ19a,LIU18,THO18,SON21}, certain symmetries become non-local. 
Alternatively, an approach based on a momentum space model has been considered~\cite{BUL20,tbgiii,XIE21,YOU19,WU20b,ISO18,LIU19a,WAN21a,BER21a}, in which correlated insulators~\cite{BUL20a,tbgiv,tbgv,LIU21,CEA20,ZHA20,LIU21a,XIE20a}, superconductivity~\cite{LIA19,WU18,GON19,LEW21,HEJ19,XIE20},
and other correlated quantum phases~\cite{KWA21,LED20,ABO20,REP20a,SHE21}
 have been identified and studied. Besides, various numerical calculations~\cite{KAN20a,SOE20,EUG20,HUA19,ZHA21,HOF21,REP20,2022arXiv221011733Z} 
have also been performed to investigate the correlated nature of the phenomena. 
However, the active phase diagram including the states at non-integer fillings is not well understood.
The exact mapping between the MATBLG and topological heavy-fermion model  constructed in Ref.~\cite{HF_MATBLG} could be used for developments in this direction. This mapping establishes a  bridge between  heavy-fermions~\cite{hewson1997kondo,coleman2015introduction,RevModPhys.56.755,QS_review,Gegenwart2008} and moir\'e systems~\cite{HF_MATBLG,Spin_MATBLG,PhysRevLett.127.026401}. The presence of localized moments in MATBG is supported by recent entropy measurements which  have found a Pomeranchuk-type transition~\cite{SAI20,ROZ21}. A large entropy observed at high-temperatures, originates from weakly interacting local moments whose fluctuations are quenched at low temperatures~\cite {SAI20,ROZ21}. Since a similar behavior is observed in heavy fermion systems~\cite{coleman2015introduction,hewson1997kondo}, where the fluctuating local moments are screened by conduction electrons (Kondo effect), this observation is suggestive of a Kondo state with screened local moments in MATBLG~\cite{hewson1997kondo,Coleman_kondo}. 

In this paper we use the KL model~\cite{Spin_MATBLG}, to describe and study the symmetric Kondo (SK) state. We focus on integer fillings $\nu =0,\pm 1,\pm 2 $, where a KL model can be constructed~\cite{Spin_MATBLG} (a KL description fails at $\nu=\pm 3$ as demonstrated in Ref.~\cite{Spin_MATBLG}). 
The SK phase preserves all symmetries;  the local moments are screened. We discuss the topology and the band structure of the SK state and extend the study to the THF model where we identify the symmetric state that is adiabatically connected to the SK state ~\cite{coleman2015introduction}. 
In order to address integer and fractional fillings on equal footing, we perform both a mean-field and a dynamical mean-field theory (DMFT) calculations of the THF defining a ``periodic Anderson model'' with a momentum-dependent hybridization between the correlated $f$- and the dispersive $c$-electrons in the non-ordered state.

 Our mean-field calculations indicate that the energy of the symmetric state is higher than that of the ordered (symmetry-breaking) states found in Ref.~\cite{HF_MATBLG} at integer filling. We thus conclude that ordered states are more energetically favored at integer fillings. 
 DMFT supports this picture as we obtain a Curie behavior of the local spin susceptibility at integer fillings, down to very low temperatures $\sim 2$K, hinting to a very small Kondo scale (lower than $\sim 2$K). Together with the mean-field results we would then expect an ordered state to be favored at low temperatures for these fillings.

Turning to the effect of doping, instead, from our mean-field analysis, we find that the energy difference between the symmetric phase and the ordered phase can be sizeably reduced. Doping hence suppresses the ordering and enhances the Kondo screening. This conclusion is 
further supported by the DMFT results at non-integer fillings. Here, we find clear deviations from the Curie behavior in the entire range from 10K down to $\sim$1K. Even though it is computationally too demanding to go further down in temperature, we point out that our evidence of a clear-cut difference in the screening properties between integer and fractional fillings is reliable. 
DMFT treats indeed local quantum fluctuations exactly~\cite{RevModPhys_dmft} and hence takes into account the many-body processes that can potentially lead to the screening of local moments at any filling.  

Since realistic samples have intrinsic strains, we finally study the effect of a $C_{3z}$-breaking strain on the symmetric phase.  Our mean-field calculations show that the order is suppressed by the strain effect and a symmetric state can be stabilized at a sufficiently large strain at $\nu=0,\pm 1, \pm 2$.

In summary, we conclude that a symmetric Kondo phase is absent at integer fillings of MATBLG, but could in principle be stabilized either at non-integer fillings or by applying strain.


{\it Topological Heavy Fermion model and the Kondo lattice model---} 
The THF model~\cite{HF_MATBLG} contains two types of electrons: topological conduction $c$-electrons ($c_{\kk,a \eta s}$) and localized $f$-electrons ($f_{\RR, \alpha \eta s}$). The operator $c_{\kk,a \eta s}$  annihilates  conduction $c$-electron with momentum $\kk$, orbital $a\in \{1,2,3,4\}$, valley $\eta \in \{+,-\}$ and spin $s\in \{\up,\dn\}$. 
At the $\Gamma_M$-point for each valley and each spin projection, $c$-electrons in the orbital $1$ and $2$ transform according to the  $\Gamma_3$ irreducible representation (of magnetic space group $P6^\prime 2^\prime 2)$~\cite{HF_MATBLG}. 
The remaining $c$-electrons ($a=3,4$) at the same valley with the same spin projection transform in the  $\Gamma_1 \oplus \Gamma_2$ reducible representation (of magnetic space group $P6^\prime 2^\prime 2)$~\cite{HF_MATBLG}. We will call them $\Gamma_3$ $c$-electrons ($a=1,2$) and $\Gamma_1\oplus \Gamma_2$ $c$-electrons ($a=3,4$) respectively. $f_{\RR,\alpha \eta s}$ is the annihilation operator of the $f$-electron at the moir\'e unit cell $\RR$ with orbital $\alpha \in \{1,2\}$, valley $\eta$ and spin $s$. The Hamiltonian of the THF model~\cite{HF_MATBLG,SM} is
\baa  
\hH_{THF} = \hH_c +\hH_{fc} +\hH_U +\hH_W +\hH_V +\hH_J 
\label{eq:THF}
\eaa  
where $\hH_c$ describes the kinetic term of conduction electrons, $\hH_{fc}$ describes the hybridization between $f$-$c$ electrons~\cite{HF_MATBLG,SM}. The interactions include
an on-site Hubbard interaction of $f$-electrons ($\hH_U$ with $U=57.95$meV), a repulsion between $f$- and $c$-electrons ($\hH_W$ with $W=48$meV), 
a Coulomb interaction between $c$-electrons ($\hH_V$ with $V(\qq=0)/\Omega_0=48.33$meV and 
$\Omega_0$ the area of moir\'e unit cell), and a ferromagnetic exchange coupling between $f$-and $c$-electrons ($\hH_J$ with $J=16.38$meV)~\cite{HF_MATBLG,SM}.

Based on the THF model~\cite{HF_MATBLG}, a KL model of MATBLG has been constructed via a generalized Schrieffer–Wolff (SW) transformation as shown in Ref.~\cite{Spin_MATBLG}. The KL model is described by  the following Hamiltonian
\baa 
\hH_{Kondo} = \hH_{c} +\hH_{cc} + \hH_{K} + \hH_J 
\label{eq:kondo_ham_main}\, . 
\eaa 
where $\hH_c,\hH_J$ come from the original THF model and $\hH_{cc},\hH_K$ emerge from the SW transformation.
$\hH_{cc}$ is the one-body scattering term of $\Gamma_3$ $c$-electrons with the form of 
\baa  
\hH_{cc} = 
 &\sum_{|\kk| <\Lambda_c}\sum_{\substack{ a,a'\in\{1,2\} \\ \eta ,s } }
e^{-|\kk|^2\lambda^2}
 :c_{\kk,a\eta s}^\dag c_{\kk, a'\eta s}: 
\bigg(  \frac{-1}{D_{\nu_c,\nu_f}}\nonumber \\
&+\frac{-1}{D_{\nu_c,\nu_f}}\bigg) 
\begin{bmatrix}
\gamma^2/2 & \gamma v_\star^\prime (\eta k_x -ik_y)   \\
\gamma v_\star^\prime (\eta k_x + ik_y) &\gamma^2/2
\end{bmatrix}_{a,a'}\, .
\label{eq:hcc}  
\eaa  
$\lambda$ is the damping factor of the $f$-$c$ hybridization in the THF model. $\gamma, v_\star^\prime$ characterize the zeroth order and linear order $f$-$c$ hybridization of the THF model with $v_\star^\prime$ characterizing a $\kk$-dependent hybridization matrix~\cite{HF_MATBLG,SM}. $D_{1,\nu_c,\nu_f}$ and $D_{2,\nu_c,\nu_f}$ are defined as
\baa  
&D_{1,\nu_c,\nu_f}=
(U-W)\nu_f -\frac{U}{2}  +(\frac{-V(0)}{\Omega_0}+W){\nu}_c
\nonumber \\
&D_{2,\nu_c,\nu_f}=(U-W)\nu_f +\frac{U}{2}   +(\frac{-V(0)}{\Omega_0}+W){\nu}_c \, , 
\label{eq:D12_def_main}
\eaa 
where
$\nu_f, \nu_c$ are the filling of $f$- and $c$-electrons determined from the calculations of the THF model at the zero-hybridization limit~\cite{Spin_MATBLG}. We point out that in the single-orbital Kondo model, the one-body scattering term merely introduces a chemical potential shift~\cite{coleman2015introduction,SW_transf} of the $c$-electrons and is usually omitted.
However, in our model, $\hH_{cc}$ cannot be ignored for two reasons. 
First, $\hH_{cc}$ is $\kk$-dependent and thus introduces additional kinetic energy to the conduction electrons. 
From Eq.~\ref{eq:hcc}, we observe the $\kk$-dependency mainly comes from the linear $\kk$ term that is proportional to $v_\star^\prime$ and can be traced back to the $\kk$-dependency of the hybridization matrix in the THF model. Secondly, even if we drop the $v_\star^\prime$ term in Eq.~\ref{eq:hcc} ($v_\star^\prime=0$ corresponding to the chiral limit~\cite{HF_MATBLG}), $\hH_{cc}$ still produces an energy shift for the $\Gamma_3$ $c$-electrons. Thus $\hH_{cc}$ leads to the energy splitting between $\Gamma_3$ and $\Gamma_1\oplus \Gamma_2$ $c$-electrons and cannot be simply treated as a shift of the chemical potential.

$\hH_K$ is the Kondo interaction between $f$- and $\Gamma_3$ $c$-electrons whose explicit form is given in Refs.~\cite{SM,Spin_MATBLG}. 
The Kondo interaction consists of two parts: the zeroth order Kondo interaction proportional to $\gamma^2/D_{\nu_c,\nu_f}$ and the first order Kondo interaction proportional to $\gamma v_\star^\prime/D_{\nu_c,\nu_f}$, where $D_{\nu_c,\nu_f}^{-1} = -D_{1,\nu_c,\nu_f}^{-1} +D_{2,\nu_c,\nu_f}^{-1}$. The zeroth order Kondo interaction term describes the antiferromagnetic interaction between the $U(8)$ moments of the $f$- and the $\Gamma_3$ $c$-electrons and has a $U(8)$ symmetry. The linear-order Kondo interaction only has a flat $U(4)$ symmetry and is $\kk$-dependent~\cite{HF_MATBLG,SM}.  
$\hH_J$ is the ferromagnetic exchange interaction between $\Gamma_1 \oplus \Gamma_2$ $c$- and $f$-electrons that already exists in the TFH model~\cite{HF_MATBLG,SM}. We also note that, for both the THF model and the KL model, ground states at filling $\nu$ and $-\nu$ are connected by a charge-conjugation transformation~\cite{HF_MATBLG}. This can be broken by other one-body terms which we did not consider here. Therefore, in what follows, we only focus on $\nu \le 0$.

\begin{figure*}
    \centering
    \includegraphics[width=1.0\textwidth]{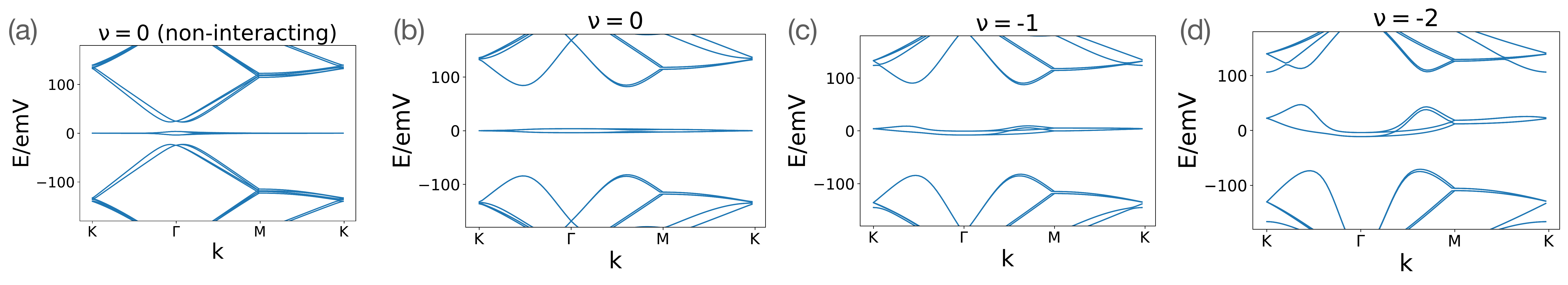}
    \caption{ (a) Band structure of the non-interaction THF model at $\nu=0$. (b), (c), (d) Band structure of the SK phase at $\nu=0,-1,-2$ respectively.}
    \label{fig:kondo_band}
\end{figure*}

{\it Mean-field Hamiltonian of the Kondo model---} 
We next perform a mean-field study of the KL model~\cite{coleman2015introduction}. 
{ This MF suppresses the RKKY interaction and essentially restores the hybridization term $\hH_{fc}$ of the original periodic Anderson model, but in a renormalized form. It becomes exact in the $N \rightarrow \infty$ limit (we have $N=4$ here which corresponds to the approximate flat $U(4)$ symmetry of the KL Hamiltonian in Eq.~\ref{eq:kondo_ham_main}).}
At the mean-field level, the Kondo interaction $\hH_K$ can be written as (see Supplementary Materials (SM))
\baa  
 \hH_K^{MF}
= &  \sum_{\RR, |\kk| <\Lambda_c}
\sum_{\alpha\eta  s}\frac{ e^{i\kk \cdot \RR -|\kk|^2 \lambda^2/2} }{\sqrt{N_M}D_{\nu_c,\nu_f} }\Bigg[ 
 -f_{\RR,\alpha \eta s}^\dag c_{\kk, a \eta s}
\nonumber \\
&
\begin{pmatrix}
   \gamma^2 V_1^* +\gamma v_\star^\prime V_2^* 
   & V_1^*(\eta k_x-ik_y)  \\ 
   V_1^*(\eta k_x+ik_y) &  \gamma^2 V_1^* +\gamma v_\star^\prime V_2^* 
\end{pmatrix}_{\alpha,a}+\text{h.c.}\bigg] \nonumber \\
&  
+N_M\bigg[\gamma^2|V_1^*|^2 +\gamma v_\star^\prime( V_1^*V^*_2 + V_2^*V^*_1 )
\bigg] +\text{H.T.}
\label{eq:ham_kondo}
\eaa  
where we have introduced the following mean fields
\baa  
V^*_1 =& \sum_{\RR, |\kk| <\Lambda_c}
\sum_{\alpha\eta s}\frac{e^{i\kk \cdot \RR -|\kk|^2 \lambda^2/2}  }{\sqrt{N_M}N_M} \langle \Psi| f_{\RR,\alpha \eta s}^\dag c_{\kk,\alpha\eta s} |\Psi\rangle  \nonumber \\ 
V^*_2 =& \sum_{\RR, |\kk| <\Lambda_c}
\sum_{\alpha a \eta s}\frac{ e^{i\kk \cdot \RR -|\kk|^2 \lambda^2/2} }{\sqrt{N_M}N_M} (\eta k_x \sigma_x +k_y\sigma_y)_{\alpha a}\nonumber \\
&
\langle \Psi| 
f_{\rr,\alpha \eta s}^\dag c_{\kk,\alpha\eta s} |\Psi\rangle  
\eaa 
with $|\Psi\rangle$ being the mean-field ground state, and $\text{H.T.}$ denotes the Hartree term 
($\langle f^\dag f\rangle, \langle c^\dag c\rangle$) whose explicit formula is in the Supplementay Materials (SM)~\cite{SM}. 
Several points are in order. 
{First, as we have mentioned above, the mean field restores the hybridization of the original Anderson model, but in a renormalized form. $V^*_1,V^*_2$ describe the renormalized hybridization between the $f$- and $\Gamma_3$ $c$-electrons driven by the Kondo interactions between two types of electrons($f$ and $\Gamma_3$ $c$)~\cite{coleman2015introduction, read1983solution}.}
Second, it is necessary to keep the Hartree contributions. 
In the canonical Kondo model, the Hartree term merely produces a chemical potential shift (in the case without symmetry breaking) and hence can be omitted. Here, Hartree contributions (see SM~\cite{SM}) are $\kk$-dependent because of the $\kk$-dependency of the Kondo interactions, and thus contribute to the dispersion of the conduction $c$-electrons. Furthermore, since only $\Gamma_3$ $c$-electrons contribute to the Kondo interaction, the Hartree term also produces an energy splitting between the $\Gamma_3$ and the $\Gamma_1\oplus\Gamma_2$ $c$-electrons.

As for $\hH_J$, we perform a similar mean-field decoupling 
\baa  
  \hH_J^{MF} 
=&J \sum_{\RR, |\kk|<\Lambda_c, \alpha \eta s } \frac{e^{i \kk \cdot\RR }}{\sqrt{N_M}}\bigg( V_3  \delta_{1,\eta (-1)^{\alpha+1}}f_{\RR,\alpha \eta s}^\dag c_{\kk,\alpha +2 \eta s} \nonumber \\
&+ 
 V_4  \delta_{-1,\eta (-1)^{\alpha+1}}\eta f_{\RR,\alpha \eta s}^\dag c_{\kk,\alpha +2 \eta s} +\text{h.c.}\bigg) \nonumber \\
 & - JN_M\bigg[ |V_3|^2 +|V_4|^2\bigg] + \text{H.T.}
 \label{eq:ham_j}
\eaa  
where we have introduced the following two mean-field averages that describe the $f$-$c$ hybridization: 
\baa  
&V_3 =  \sum_{\RR, |\kk| <\Lambda_c}
\sum_{\alpha\eta, s}\frac{e^{i\kk \cdot \RR  }\delta_{1, \eta (-1)^{\alpha+1}}}{\sqrt{N_M}N_M} \langle \Psi| f_{\RR,\alpha \eta s}^\dag c_{\kk,\alpha+2\eta s} |\Psi\rangle 
\nonumber  \\ 
&V_4 =  \sum_{\RR, |\kk| <\Lambda_c}
\sum_{\alpha\eta, s}\frac{e^{i\kk \cdot \RR  }\delta_{-1, \eta (-1)^{\alpha+1}}}{\sqrt{N_M}N_M} \langle \Psi| \eta f_{\RR,\alpha \eta s}^\dag c_{\kk,\alpha+2\eta s} |\Psi\rangle   \, , 
\eaa 

To impose the filling of the $f$-electrons to be $\nu_f$, we introduce the Lagrange multiplier ~\cite{Coleman_kondo,read1983solution,SM}:
\baa  
\hH_{\lambda_f}  = \sum_{\RR, \alpha \eta s }\lambda_f\bigg( :f_{\RR,\alpha \eta s} ^\dag f_{\RR,\alpha \eta s}: - \nu_f \bigg) 
\label{eq:ham_lamf}
\eaa 
with $\lambda_f$ to be determined self-consistently~\cite{SM}. 
Finally, we introduce a chemical potential $\mu_c$ to the $c$-electrons 
\baa 
\hH_{\mu_c} = -\mu_c \sum_{|\kk|<\Lambda_c, a\eta s} :c_{\kk,a\eta s}^\dag c_{\kk,a \eta s }: \, . 
\label{eq:ham_muc} 
\eaa 
In the calculation, we tune $\mu_c$ and $\lambda_c$ together to fix both the total filling $\nu = \nu_f+\nu_c$ and the filling of $f$-electrons~\cite{SM}. 
The final mean-field Hamiltonian of the KL model now is
\baa  
\hH_{Kondo}^{MF} = \hH_{c} +\hH_{cc} +\hH_{K}^{MF} +\hH_J^{MF} +\hH_{\lambda_f} +\hH_{\mu_c} \, . 
\label{eq:mf_ham}
\eaa  

We then self-consistently solve the mean-field equations (see SM~\cite{SM}).  
At $\nu=\nu_f=0,-1,-2$ (where a KL model can be constructed), we identify a SK state that preserves all the symmetries and is characterized by $V^*_1\ne 0, V^*_2\ne 0, V^*_3 = 0,V^*_4=0 $~\cite{SM}. 
We comment that the exchange interaction $\hH_J$~\cite{HF_MATBLG}
between $f$- and $\Gamma_1\oplus \Gamma_2 $ $c$-electrons 
is ferromagnetic, and hence disfavors the singlet formation or hybridization ($V_3,V_4$) between $f$- and $\Gamma_1 \oplus \Gamma_2$ $c$-electrons. We find that $V_3,V_4$ vanish (their numerical amplitudes are smaller than $10^{-5}$). In fact, $\hH_J$ favors the triplet formation or pairing formation ($f^\dag c^\dag$), where both lead to a symmetry-breaking state at the mean-field level and are beyond our current consideration of SK state.


{\it Properties of the symmetric Kondo phase---}
In Fig.~\ref{fig:kondo_band}, we plot the band structure of the SK phase and compare it with the non-interacting THF model. 
 We find the hybridization in the SK state defined in Eq.~\ref{eq:ham_kondo} to be enhanced compared to the non-interacting limit of THF model, which is clear from the increase of the gap of  the $\Gamma_3$ states at the $\Gamma$ point~\cite{HF_MATBLG} from its non-interacting value $24.75$meV  at $\nu=0$, 
 to 
 $168$meV, $190$meV, $213$meV at $\nu=0,-1,-2$ respectively.
We also find that in the SK phase the bandwidths of the flat bands at $\nu=-1,-2$ become $16$meV, $53$meV, which are (much) larger than the non-interacting flat-band bandwidth ($=7.4$meV) of the THF model (Fig.~\ref{fig:kondo_band}). 
However, at $\nu=0$, the flat-band bandwidth is the same as the non-interacting flat-band bandwidth.
This is because, at $\nu=0$, the one-body scattering term and the Hartree contributions from $\hH_K,\hH_J$ all vanish~\cite{SM}, and the enhanced hybridization pushes the remote bands away from the Fermi energy and does not change much the band structures of the flat bands. In addition, unlike the non-interacting case, here we found the flat bands are mostly formed by $\Gamma_1\oplus\Gamma_2$ $c$-electrons with orbital weights larger than $70\%$ at $\nu=0,-1,-2$. This is because the large $f$-$c$ hybridization induced by $V_1,V_2$ (Eq.~\ref{eq:ham_kondo}) pushes the energy of $\Gamma_3$ $c$- and $f$-electrons away from the Fermi energy and reduces their orbital weights~\cite{SM}.

The flat bands in the SK phase form $\Gamma_1\oplus\Gamma_2$, $M_1\oplus M_2$ and $K_2K_3$ representations at $\Gamma_M,M_M,K_M$ respectively, and have the same topology as the flat bands in the non-interacting THF model~\cite{HF_MATBLG}. More explicitly, the flat bands for each valley and each spin projection belong to a fragile topology~\cite{HF_MATBLG} at $\nu=-1,-2$. At $\nu=0$, due to the additional particle-hole symmetry, flat bands have a stable topology~\cite{HF_MATBLG,SON19,SON21,SM}, which is characterized by the odd winding number of the Wilson loop as shown in supplementary material~\cite{SM}. 
We mention that the interplay between Kondo effect and the topological bands has also been studied in various other systems~\cite{dzero2010topological,lai2018weyl,hu2022coupled, chen2022emergent,Coleman_TKI, LEI21}.

{\it Symmetric phase in the topological heavy-fermion model--- }
We next investigate the similar symmetric phase in the THF model Eq.~\ref{eq:THF}.  We first focus on integer fillings $\nu=0,-1,-2,-3$ and perform the mean-field calculations of THF as introduced in Ref.~\cite{HF_MATBLG,SM}. By enforcing the mean-field Hamiltonian to preserve all the symmetries, we are able to identify a symmetric state that preserves all the symmetries at $\nu=0,-1,-2,-3$. To observe the stability of the symmetric phase, we compare its energy ($E_{sym}$) with the energy ($E_{order}$) of the ordered (symmetry-breaking) ground states derived in Ref.~\cite{HF_MATBLG}. The ordered ground states in Ref.~\cite{HF_MATBLG} are a Kramers inter-valley-coherent (KIVC) state at $\nu=0$, a KIVC+valley polarized (VP) state at $\nu=-1$, a KIVC state at $\nu=-2$ and a VP state at $\nu=-3$. We point out that at $\nu=-3$ other states with lower energy exist~\cite{xie2022phase}.  In our numerical calculations, we find $\Delta E = E_{sym}-E_{order} =47\mathrm{meV}, 40\mathrm{meV}, 33\mathrm{meV}, 23\mathrm{meV}$ at $\nu=0,-1,-2,-3$ respectively. In all integer filling cases, the symmetric states have higher energy, and the ground states cannot be the symmetric state, which is consistent with the previous calculations of Ref.~\cite{Spin_MATBLG,HF_MATBLG,tbgv}. 
Note that our mean-field calculation does not include a Gutzwiller projection to fix the filling of $f$-electrons at each site, and hence we expect the energy of projected symmetric states will be lower. However, as we show later, after including the effect of local correlations via the DMFT approach, we confirm that the Kondo phase, which is adiabatically connected to the symmetric phase in the mean-field calculations, is strongly suppressed 
at integer fillings (down to temperatures $\sim 1$-$2$K). This further supports the development of ordering at integer fillings. 

\begin{figure}
    \centering
    \includegraphics[width=0.5\textwidth]{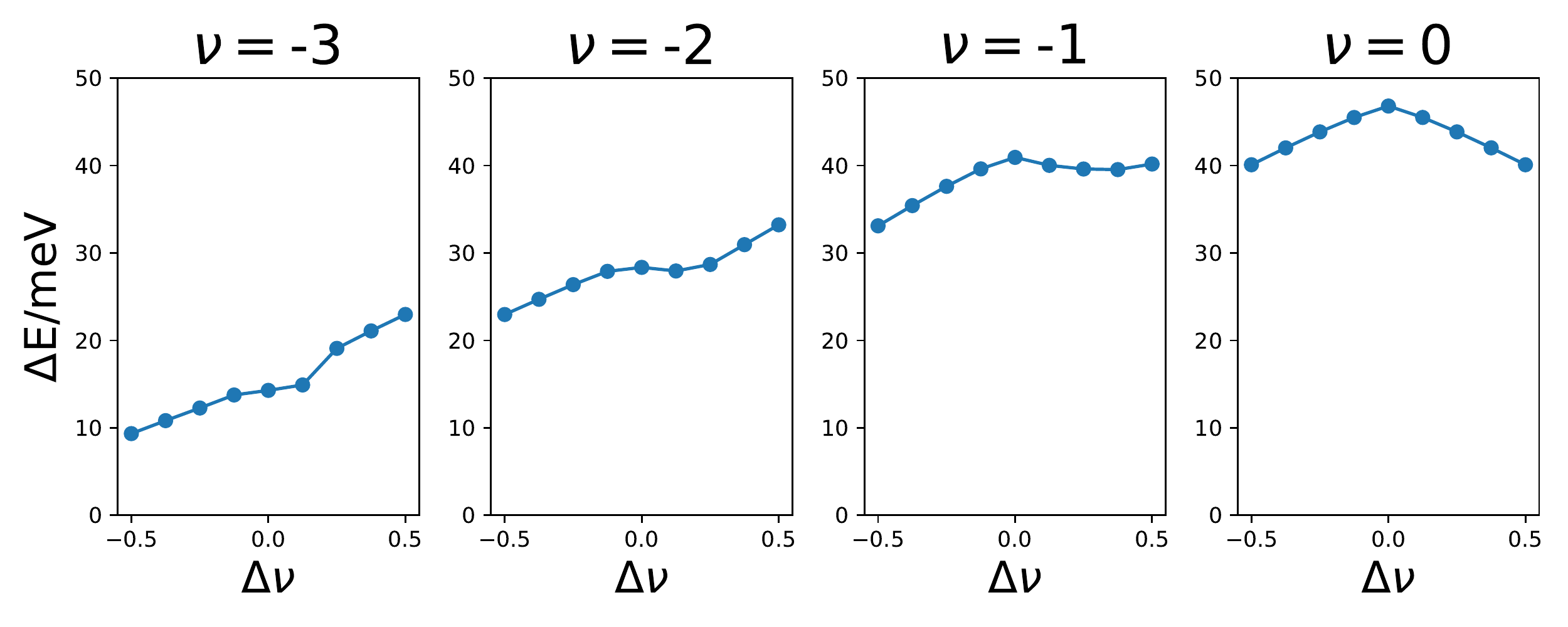}
    \caption{Doping dependence of the ground state energy difference $\Delta E= E_{sym}-E_{order}$ near integer fillings $\nu_t=0,-1,-2,-3$.
    }
    \label{fig:doping}
\end{figure}

{\it Effects of doping---} We next investigate the effects of doping, first at the level of mean-field theory.
We stick to a narrow region $\nu \in [\nu_{int}-0.5, \nu_{int}+0.5]$ near each integer filling $\nu_{int}=0,-1,-2,-3$ and  compare the energies of the ordered states $E_{ord}$ and the symmetric states $E_{sym}$ in the THF model. To find the ordered state solutions, we first initialize the calculations with the mean-field solutions at integer filling $\nu_{int}$ and fill the mean-field bands up to current filling $\nu$. We then self-consistently solve the mean-field equations and calculate the energy of the resulting states. 
We obtain the symmetric-state solution in a similar manner but take the symmetric solution at $\nu_{int}$ as initialization and enforce the symmetry of the mean-field Hamiltonian during the calculations. 
Fig.~\ref{fig:doping} displays a plot of the difference of the ground state energies $\Delta E = E_{sym}-E_{order}$ as a function of doping $\Delta \nu = \nu-\nu_{int}$ near $\nu_{int}=0,-1,-2,-3$. We observe that hole doping at $\nu=0,-1,-2,-3$ and electron doping at $\nu=0$ decreases the $\Delta E$. Doping holes at $\nu=0,-1,-2,-3$ and doping electrons at $\nu=0$ to the ordered states  is equivalent to doping the light (dispersive) bands which are mostly formed by conduction $c$-electrons. After doping, the conduction electrons will stay close to the Fermi energy, and then enhance the tendency towards the Kondo effect. 

However, doping electrons at $\nu=-1,-2$ to the ordered states is equivalent to doping heavy (flat) bands which mostly come from the $f$-electrons. Because of the flatness of the band, we find the nature of the ordered states will change with doping (see SM ~\cite{SM}). 
For example at $\nu=2$, the KIVC order is suppressed by the electron doping (see SM~\cite{SM}). Thus, $\Delta E$ will be affected by both, changes of ordering and doping. 
However, a sizeable change of the order parameters is not observed for hole doping at $\nu=0,-1,-2,-3$ and also electron doping at $\nu=0$ (see SM~\cite{SM}), because we are doping conduction $c$-electrons in both cases.
We also point out the complexity of $\nu=-3$. First, other states that break translational invariant~\cite{xie2022phase} could have lower energy than the VP state we currently considered. Second, even for the VP state, doping electrons is equivalent to doping both heavy and light bands~\cite{HF_MATBLG}, since both light and heavy bands appear in the electron doping case~\cite{HF_MATBLG}. 
In practice, as we increase $\Delta \nu$, we find that, at $\nu=-1,-2$, $\Delta E$ will first decrease and then increase and, at $\nu=-3$, $\Delta E$ will always increase. 

In summary, we conclude that hole doping can suppress the long-range order and enhance the tendency towards the Kondo effect near $\nu =0,-1,-2$. Electron doping, depending on the fillings, could also enhance the tendency toward the Kondo effect. However, on the electron doping side, the change of order moments indicates the importance of the correlation effect which could be underestimated in the mean-field approach. In the next section, we provide a more comprehensive study of the doping effect using the DMFT calculation.



\begin{figure*}
    \centering
 \includegraphics{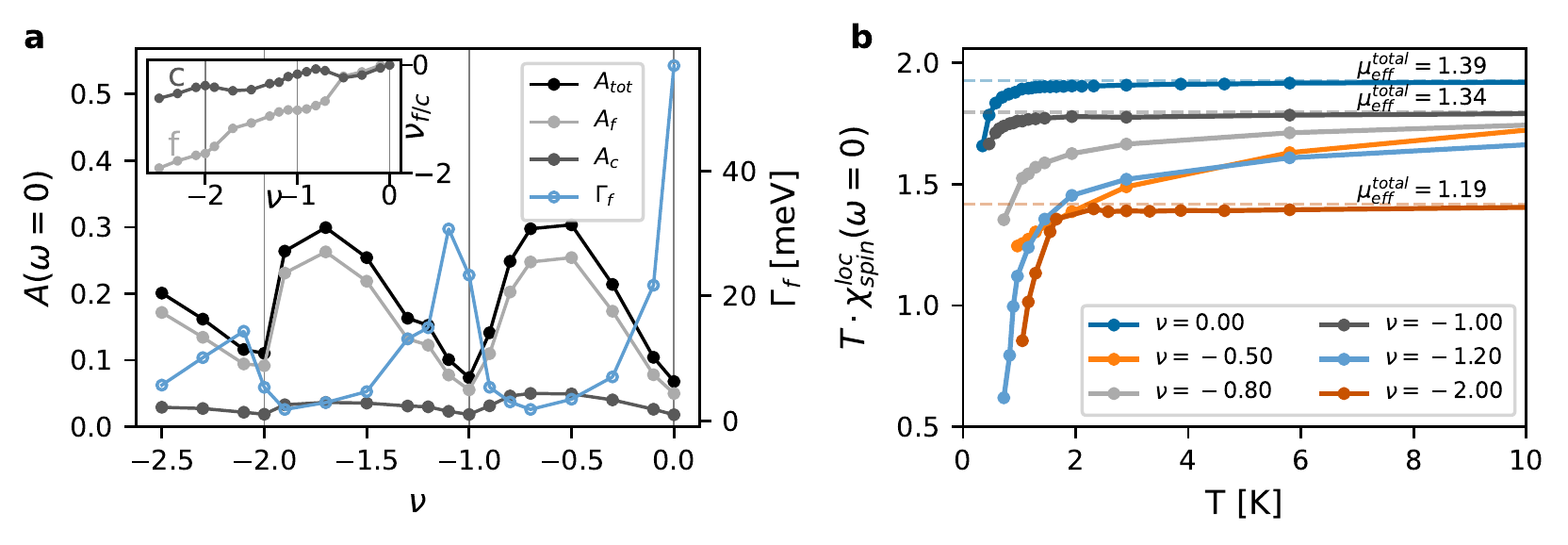}
    \caption{DMFT solution of the THF model. (a) Doping $\nu$ dependent low-energy spectral function at the Fermi level ($A(\omega= 0)$) for the full system $A_{tot}$, the c- ($A_c$) and the f-electrons ($A_f$) at $11.6$~K. Also shown is the scattering rate $\Gamma_f$ as extracted from the local f-electron self-energy. (b) Effective local moment $T \cdot \chi_\text{spin}^\text{loc}(\omega=0)$ as a function of temperature $T$ for different doping levels $\nu$.}
    \label{fig:DMFT}
\end{figure*}

{\it Dynamical mean-field theory results of the THF model---}

In the following, we present the dynamical mean-field theory resultss of the THF model (Eq.~\ref{eq:THF}), where we describe the local quantum many-body effects of the density-density Hubbard term $\hH_U$ within the $f$-subspace. The $\hH_W$ and $\hH_V$ interactions involving density fluctuations on the $c$ orbitals are accounted for at the static mean-field level. We neglect ordered phases and perform calculations in the non-ordered one. There, we focus in particular on lifetime effects, quasiparticle weights and exploit the ability of DMFT to take local vertex corrections to the spin-spin correlation function into account. 

First, DMFT finds a qualitative difference between the strong quasiparticle renormalization when the $f$+$c$ manifold is occupied with an integer number of electrons and a lighter Fermi liquid at fractional fillings: this can be seen from the scattering rate $\Gamma_f = -\text{Im} \Sigma_f(\omega=0)$ which is shown as a function of the total filling $\nu$ at $T=11.6$K (light blue empty circles) in Fig.\ref{fig:DMFT}(a). The largest scattering rates are found close to $\nu=0.0$, -1.0 and -2.0, progressively decreasing as one moves away from the charge neutrality point. Correspondingly, the spectral weight at the Fermi level (black and grey solid circles) is suppressed at these fillings, with a residual nonzero value due to the finite temperature on the one hand and the resilient $f$/$c$ hybridization on the other.

Fig.\ref{fig:DMFT}(b) illustrates the temperature-dependent screening of the local magnetic moment on the $f$ orbitals at different fillings. A flat $T \cdot \chi_\text{spin}^\text{loc}(\omega=0)$ indicates Curie behavior and a well-defined effective local moment, while deviations signal the onset of screening and a crossover towards a renormalized Pauli-like behavior, in agreement with the general expectation of zero-temperature Fermi-liquid in the periodic Anderson model \cite{AntoinePRL}. While at $\nu=0.0$, -1.0 and -2.0 the $1/T$ local spin susceptibility persists down to 1-2 K, the fractional fillings deviate from Curie at much higher temperatures, in line with the better Fermi-liquid nature signaled by the single-particle quantities in Fig.\ref{fig:DMFT}(a).

As in the Hartree-Fock treatment of the THF model, DMFT confirms the difference between electron doping and hole doping (particle-hole asymmetry) near integer filling $\nu=-1$ and -2. Here, DMFT reveals particle-hole asymmetric scattering rates (Fig. \ref{fig:DMFT}(a)) and also in the difference of effective local moments at $\nu=-0.8$ and $-1.2$(Fig. \ref{fig:DMFT}(b)). 

In summary, our DMFT calculations confirm that the Kondo phase is strongly suppressed at integer fillings $\nu=0,-1,-2$, increasing the propensity towards long-range order at these fillings. However, by doping the system, the development of Kondo screening (starting from $\sim 10$K) is observed, which suggests that doping could enhance the Kondo effect. This picture is consistent with our mean-field calculations.

 \begin{figure}
     \centering
     \includegraphics[width=0.4\textwidth]{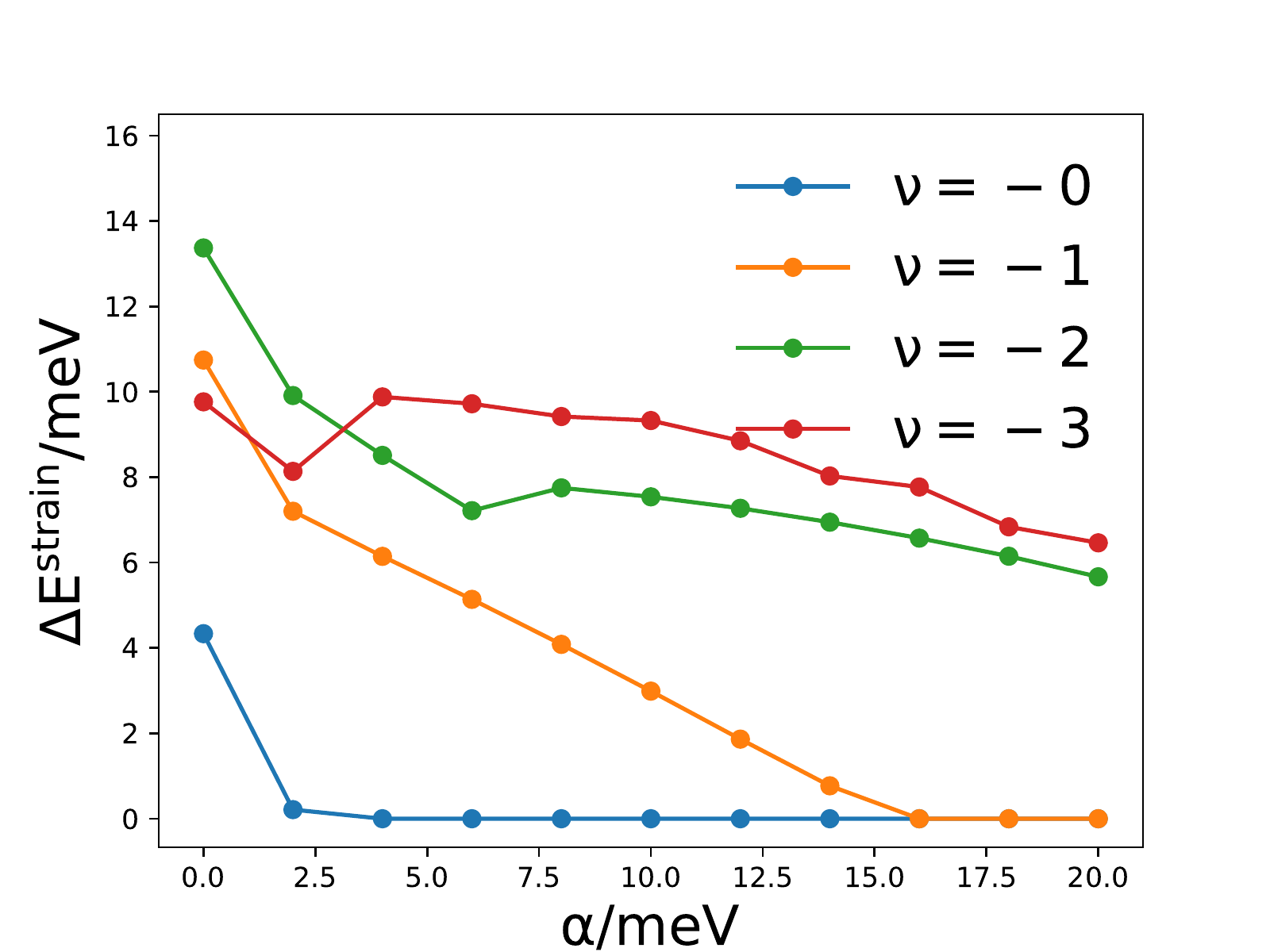}
     \caption{Energy difference $\Delta E^{strain}=E_{sym}^{strain}-E_{ord}^{strain}$ between the symmetric state that only breaks $C_{3z}$ symmetry ($E_{sym}^{strain}$) and the  ordered state ($E_{ord}^{strain}$) as a function of $\alpha$ - a parameter  characterizing the strain amplitude. We note that even at zero strain $\alpha=0$, a symmetric state that only breaks $C_{3z}$ symmetry has lower energy than the fully symmetric state. Thus $\Delta E^{strain}$ at $\alpha=0$ is smaller than the corresponding $\Delta E$ in Fig.~\ref{fig:doping}.}
     \label{fig:strain}
 \end{figure}
{\it Effects of strain---  } 
Since twisted bilayer graphene samples exhibit intrinsic strain~\cite{zhang2022electronic} and the ordered states are disfavored by strain, we investigate the effect of strain on the symmetric state of THF model via mean-field approach. We focus on $\nu=0,-1,-2,-3$ and introduce the following Hamiltonian~\cite{SM} that qualitatively characterizes the effect of strain
\ba  
\hH_{strain} = \alpha \sum_{\RR,\eta s}(f_{\RR,1\eta s}^\dag f_{\RR,2\eta s} +\text{h.c.}) 
\ea
where $\alpha$ is proportional to the strain amplitude (we leave the construction of a realistic strain Hamiltonian~\cite{koshino_22,Koshino_strain,vafek2022continuum} for a future study).  A non-zero $\alpha$ breaks the $C_{3z}$ symmetry but preserves all other symmetries~\cite{SM}. We compare the ground state energies of the symmetric states ($E^{strain}_{sym}$) and the ordered states ($E^{strain}_{ord}$) at non-zero strain. 
To obtain the symmetric state solution, we solve the mean-field equations by requiring the mean-field Hamiltonian to satisfy all symmetries except for the $C_{3z}$. We obtain the solution of the ordered states by initializing the mean-field calculations with the ordered ground states at zero strain and then perform self-consistent calculations. 
In Fig.~\ref{fig:strain}, we plot the difference between the ground state energies of the symmetric and the ordered states $\Delta E^{strain}= E_{sym}^{strain}-E^{strain}_{order}$ as a function of the effective strain amplitude $\alpha$ with $0\text{meV}<\alpha <20$meV at $\nu=0,-1,-2,-3$. We observe that $\Delta E$ at $\nu=0,-1$ vanishes at sufficiently large strain. A detailed analysis~\cite{SM} of the wavefunction shows that the ordered state cannot be stabilized and converged to a $C_{3z}$ broken symmetric solution at large strain. By further increasing strain, we find a symmetric state  at $\nu=-2$ 
 can also be stabilized at $\alpha \sim 45$meV (see SM~\cite{SM}).
We conclude that a 
symmetric phase can be stabilized by sufficiently large strain at $\nu=0,-1,-2$. As for $\nu=-3$, we mention that other ordered states, that break translational symmetry and have lower energy than the VP state, exist even at zero strain. We leave a systematical analysis of $\nu=-3$ for future study. Finally, we comment that even at zero strain, a symmetric state that breaks $C_{3z}$ symmetry has lower energy than the fully symmetric state that preserves all the symmetries (including $C_{3z}$). Therefore, $\Delta E^{strain}$ (energy difference between a symmetric state that only breaks $C_{3z}$ and an ordered state) at zero strain $\alpha=0$ in Fig.~\ref{fig:strain} is smaller than the corresponding $\Delta E$ (energy difference between a fully symmetric state and an ordered state) in Fig.~\ref{fig:doping}. 

{\it Discussion and summary---} 
Our main result is that an ordered state, instead of a SK state, will be the ground state of the system at integer filling $\nu=0,-1,-2,-3$. 
{Our mean-field calculations of THF model indicate ground state energy of the symmetric state is higher than the one of the ordered states at these fillings. Via DMFT calculations, we find the Kondo temperature to be substantially smaller than 2K. Thus, we conclude the Kondo effect is suppressed at integer filling $\nu=0,-1,-2,-3$, and the ground state is likely to be an ordered state. However, our mean-field calculations suggest doping can reduce the energy difference between the symmetric state and the ordered state enhancing the tendency towards the SK state. This has also been confirmed by the DMFT calculations which show a strong deviation from the Curie Weiss law at non-integer fillings $\nu=-0.5,-0.8,-1.2$ already around 10K. Furthermore, we show that a sufficiently large $C_{3z}$ breaking strain could also stabilize a symmetric state that only breaks the $C_{3z}$ symmetry at $\nu=0,-1,-2$. Therefore, we conclude both doping and strain enhance the Kondo effect and could, in principle, stabilize a SK state. 
Our results may explain 
the recent entropy experiments~\cite{Saito2021, ROZ21} which reveal a high-temperature phase with fluctuating moments and a low-temperature Fermi liquid phase with unpolarized isospins. This could be understood as a sign of  screening of the local moments and the development of the SK phase at low temperatures. 

{As far as the SK state is concerned, we have performed a systematic study of its band structure and topology. Via the mean-field approach, we successfully identified the SK state in the KL model, and a symmetric state, that is adiabatically connected to the SK state, in the THF model. 
For the SK state in the KL model, we find that the $\Gamma_3$ states near the $\Gamma_M$ point have been pushed away, and the bandwidth of the flat bands is enlarged at $\nu=-1,-2$. The hybridization between $f$-electrons and $\Gamma_3$ $c$-electons is enhanced by the Kondo interactions. Consequently, the flat bands are mostly formed by $\Gamma_1\oplus \Gamma_2$ $c$-electrons. The topology of the flat bands remains the same as in the non-interacting case. However, for the symmetric state in the THF model, the enhanced $f$-$c$ hybridization does not appear. We mention that the mean-field solution of the symmetric state in the THF model underestimates the correlation effect, which could be the origin of the weak $f$-$c$ hybridization. We expect introducing a Gutzwiller projector will give a more precise description of the symmetric state in the THF model.}

{\it Note added---}
After finishing this work, we learned that related, but not identical, results have also recently been obtained by the S. Das Sarma's ~\cite{das_sarma_kondo},
P. Coleman's~\cite{Coleman_kondo}, 
and Z. Song's groups~\cite{song_kondo}. We also mention that results from Z. Song's group are compatible with our DMFT results.

{\it Acknowledgements---}
B. A. B.'s work was primarily supported by the DOE Grant No. DE-SC0016239, the Simons Investigator Grant No. 404513. H. H. was supported by the European Research Council (ERC) under the European Union’s
Horizon 2020 research and innovation program (Grant Agreement No. 101020833). J. H. A. was supported by a Hertz Fellowship. 
D. C. was supported by the DOE Grant No. DE-SC0016239 and the Simons Investigator Grant No. 404513.
A. M. T. was supported by the Office of Basic Energy Sciences, Material Sciences and Engineering Division, U.S. Department of Energy (DOE) under Contract No. DE-SC0012704. 
G. R, L. K, T. W., G. S. and R. V. thank the Deutsche Forschungsgemeinschaft (DFG,
German Research Foundation) for funding through QUAST FOR
5249-449872909 (Projects P4 and P5). G.R. acknowledges funding from the European Commission via the Graphene
Flagship Core Project 3 (grant agreement ID: 881603). T.W. acknowledges support by the Cluster of Excellence “CUI: Advanced Imaging of Matter” of the Deutsche Forschungsgemeinschaft (DFG)–EXC 2056–Project ID No. 390715994.

\bibliography{ref,DMFT_refs}

\clearpage 
\onecolumngrid
\begin{center}
\textbf{Supplementary Materials}
\end{center}

\newcommand{\hh}{\textcolor{black}}
\newenvironment{hhc}{\par\color{black}}{\par}
\newenvironment{hbb}{\par\color{black}}{\par}

\newcommand{\hhb}{\textcolor{black}}

\newcommand{\hb}{\textcolor{black}}

\newcommand{\bh}{\textcolor{black}}
\renewcommand{\thefigure}{S\arabic{figure}}

\renewcommand{\thetable}{S\arabic{table}}

\renewcommand{\thesection}{S\arabic{section}}

\renewcommand{\theequation}{S\arabic{equation}}

\tableofcontents

\clearpage

\section{Toplogical heavy-fermion model} 
The topological heavy-fermion (THF) model introduced in Ref.~\cite{HF_MATBLG} takes the following Hamiltonian 
\baa  
\hH = \hH_c  +\hH_{fc} + \hH_U +\hH_J + \hH_W + \hH_V +\hH_{\mu}
\label{eq:thf_ham}
\eaa  
The single-particle Hamiltonian of conduction $c$-electrons has the form of 
\baa 
 &\hat{H}_c = \sum_{\eta,s,a,a',|\kk |<\Lambda_c} 
 H_{a,a'}^{(c,\eta)}(\kk )c_{\kk a\eta s}^\dag c_{\kk a'\eta s} 
 \quad ,\quad 
 H^{(c,\eta)}(\kk ) = \begin{bmatrix}
 0_{2\times 2 } & v_\star(\eta k_x\sigma_0+ik_y\sigma_z)\\
 v_\star(\eta k_x\sigma_0-ik_y\sigma_z) & M\sigma_x .
 \end{bmatrix}  
 \label{eq:hc_def}
\eaa  
where $\sigma_{0,x,y,z}$ are identity and Pauli matrices.
$c_{\kk a\eta s}$ represents the annihilation operator of the $a(=1,2,3,4)$-th conduction band basis of the valley $\eta (=\pm)$ and spin $s(=\up,\dn)$ at the moir\'e momentum $\kk$. At $\Gamma_M$ point ($\kk=0$) of the moir\'e Brillouin zone (MBZ), $c_{\kk 1 \eta s},c_{\kk 2 \eta s}$ form a $\Gamma_3$ irreducible representation (of $P6'2'2$ group),
$c_{\kk 3 \eta s},c_{\kk 4 \eta s}$ form a $\Gamma_1\oplus \Gamma_2$ reducible (into $\Gamma_1$ and $ \Gamma_2$ - as they are written, the  $c_{\kk 3 \eta s},c_{\kk 4 \eta s}$ are just the $\sigma_x$ linear combinations of $\Gamma_1 \pm  \Gamma_2$ ) representation (of $P6'2'2$ group). $\Lambda_c$ is the momentum cutoff for the $c$-electrons. $N$ is the total number of moir\'e unit cells. 
The parameter values are 
$v_\star = -4.303 \mathrm{eV}\cdot\mathrm{\mathring{A}}$, $M=3.697$meV.  

The hybridization between $f$ and $c$ electrons has the form of 
\baa  
\hat{H}_{fc} =\frac1{\sqrt{N_M}} \sum_{\substack{|\kk|<\Lambda_c\\ \RR}} \sum_{\alpha a \eta s} \bigg( e^{i\kk\cdot\RR -\frac{|\kk|^2\lambda^2}2} H^{(fc,\eta)}_{\alpha a} (\kk)  f_{\RR \alpha\eta s}^\dag c_{\kk a\eta s} + h.c. \bigg) \ ,
\label{eq:h_fc}
\eaa  
where $f_{\RR \alpha\eta s}$ represents the annihilation operators of the $f$ electrons with orbital index $\alpha(=1,2)$, valley index $\eta(=\pm)$ and spin $s(=\up,\dn)$ at the moir\'e unit cell $\RR$. $N_M$ is the number of moir\'e unit cells and $\lambda=0.3376a_M$ is the damping factor, where $a_M$ is the moir\'e lattice constant.
The hybridization matrix $H^{(fc,\eta)}$ has the form of 
\begin{equation} 
   H^{(fc,\eta)}(\kk) = \begin{pmatrix}
     \gamma \sigma_0 + v_\star'(\eta k_x\sigma_x + k_y\sigma_y), & 0_{2\times 2} 
    \end{pmatrix}  
    \label{eq:def_fc_mat}
\end{equation}
which describe the hybridization between $f$ electrons and $\Gamma_3$ $c$ electrons $(a=1,2)$. The parameter values are $\gamma=-24.75$meV, $v_\star'=1.622 \mathrm{eV}\cdot\mathrm{\mathring{A}}$.

$\hH_{U}$ ($U = 57.89\mathrm{meV}$) describes the on-site interactions of $f$-electrons.
    \begin{equation}
    \hH_{U} =  \frac{U}2 \sum_{\RR} :n^f_{\RR}: :n^f_{\RR}:  ,
    \label{eq:hu_def}
    \end{equation} 
     where $n^f_{\RR} = \sum_{\alpha\eta s} f_{\RR \alpha\eta s}^\dagger f_{\RR \alpha\eta s}$ is the $f$-electrons density and the colon symbols represent the normal ordered operator with respect to the normal state: $:f^\dagger_{\RR\alpha_1 \eta_1 s_1} f_{\RR \alpha_2 \eta_2 s_2}: = f^\dagger_{\RR\alpha_1 \eta_1 s_1} f_{\RR \alpha_2 \eta_2 s_2} - \frac{1}{2} \delta_{\alpha_1 \eta_1 s_1; \alpha_2 \eta_2 s_2}$. 

The ferromagnetic exchange interaction between $f$ and $c$ electrons $\hH_J$ is defined as 
\baa  
 H_J = - \frac{J}{2N_M} \sum_{\RR s_1 s_2} \sum_{\alpha\alpha'\eta\eta'} \sum_{|\kk_1|,|\kk_2|<\Lambda_c }   
     e^{i( \kk_1- \kk_2 )\cdot\RR } 
     ( \eta\eta' + (-1)^{\alpha+\alpha'} )
     :f_{\RR \alpha \eta s_1}^\dagger f_{\RR \alpha' \eta' s_2}:  :c_{\kk_2, \alpha'+2, \eta' s_2}^\dagger  c_{ \kk_1, \alpha+2, \eta s_1}: 
     \label{eq:hj_def}
\eaa  
where $J=16.38$meV and $ :c_{\kk_2, \alpha'+2, \eta' s_2}^\dagger  c_{ \kk_1, \alpha+2, \eta s_1}: =  c_{\kk_2, \alpha'+2, \eta' s_2}^\dagger  c_{ \kk_1, \alpha+2, \eta s_1}-\frac{1}{2}\delta_{\kk_1,\kk_2}\delta_{\alpha,\alpha'}\delta_{\eta,\eta'} \delta_{s_1,s_2} $

The repulsion between $f$ and $c$ electrons $\hH_W$ has the form of 
\baa  
&\hat{H}_W = \sum_{\eta,s,\eta',s',a,\alpha}\sum_{|\kk |<\Lambda_c, |\bm{\kk+\qq}|<\Lambda_c } 
W_ae^{-i\qq \cdot \RR} :f_{\RR, a\eta s}^\dag f_{\RR ,a\eta s}:
:c_{\kk+\qq, a\eta's'}^\dag c_{\kk,a\eta's'} :
\label{eq:hw_def}
\eaa 
where we take $W_1=W_2=44.03$meV and $W_3=W_4=50.20$meV.

The Coulomb interaction between $c$ electrons has the form of
\baa  
&     \hH_{V} = \frac{1}{2\Omega_0 N} \sum_{\eta_1 s_1 a_1} \sum_{\eta_2 s_2 a_2} \sum_{|\kk_1|, |\kk_2|<\Lambda_c} \sum_{\substack{\qq \\ |\kk_1+\qq|,|\kk_2+\qq|<\Lambda_c}} V(\qq) 
        :c_{\kk_1 a_1\eta_1 s_1}^\dagger c_{\kk_1+\qq a_1 \eta_1 s_1}:
        :c_{\kk_2+\qq a_2 \eta_2 s_2}^\dagger c_{\kk_2 a_2 \eta_2 s_2}:
        \label{eq:hv_def}
\eaa 
where $\Omega_0$ is the area of the moir\'e unit cell and $V(\qq=0)/\Omega_0= 48.33\mathrm{meV}$. 
We will always treat $\hH_V$ at mean-field level (int both the THF model and the Kondo lattice (KL) model)~\cite{HF_MATBLG}
\baa  
\hH_V \approx \frac{V(0)}{\Omega_0} \nu_c \sum_{|\kk|<\Lambda_c, a , \eta , s} c_{\kk,a\eta s}^\dag c_{\kk, a\eta s} - \frac{V(0)}{2\Omega_0}N_M\nu_c^2 +\frac{V(0)}{\Omega_0} \sum_{|\kk|<\Lambda_c} 8\nu_c 
\label{eq:mf_hV}
\eaa 
where $\nu_c$ is the filling of $c$ electrons 
$
\nu_c = \frac{1}{N_M} \sum_{|\kk|<\Lambda_c, a , \eta , s}\langle  \Psi| :c_{\kk,a\eta s}^\dag c_{\kk, a\eta s}: |\Psi \rangle 
$
with $|\psi\rangle$ the ground state. 

Finally, we introduce a chemical potential term
\baa 
\hH_\mu = -\mu \sum_{|\kk|<\Lambda_c,a\eta s}c_{\kk,a\eta s}^\dag c_{\kk,a \eta s} -\mu \sum_{\RR, \alpha \eta s}f_{\RR,\alpha \eta s}^\dag f_{\RR,\alpha \eta s}\,. 
\label{eq:thf_mu}
\eaa

\section{Kondo lattice model} 
The Kondo lattice model is derived by performing a generalized Schrieffer-Wolff (SW) transformation on the topological heavy fermion model (detailed derivation in Ref.~\cite{Spin_MATBLG}). The Hamiltonian has the form of 
\baa 
\hH_{Kondo} = \hH_{c}  +\hH_V +\hH_W +  \hH_J + \hH_{K} +\hH_{cc} -\hH_{\mu_c}
\label{eq:kondo_ham_sm}
\eaa  
where $\hH_c$ (Eq.~\ref{eq:hc_def}), $\hH_V$ (Eq.~\ref{eq:hv_def}) and $\hH_W$ (Eq.~\ref{eq:hw_def}) and $\hH_J$ (Eq.~\ref{eq:hj_def}) come from the original TFH model. The Kondo interactions and the one-body scattering term are 
\baa  
\hH_K =&\sum_{\RR, |\kk| <\Lambda_c,|\kk'| <\Lambda_c}
\sum_{\alpha,\alpha',a,a',\eta,\eta',s,s'}
\frac{ e^{i(\kk-\kk') \RR - |\kk|^2 \lambda^2/2 - |\kk'|^2\lambda^2/2}}{N_MD_{\nu_c,\nu_f}} 
:f_{\RR,\alpha \eta s}^\dag  f_{\RR,\alpha'\eta' s'}: :c_{\kk',a'\eta' s'}^\dag c_{\kk, a\eta s}: \nonumber \\ 
&
\bigg[ 
\gamma ^2 \delta_{\alpha',a'} \delta_{\alpha, a}
 +
\gamma v_\star^\prime \delta_{\alpha, a}[\eta' k'_x\sigma_x - k'_y\sigma_y]_{\alpha'a'}
 +
\gamma v_\star^\prime \delta_{\alpha', a'}[\eta k_x\sigma_x + k_y\sigma_y]_{\alpha a}
\bigg] \nonumber  \\
\eaa 
and 
\baa   
\hH_{cc} = 
 &-\sum_{|\kk| <\Lambda_c,\eta,s}\sum_{a,a'\in\{1,2\}} e^{-|\kk|^2\lambda^2}
\bigg( 
\frac{1}{D_{1,\nu_c,\nu_f}}+\frac{1}{D_{2,\nu_c,\nu_f}}
\bigg)
\begin{bmatrix}
\gamma^2/2 & \gamma v_\star^\prime (\eta k_x -ik_y)   \\
\gamma v_\star^\prime (\eta k_x + ik_y) &\gamma^2/2
\end{bmatrix}_{a,a'}
:c_{\kk,a\eta s}^\dag c_{\kk, a'\eta s}: 
\label{eq:hcc} 
\eaa  
where 
\baa  
&D_{1,\nu_c,\nu_f}=
(U-W)\nu_f -\frac{U}{2}  +(\frac{-V_0}{\Omega_0}+W){\nu}_c
\quad,\quad   D_{2,\nu_c,\nu_f}=(U-W)\nu_f +\frac{U}{2}   +(\frac{-V_0}{\Omega_0}+W){\nu}_c \nonumber \\
& D_{\nu_c,\nu_f} = \bigg[-\frac{1}{D_{1,\nu_c,\nu_f}} +\frac{1}{D_{2,\nu_c,\nu_f} } \bigg]^{-1}
\,.
\eaa 
We point out that, at $\nu=\nu_f=\nu_c=0$, $D_{1,\nu_c,\nu_f} = -D_{2,\nu_c,\nu_f}$ and the on-body term $\hH_{cc}(=0)$ vanishes.

We note that in the Kondo model the filling of $f$ electron at each site is fixed to be $\nu_f$. Then we can replace $\sum_{\alpha \eta s}:f_{\RR,\alpha\eta s}^\dag f_{\RR,\alpha\eta s}: $ with $\nu_f$ and $\hH_W$ becomes 
\baa  
\hH_W = \sum_{|\kk| < \Lambda_c,|\kk'| < \Lambda_c, a \eta s }\sum_{\QQ} W \nu_f :c_{\kk,a \eta' s'}^\dag c_{\kk', a \eta s }\delta_{\kk, \kk'+\QQ}
\eaa  
where $\QQ \in \{ m \bm{b}_{M1}+m \bm{b}_{M2}|m,n\in \mathbb{Z} \} $ and $\mathbf{b}_{M1}$,  $\mathbf{b}_{M2}$ are the reciprocal lattice vectors. If we focus on the conduction electrons within the first MBZ, we can replace $\delta_{\kk,\kk'+\QQ}$ by $\delta_{\kk,\kk'}$ and 
\baa  
\hH_W = \sum_{|\kk| < \Lambda_c, a \eta s }\sum_{\QQ} W \nu_f :c_{\kk,a \eta' s'}^\dag c_{\kk, a \eta s } 
\eaa  
which is a chemical shift of conduction electrons. We also set $W_1=W_2=W_3=W_4=W=47.12$meV in $\hH_W$ to simplify the SW transformation. The realistic values of $W_{1,2,3,4}$ are not identical but the difference is about $15\%$.  


Finally, we introduce a chemical potential $\mu_c$ to tune the filling of the system 
\baa 
\hH_{\mu_c} = -\mu_c \sum_{|\kk|<\Lambda_c, a\eta s} :c_{\kk,a\eta s}^\dag c_{\kk,a \eta s }:
\label{eq:hmuc}
\eaa

\section{Symmetry}
We now provide the symmetry transformation of electron operators. For a given symmetry operation $g$, we let $D^f(g), D^{c'}(g), D^{c''}(g)$ denote the  representation matrix of $f$-, $\Gamma_3$ $c$- and $\Gamma_1\oplus \Gamma_2$ $c$-electrons:
\baa  
&g f_{\RR,\alpha \eta s}^\dag g^{-1} = \sum_{\alpha'\eta's'} f_{g\RR,\alpha'\eta's'}^\dag  D^f(g)_{\alpha'\eta' s',\alpha \eta s}  \nonumber \\
&g c_{\kk,a \eta s}^\dag g^{-1} = \sum_{a' \in \{1,2\}, \eta's'} c_{g\kk,a'\eta's'}^\dag  D^{c'}(g)_{a'\eta' s',a \eta s},\quad a\in\{1,2\}  \nonumber \\
&g c_{\kk,a \eta s}^\dag g^{-1} = \sum_{a' \in \{3,4\}, \eta's'} c_{g\kk,a'\eta's'}^\dag  D^{c''}(g)_{a'+2\eta' s',a+2 \eta s},\quad a\in\{3,4\} 
\eaa  
We consider the following symmetry operations as given in Ref.~\cite{HF_MATBLG}. 
\baa  
T,C_{3z},C_{2x},C_{2z}T
\label{eq:desc_sym_op}
\eaa  
with the following representation matrices
\baa  
T&:\quad D^f(T) = \sigma_0\tau_x \varsigma_0,\quad 
D^{c'}(T) = \sigma_0\tau_x \varsigma_0,\quad D^{c''}(T) = \sigma_0\tau_x \varsigma_0
\nonumber \\ 
C_{3z}&:\quad D^f(C_{3z}) = e^{i\frac{2\pi}{3}\sigma_z\tau_z} \varsigma_0,\quad  D^{c'}(C_{3z}) = e^{i\frac{2\pi}{3}\sigma_z\tau_z} \varsigma_0
,\quad D^{c''}(C_{3z}) = \sigma_0\tau_0 \varsigma_0
\nonumber \\ 
C_{2x}&:\quad D^f(C_{2x}) = \sigma_x\tau_0 \varsigma_0,\quad D^{c'}(C_{2x}) = \sigma_x\tau_0 \varsigma_0 ,
\quad 
D^{c''}(C_{2x}) = \sigma_x\tau_0 \varsigma_0
\nonumber \\ 
C_{2z}T&:\quad D^f(C_{2x}T) = \sigma_x\tau_0 \varsigma_0,\quad  D^{c'}(C_{2x}T) = \sigma_x\tau_0 \varsigma_0,\quad 
D^{c''}(C_{2z}T) = \sigma_x\tau_0 \varsigma_0
\label{eq:desc_sym}
\eaa  
where $\sigma_{x,y,z,0}, \tau_{x,y,z,0},\varsigma_{x,y,z,0}$ are Pauli or identity matrices of orbital, valley and spin degrees of freedom respectively. 

At $M\ne 0, v_\star^\prime \ne 0$, we also have $U(1)_c$ charge symmetry, $U(1)_v$ valley symmetry and $SU(2)_\eta$ spin symmetry for each valley $\eta$. We also mention that at $M=0$, we have an enlarged flat $U(4)$ symmetry and at $v_\star^\prime =0$ we have an enlarged chiral $U(4)$ symmetry~\cite{HF_MATBLG,Spin_MATBLG}. At $M=0,v_\star^\prime =0$, we have a $U(4)\times U(4)$ symmetry~\cite{HF_MATBLG,Spin_MATBLG}. 
Here, we consider the case of $M\ne 0, v_\star^\prime \ne 0$, where we only have a $U(1)_c\times U(1)_v\times SU(2)_{\eta=+}\times SU(2)_{\eta=-}$ symmetry. We comment that $M=3.698$meV is relatively small and we have an approximate flat $U(4)$ symmetry. 
Under $U(1)_c$ transformation $g_{U(1)_c}({\theta_c})$ (characterized by a real number $\theta_c$), 
$U(1)_v$ transformation $g_{U(1)_v}({\theta_v})$ (characterized by a real number $\theta_v$) 
and $SU(2)_\eta$ spin transformation $g_{SU(2)_\eta}({\theta^\mu_\eta})$
(characterized by three real numbers $\theta^\mu_\eta, \mu \in \{x,y,z\}$ ), we have 
\baa  
U(1)_c:&\quad D^f( g_{U(1)_c}((\theta_c) ) = e^{-i\theta_c }\sigma_0\tau_0\varsigma_0 ,\quad
 D^{c'}( g_{U(1)_c}((\theta_c) ) = e^{-i\theta_c }\sigma_0\tau_0\varsigma_0 ,\quad 
 D^{c''}( g_{U(1)_c}((\theta_c) ) = e^{-i\theta_c }\sigma_0\tau_0\varsigma_0 \nonumber \\
U(1)_v:&\quad D^f( g_{U(1)_v}((\theta_v) ) = \sigma_0e^{-i\theta_v \tau_z  }\varsigma_0 ,\quad
 D^{c'}( g_{U(1)_v}((\theta_v) ) = \sigma_0e^{-i\theta_v \tau_z  }\varsigma_0 ,\quad 
 D^{c''}( g_{U(1)_v}((\theta_v) ) = \sigma_0e^{-i\theta_v \tau_z  }\varsigma_0 \nonumber \\
SU(2)_{\eta}:&\quad  D^{f}(g_{SU(2)_\eta}(\theta_\eta^\mu) )=\sigma_0 e^{ -i \sum_\mu \theta_\mu^\eta  
\frac{\tau_0+\eta \tau_z}{4} \varsigma_\mu },\quad  
D^{c'}(g_{SU(2)_\eta}(\theta_\eta^\mu) )=\sigma_0 e^{ -i \sum_\mu \theta^\eta_\mu  
\frac{\tau_0+\eta \tau_z}{4} \varsigma_\mu }, \quad  \nonumber \\
&\quad 
D^{c''}(g_{SU(2)_\eta}(\theta_\eta^\mu) )=\sigma_0 e^{ -i \sum_\mu \theta_\mu^\eta
\frac{\tau_0+\eta \tau_z}{4} \varsigma_\mu }
\label{eq:cont_sym}
\eaa

\section{Mean-field solutions of the Kondo lattice model}
The Kondo Hamiltonian in Eq.~\ref{eq:kondo_ham_sm} contains two single-particle term $\hH_c$ and $\hH_{cc}$ and two interaction terms $\hH_K+\hH_J$. We now discuss the mean-field decoupling of $\hH_K,\hH_J$. 

\subsection{Mean-field decoupling of $\hH_K$}
\label{sec:kondo_mf_hk}
We treat the interaction terms via mean-field decoupling 
\baa  
\hH_K \approx &\hH_K^{MF} \nonumber \\ 
= &
\sum_{\RR, |\kk| <\Lambda_c,|\kk'| <\Lambda_c}
\sum_{\alpha,\alpha',a,a',\eta,\eta',s,s'}
\frac{ e^{i(\kk-\kk') \RR - |\kk|^2 \lambda^2/2 - |\kk'|^2\lambda^2/2}}{N_MD_{\nu_c,\nu_f}} \nonumber \\
&\bigg[ 
\gamma ^2 \delta_{\alpha',a'} \delta_{\alpha, a}
 +
\gamma v_\star^\prime \delta_{\alpha, a}[\eta' k'_x\sigma_x - k'_y\sigma_y]_{\alpha'a'}
 +
\gamma v_\star^\prime \delta_{\alpha', a'}[\eta k_x\sigma_x + k_y\sigma_y]_{\alpha a} 
\bigg] \nonumber \\
&\bigg\{ 
\langle f_{\RR,\alpha \eta s}^\dag c_{\kk, a\eta s} \rangle \langle c_{\kk',a'\eta' s'}^\dag f_{\RR,\alpha'\eta' s'}\rangle  
- \langle f_{\RR,\alpha \eta s}^\dag c_{\kk, a\eta s} \rangle c_{\kk',a'\eta' s'}^\dag f_{\RR,\alpha'\eta' s'}
- f_{\RR,\alpha \eta s}^\dag c_{\kk, a\eta s}  \langle c_{\kk',a'\eta' s'}^\dag f_{\RR,\alpha'\eta' s'}\rangle  \nonumber \\ 
&- 
\langle :f_{\RR,\alpha \eta s}^\dag  f_{\RR,\alpha'\eta' s'}: \rangle \langle :c_{\kk',a'\eta' s'}^\dag c_{\kk, a\eta s}:\rangle 
+ \langle :f_{\RR,\alpha \eta s}^\dag  f_{\RR,\alpha'\eta' s'}: \rangle :c_{\kk',a'\eta' s'}^\dag c_{\kk, a\eta s}:
+  :f_{\RR,\alpha \eta s}^\dag  f_{\RR,\alpha'\eta' s'}: 
\langle :c_{\kk',a'\eta' s'}^\dag c_{\kk, a\eta s}:\rangle \bigg\} 
\eaa 
where for an operator $O$, $\langle O\rangle = \langle \Psi|O|\Psi\rangle $ with $|\Psi\rangle$ the mean-field ground state. 
\subsubsection{Fock term}
We first consider the Fock term (F.T.), which takes the form of 
\baa  
\text{F.T.}\nonumber 
=&\sum_{\RR, |\kk| <\Lambda_c,|\kk'| <\Lambda_c}
\sum_{\alpha,\alpha',a,a',\eta,\eta',s,s'}
\frac{ e^{i(\kk-\kk') \RR - |\kk|^2 \lambda^2/2 - |\kk'|^2\lambda^2/2}}{N_MD_{\nu_c,\nu_f}} \nonumber \\
&\bigg[ 
\gamma ^2 \delta_{\alpha',a'} \delta_{\alpha, a}
 +
\gamma v_\star^\prime \delta_{\alpha, a}[\eta' k'_x\sigma_x - k'_y\sigma_y]_{\alpha'a'}
 +
\gamma v_\star^\prime \delta_{\alpha', a'}[\eta k_x\sigma_x + k_y\sigma_y]_{\alpha a} 
\bigg] \nonumber \\
&\bigg\{ 
\langle f_{\RR,\alpha \eta s}^\dag c_{\kk, a\eta s} \rangle \langle c_{\kk',a'\eta' s'}^\dag f_{\RR,\alpha'\eta' s'}\rangle  
- \langle f_{\RR,\alpha \eta s}^\dag c_{\kk, a\eta s} \rangle c_{\kk',a'\eta' s'}^\dag f_{\RR,\alpha'\eta' s'}
- f_{\RR,\alpha \eta s}^\dag c_{\kk, a\eta s}  \langle c_{\kk',a'\eta' s'}^\dag f_{\RR,\alpha'\eta' s'}\rangle \bigg\} \nonumber \\
= &
\sum_{\RR}
\frac{ 1 }{D_{\nu_c,\nu_f}}\bigg\{
\gamma ^2  \langle \sum_{|\kk|<\Lambda_c}\sum_{\alpha\eta s}\frac{ e^{i\kk\cdot\RR- |\kk|^2\lambda^2/2}}{\sqrt{N_M}} f_{\RR,\alpha \eta s}^\dag c_{\kk, a\eta s} \rangle
\langle \sum_{|\kk'|<\Lambda_c}\sum_{\alpha'\eta's'}\frac{e^{-i\kk'\cdot \RR -|\kk'|^2\lambda^2/2} }{\sqrt{N_M}} c_{\kk',\alpha'\eta's'}^\dag f_{\RR,\alpha'\eta's'}\rangle   \nonumber \\ 
&- \bigg[ \gamma ^2 \sum_{|\kk|<\Lambda_c}\sum_{\alpha\eta s}\frac{ e^{i\kk\cdot\RR- |\kk|^2\lambda^2/2}}{\sqrt{N_M}} f_{\RR,\alpha \eta s}^\dag c_{\kk, a\eta s} 
\langle \sum_{|\kk'|<\Lambda_c}\sum_{\alpha'\eta's'}\frac{e^{-i\kk'\cdot \RR -|\kk'|^2\lambda^2/2} }{\sqrt{N_M}} c_{\kk',\alpha'\eta's'}^\dag f_{\RR,\alpha'\eta's'}\rangle  + \text{h.c.} \bigg] \nonumber \\ 
 &+\gamma v_\star^\prime \langle \sum_{|\kk|<\Lambda_c}\sum_{\alpha \eta s}\frac{e^{i\kk\cdot\RR -|\kk|^2\lambda^2/2} }{\sqrt{N_M}}f_{\RR,\alpha \eta s}^\dag c_{\kk,\alpha \eta s} \rangle \langle \sum_{|\kk'|<\Lambda_c}\sum_{a'\alpha'\eta's'} 
 \frac{e^{-i\kk'\cdot \RR -|\kk'|^2\lambda^2/2} [\eta' k'_x\sigma_x - k'_y\sigma_y]_{\alpha'a'}}{\sqrt{N_M}} c_{\kk',a'\eta's'}^\dag f_{\RR,\alpha'\eta's'} 
 \rangle \nonumber \\ 
  &+\gamma v_\star^\prime \langle \sum_{|\kk|<\Lambda_c}\sum_{\alpha a \eta s}\frac{e^{i\kk\cdot\RR -|\kk|^2\lambda^2/2}
  [\eta k_x\sigma_x + k_y\sigma_y]_{\alpha a}} 
  {\sqrt{N_M}}f_{\RR,\alpha \eta s}^\dag c_{\kk,\alpha \eta s} \rangle \langle \sum_{|\kk'|<\Lambda_c}\sum_{\alpha'\eta's'} 
 \frac{e^{-i\kk'\cdot \RR -|\kk'|^2\lambda^2/2}}{\sqrt{N_M}} c_{\kk',\alpha'\eta's'}^\dag f_{\RR,\alpha'\eta's'} 
 \rangle \nonumber \\ 
 &-\gamma v_\star^\prime  \bigg[ \sum_{|\kk|<\Lambda_c}\sum_{\alpha a \eta s}\frac{e^{i\kk\cdot\RR -|\kk|^2\lambda^2/2}
  [\eta k_x\sigma_x + k_y\sigma_y]_{\alpha a}} 
  {\sqrt{N_M}}f_{\RR,\alpha \eta s}^\dag c_{\kk,\alpha \eta s}  \langle \sum_{|\kk'|<\Lambda_c}\sum_{\alpha'\eta's'} 
 \frac{e^{-i\kk'\cdot \RR -|\kk'|^2\lambda^2/2}}{\sqrt{N_M}} c_{\kk',\alpha'\eta's'}^\dag f_{\RR,\alpha'\eta's'} 
 \rangle \nonumber \\
 &+\sum_{|\kk|<\Lambda_c}\langle \sum_{\alpha a \eta s}\frac{e^{i\kk\cdot\RR -|\kk|^2\lambda^2/2}
  [\eta k_x\sigma_x + k_y\sigma_y]_{\alpha a}} 
  {\sqrt{N_M}}f_{\RR,\alpha \eta s}^\dag c_{\kk,\alpha \eta s}  \rangle \sum_{|\kk'|<\Lambda_c}\sum_{\alpha'\eta's'} 
 \frac{e^{-i\kk'\cdot \RR -|\kk'|^2\lambda^2/2}}{\sqrt{N_M}} c_{\kk',\alpha'\eta's'}^\dag f_{\RR,\alpha'\eta's'}  +\text{h.c.} \bigg]
 \bigg\} 
 \label{eq:FT_int}
\eaa 
We introduce the following mean-field expectation values 
\baa  
&V_1 = \sum_{\RR, |\kk| <\Lambda_c}
\sum_{\alpha\eta s}\frac{e^{i\kk \cdot \RR -|\kk|^2 \lambda^2/2}  }{N_M\sqrt{N_M}} \langle \Psi| f_{\RR,\alpha \eta s}^\dag c_{\kk,\alpha\eta s} |\Psi\rangle  \nonumber \\
&V_2 = \sum_{\RR, |\kk| <\Lambda_c}
\sum_{\alpha a \eta s}\frac{ e^{i\kk \cdot \RR -|\kk|^2 \lambda^2/2} }{N_M\sqrt{N_M}} (\eta k_x \sigma_x +k_y\sigma_y)_{\alpha a}
\langle \Psi| 
f_{\RR,\alpha \eta s}^\dag c_{\kk,a \eta s} |\Psi\rangle  \label{eq:sc_eq_1_v} 
\eaa 
and assume the ground state is translational invariant such that
\baa  
&\sum_{|\kk| <\Lambda_c}\sum_{\alpha\eta s}\frac{e^{i\kk \cdot \RR -|\kk|^2 \lambda^2/2}  }{\sqrt{N_M}} \langle \Psi| f_{\RR,\alpha \eta s}^\dag c_{\kk,\alpha\eta s} |\Psi\rangle  = 
\frac{1}{N_M}\sum_{\RR, |\kk|<\Lambda_c}\sum_{\alpha\eta s}\frac{e^{i\kk \cdot \RR -|\kk|^2 \lambda^2/2}  }{\sqrt{N_M}} \langle \Psi| f_{\RR,\alpha \eta s}^\dag c_{\kk,\alpha\eta s} |\Psi\rangle  = V_1 \nonumber \\ 
&\sum_{|\kk| <\Lambda_c}
\sum_{\alpha a \eta s}\frac{ e^{i\kk \cdot \RR -|\kk|^2 \lambda^2/2} }{\sqrt{N_M}} (\eta k_x \sigma_x +k_y\sigma_y)_{\alpha a}
\langle \Psi| 
f_{\rr,\alpha \eta s}^\dag c_{\kk,\alpha\eta s} |\Psi\rangle 
\nonumber \\
=&\frac{1}{N_M}\sum_{\RR,|\kk| <\Lambda_c}
\sum_{\alpha a \eta s}\frac{ e^{i\kk \cdot \RR -|\kk|^2 \lambda^2/2} }{\sqrt{N_M}} (\eta k_x \sigma_x +k_y\sigma_y)_{\alpha a}
\langle \Psi| 
f_{\rr,\alpha \eta s}^\dag c_{\kk,\alpha\eta s} |\Psi\rangle 
= V_2 
\eaa 

Then the Fock term (Eq.~\ref{eq:FT_int}) becomes
\baa  
\text{F.T.} =& - \frac{\gamma^2}{D_{\nu_c,\nu_f}}\sum_{\RR, |\kk| <\Lambda_c}
\sum_{\alpha\eta, s}\frac{e^{i\kk \cdot \RR -|\kk|^2 \lambda^2/2} }{\sqrt{N_M}} \bigg(  V^* _{1}f_{\RR,\alpha \eta s}^\dag c_{\kk,\alpha\eta s} +\text{h.c.}
\bigg) 
  + \frac{N_M\gamma^2 |V_1|^2 }{D_{\nu_c,\nu_f}}\nonumber \\ 
& - \frac{\gamma v_\star^\prime }{D_{\nu_c,\nu_f}}\sum_{\RR, |\kk| <\Lambda_c}
\sum_{\alpha,a,\eta, s}\frac{e^{i\kk \cdot \RR -|\kk|^2 \lambda^2/2} }{\sqrt{N_M}}   \bigg( V^* _{1} (\eta k_x \sigma_x +k_y\sigma_y)_{\alpha a}f_{\RR,\alpha \eta s}^\dag c_{\kk,\alpha\eta s} +\text{h.c.}\bigg) 
 + \frac{N_M\gamma v_\star^\prime V_1^* V_2 }{D_{\nu_c,\nu_f}}\nonumber \\ 
&
- \frac{\gamma v_\star^\prime }{D_{\nu_c,\nu_f}}\sum_{\RR, |\kk| <\Lambda_c}
\sum_{\alpha, \eta, s}\frac{e^{i\kk \cdot \RR -|\kk|^2 \lambda^2/2} }{\sqrt{N_M}}
\bigg( V^* _{2}f_{\RR,\alpha \eta s}^\dag c_{\kk,\alpha\eta s}  +\text{h.c.}\bigg) +\frac{N_M\gamma v_\star^\prime V_2^* V_1 }{D_{\nu_c,\nu_f}}
\label{eq:hk_fock}
\eaa  

\subsubsection{Hartree term}
For the Hartree term (\text{H.T.}), we introduce the following density matrices $O^f,O^{c',1},O^{c',2}$, where $O^f$ have also been used in the mean-field calculations of the THF model as shown in Ref.~\cite{HF_MATBLG} (however, $O^{c',1},O^{c',2},V_1,V_2$ are absent in the THF model) 
\baa 
&O^f_{\alpha \eta s,\alpha'\eta' s'} = \frac{1}{N_M}\sum_\RR \langle \Psi| :f_{\RR,\alpha \eta s}^\dag f_{\RR,\alpha' \eta' s'} : |\Psi\rangle 
\nonumber \\ 
&{O}^{c',1}_{a \eta s,a'\eta's'} = \frac{1}{N_M}\sum_{|\kk|<\Lambda_c}  e^{-|\kk|^2\lambda^2}\langle \Psi| :c_{\kk,a \eta s}^\dag c_{\kk, a' \eta' s'} : |\Psi\rangle, \quad a,a'\in \{1,2\} \nonumber \\
&{O}^{c',2}_{a' \eta' s',\alpha \eta s } = \frac{1}{N_M}\sum_{|\kk|<\Lambda_c} \sum_{a=1,2} e^{-|\kk|^2\lambda^2}(\eta k_x \sigma_x +k_y\sigma_y)_{\alpha a}\langle \Psi| :c_{\kk,a'\eta' s'}^\dag c_{\kk, a\eta s} : |\Psi\rangle, \quad a',\alpha\in \{1,2\} 
\label{eq:sc_eq_1} \, . 
\eaa  
We then assume the ground state is translational invariance such that
\baa  
&\langle \Psi| :f_{\RR,\alpha \eta s}^\dag f_{\RR,\alpha' \eta' s'} : |\Psi\rangle  = \frac{1}{N_M} \sum_\RR \langle \Psi| :f_{\RR,\alpha \eta s}^\dag f_{\RR,\alpha' \eta' s'} : |\Psi\rangle  = O^f_{\alpha \eta s,\alpha'\eta's'} \,. 
\label{eq:sc_eq_1_translation}
\eaa
Using Eq.~\ref{eq:sc_eq_1} and Eq.~\ref{eq:sc_eq_1_translation}, the Hartree term can be written as  
\baa  
&\text{H.T.}\nonumber \\ 
= &\sum_{\substack{\RR, \\ |\kk| <\Lambda_c,|\kk'| <\Lambda_c}} 
\sum_{\substack{ \alpha,\alpha',a,a',\\ \eta,\eta',s,s'} }
\frac{ e^{i(\kk-\kk') \RR - |\kk|^2 \lambda^2/2 - |\kk'|^2\lambda^2/2}}{N_MD_{\nu_c,\nu_f}} \bigg[ 
\gamma ^2 \delta_{\alpha',a'} \delta_{\alpha, a}
+
\gamma v_\star^\prime \delta_{\alpha, a}[\eta' k'_x\sigma_x - k'_y\sigma_y]_{\alpha'a'}+
\gamma v_\star^\prime \delta_{\alpha', a'}[\eta k_x\sigma_x + k_y\sigma_y]_{\alpha a}
\bigg] \nonumber \\
&\bigg\{  - 
\langle :f_{\RR,\alpha \eta s}^\dag  f_{\RR,\alpha'\eta' s'}: \rangle \langle :c_{\kk',a'\eta' s'}^\dag c_{\kk, a\eta s}:\rangle 
+ \langle :f_{\RR,\alpha \eta s}^\dag  f_{\RR,\alpha'\eta' s'}: \rangle :c_{\kk',a'\eta' s'}^\dag c_{\kk, a\eta s}:
+  :f_{\RR,\alpha \eta s}^\dag  f_{\RR,\alpha'\eta' s'}: 
\langle :c_{\kk',a'\eta' s'}^\dag c_{\kk, a\eta s}:\rangle \bigg\}  \nonumber \\
= &
\sum_{\substack{ \alpha,\alpha',\\ \eta,\eta',s,s'} }
\frac{N_M}{D_{\nu_c,\nu_f}}
\bigg[ -\gamma^2 O^f_{\alpha \eta s,\alpha \eta's'} O^{c'1}_{\alpha'\eta's',\alpha\eta s} -\bigg( \gamma v_\star^\prime O^f_{\alpha \eta
s,\alpha'\eta's'}O^{c',2}_{\alpha'\eta's',\alpha \eta s} +\text{h.c.}\bigg) \bigg]\nonumber \\ 
 & + \sum_{|\kk|<\Lambda_c}  
 \sum_{\substack{ \alpha,\alpha',\\ \eta,\eta',s,s'} }
 \bigg\{ 
 O^f_{\alpha \eta s,\alpha'\eta's'}e^{-|\kk|^2\lambda^2}:c_{\kk,a' \eta's'}^\dag c_{\kk, a \eta s}: \delta_{\alpha,a}\delta_{\alpha',a'} 
 +\bigg[ \gamma v_\star^\prime O^f_{\alpha \eta s,\alpha'\eta's'}\delta_{\alpha,a} [\eta' k_x\sigma_x -k_y\sigma_y]_{\alpha' a'}e^{-|\kk|^2 \lambda^2}
 \nonumber \\
 &
: c^\dag_{\kk,a'\eta' s}c_{\kk,a\eta s} : +\text{h.c.} \bigg] \bigg\} 
 + \sum_{\RR}  
 \sum_{\substack{ \alpha,\alpha',\\ \eta,\eta',s,s'} }
 \bigg\{ 
 :f_{\RR,\alpha \eta s}^\dag f_{\RR,\alpha'\eta's'} :O^{c',1}_{\alpha'\eta's',\alpha \eta s} 
 +\bigg[\gamma v_\star^\prime :f_{\RR,\alpha \eta s}^\dag f_{\RR',\alpha'\eta's'} :O^{c',2}_{\alpha'\eta's',\alpha\eta s}  +\text{h.c.} \bigg] \bigg\} 
 \label{eq:hk_hartree}
\eaa 

\subsubsection{Fock and Hartree terms}
Combining Fock and Hartree (Eq.~\ref{eq:hk_fock} and Eq.~\ref{eq:hk_fock}) terms, we have 
\baa  
\hH_K \approx &\hH_K^{MF} \nonumber \\
=&\text{F.T.}+\text{H.T.} \nonumber \\
=& - \frac{\gamma^2}{D_{\nu_c,\nu_f}}\sum_{\RR, |\kk| <\Lambda_c}
\sum_{\alpha\eta, s}\frac{e^{i\kk \cdot \RR -|\kk|^2 \lambda^2/2} }{\sqrt{N_M}} \bigg[\bigg(  V^* _{1}f_{\RR,\alpha \eta s}^\dag c_{\kk,\alpha\eta s} +\text{h.c.}
\bigg) 
\bigg]  + \frac{N_M\gamma^2 |V_1|^2 }{D_{\nu_c,\nu_f}}\nonumber \\ 
& - \frac{\gamma v_\star^\prime }{D_{\nu_c,\nu_f}}\sum_{\RR, |\kk| <\Lambda_c}
\sum_{\alpha,a,\eta, s}\frac{e^{i\kk \cdot \RR -|\kk|^2 \lambda^2/2} }{\sqrt{N_M}}  \bigg[ \bigg( V^* _{1} (\eta k_x \sigma_x +k_y\sigma_y)_{\alpha a}f_{\RR,\alpha \eta s}^\dag c_{\kk,\alpha\eta s} +\text{h.c.}\bigg) 
\bigg]  + \frac{N_M\gamma v_\star^\prime V_1^* V_2 }{D_{\nu_c,\nu_f}}\nonumber \\ 
&
- \frac{\gamma v_\star^\prime }{D_{\nu_c,\nu_f}}\sum_{\RR, |\kk| <\Lambda_c}
\sum_{\alpha, \eta, s}\frac{e^{i\kk \cdot \RR -|\kk|^2 \lambda^2/2} }{\sqrt{N_M}} \bigg[
\bigg( V^* _{2}f_{\RR,\alpha \eta s}^\dag c_{\kk,\alpha\eta s}  +\text{h.c.}\bigg) -V_2^* V_1 \bigg] +\frac{N_M\gamma v_\star^\prime V_2^* V_1 }{D_{\nu_c,\nu_f}} \nonumber \\
+&
\sum_{\substack{ \alpha,\alpha',\\ \eta,\eta',s,s'} }
\frac{N_M}{D_{\nu_c,\nu_f}}
\bigg[ -\gamma^2 O^f_{\alpha \eta s,\alpha \eta's'} O^{c'1}_{\alpha'\eta's',\alpha\eta s} -\bigg( \gamma v_\star^\prime O^f_{\alpha \eta 
 s,\alpha'\eta's'}O^{c',2}_{\alpha'\eta's',\alpha \eta s} +\text{h.c.}\bigg) \bigg]\nonumber \\ 
 & + \sum_{|\kk|<\Lambda_c}  
 \sum_{\substack{ \alpha,\alpha',\\ \eta,\eta',s,s'} }
 \bigg\{ 
 O^f_{\alpha \eta s,\alpha'\eta's'}e^{-|\kk|^2\lambda^2}:c_{\kk,a' \eta's'}^\dag c_{\kk, a \eta s}: \delta_{\alpha,a}\delta_{\alpha',a'} 
 +\bigg[ \gamma v_\star^\prime O^f_{\alpha \eta s,\alpha'\eta's'}\delta_{\alpha,a} [\eta' k_x\sigma_x -k_y\sigma_y]_{\alpha' a'}e^{-|\kk|^2 \lambda^2}
 \nonumber \\
 &
 :c^\dag_{\kk,a'\eta' s}c_{\kk,a\eta s} : +\text{h.c.} \bigg] \bigg\} 
 + \sum_{\RR}  
 \sum_{\substack{ \alpha,\alpha',\\ \eta,\eta',s,s'} }
 \bigg\{ 
 :f_{\RR,\alpha \eta s}^\dag f_{\RR,\alpha'\eta's'} :O^{c',1}_{\alpha'\eta's',\alpha \eta s} 
 +\bigg[\gamma v_\star^\prime :f_{\RR,\alpha \eta s}^\dag f_{\RR',\alpha'\eta's'} :O^{c',2}_{\alpha'\eta's',\alpha\eta s}  +\text{h.c.} \bigg] \bigg\} 
\label{eq:mf_kondo_int}
 \eaa  
$V_1,V_2$ describes the Fock contribution that characterize the hybridization between $f$- and $\Gamma_3$ $c
$-electrons. $O^f,O^{c',1},O^{c',2}$ are the mean fields taking the form of $\langle f^\dag f\rangle, \langle c^\dag c\rangle$ which represent the Fock contribution.

\subsection{Mean-field decoupling of $\hH_J$}
\label{sec:kondo_mf_hj}
We now perform a mean-field decoupling of the ferromagnetic exchange coupling term~\cite{HF_MATBLG}
\baa  
\hH_J \approx & \hH_J^{MF} \nonumber \\ 
=& - \frac{J}{2N_M} \sum_{\RR } \sum_{\alpha\alpha'\eta\eta',ss'} \sum_{|\kk|,|\kk'|<\Lambda_c }   
     e^{i( \kk- \kk' )\cdot\RR } 
     ( \eta\eta' + (-1)^{\alpha+\alpha'} )
     \bigg\{     
    \langle  f_{\RR , \alpha \eta s}^\dag  c_{ \kk', \alpha+2, \eta s} \rangle \langle c_{\kk', \alpha'+2, \eta' s'}^\dagger   f_{\RR ,\alpha' \eta' s'}\rangle   \nonumber \\ 
    &
    -\langle  f_{\RR, \alpha \eta s}^\dag  c_{ \kk, \alpha+2, \eta s_1} \rangle  c_{\kk', \alpha'+2, \eta' s'}^\dagger   f_{\RR, \alpha' \eta' s'}
    -  f_{\RR ,\alpha \eta s_1}^\dag  c_{ \kk, \alpha+2, \eta s} \langle   c_{\kk', \alpha'+2, \eta' s'}^\dagger   f_{\RR, \alpha' \eta' s'}\rangle\nonumber \\ 
    & 
     -\langle  :f_{\RR, \alpha \eta s}^\dag  f_{\RR ,\alpha' \eta' s'} :\rangle \langle : c_{\kk', \alpha'+2, \eta' s'}^\dagger  c_{ \kk, \alpha+2, \eta s}: \rangle  
     +: f_{\RR ,\alpha \eta s}^\dag  f_{\RR, \alpha' \eta' s'} : \langle : c_{\kk', \alpha'+2, \eta' s'}^\dagger  c_{ \kk, \alpha+2, \eta s}: \rangle   \nonumber \\ 
     &
     + \langle  :f_{\RR ,\alpha \eta s}^\dag  f_{\RR, \alpha' \eta' s'} :\rangle : c_{\kk', \alpha'+2, \eta' s'}^\dagger  c_{ \kk, \alpha+2, \eta s}: 
\bigg\} 
\eaa 
\subsubsection{Fock term}
The Fock term takes the form of 
\baa  
\text{F.T.} =& - \frac{J}{2N_M} \sum_{\RR } \sum_{\alpha\alpha'\eta\eta',ss'} \sum_{|\kk|,|\kk'|<\Lambda_c }   
     e^{i( \kk- \kk' )\cdot\RR } 
     ( \eta\eta' + (-1)^{\alpha+\alpha'} )
     \bigg\{     
    \langle  f_{\RR , \alpha \eta s}^\dag  c_{ \kk', \alpha+2, \eta s} \rangle \langle c_{\kk', \alpha'+2, \eta' s'}^\dagger   f_{\RR ,\alpha' \eta' s'}\rangle   \nonumber \\ 
    &
    -\langle  f_{\RR, \alpha \eta s}^\dag  c_{ \kk, \alpha+2, \eta s_1} \rangle  c_{\kk', \alpha'+2, \eta' s'}^\dagger   f_{\RR, \alpha' \eta' s'}
    -  f_{\RR ,\alpha \eta s_1}^\dag  c_{ \kk, \alpha+2, \eta s} \langle   c_{\kk', \alpha'+2, \eta' s'}^\dagger   f_{\RR, \alpha' \eta' s'}\rangle \bigg\}\nonumber \\ 
    =& 
    - J \sum_{\RR } \sum_{\xi = \pm } 
     \bigg\{     
    \langle   \sum_{|\kk'|<\Lambda_c,\alpha \eta s}\frac{e^{i(-\kk')\cdot \RR}}{\sqrt{N_M}} \delta_{\xi,\eta (-1)^{\alpha+1}} f_{\RR , \alpha \eta s}^\dag  c_{ \kk', \alpha+2, \eta s} \rangle 
    \langle \sum_{|\kk|<\Lambda_c,\alpha'\eta' s'}\delta_{\xi,\eta'(-1)^{\alpha'+1}} \frac{e^{i\kk\cdot \RR}}{\sqrt{N_M}} c_{\kk', \alpha'+2, \eta' s'}^\dagger   f_{\RR ,\alpha' \eta' s'}\rangle   \nonumber \\ 
    &-\sum_{|\kk|<\Lambda_c,\alpha \eta s} \frac{e^{i(-\kk')\cdot \RR}}{\sqrt{N_M}}  \delta_{\xi,\eta (-1)^{\alpha+1}} f_{\RR , \alpha \eta s}^\dag  c_{ \kk', \alpha+2, \eta s} 
    \langle \sum_{|\kk'|<\Lambda_c,\alpha\eta  s}\delta_{\xi,\eta'(-1)^{\alpha'+1}} \frac{e^{i\kk\cdot \RR}}{\sqrt{N_M}}\sum_{\alpha'\eta' s'} c_{\kk', \alpha'+2, \eta' s'}^\dagger   f_{\RR ,\alpha' \eta' s'}\rangle \nonumber \\
   &- \langle  \sum_{\kk',\alpha \eta s}\frac{e^{i(-\kk')\cdot \RR}}{\sqrt{N_M}}\delta_{\xi,\eta (-1)^{\alpha+1}} f_{\RR , \alpha \eta s}^\dag  c_{ \kk', \alpha+2, \eta s} \rangle 
   \sum_{|\kk|<\Lambda_c, \alpha'\eta' s'}
    \frac{e^{i\kk\cdot \RR}}{\sqrt{N_M}} \delta_{\xi,\eta'(-1)^{\alpha'+1}} c_{\kk', \alpha'+2, \eta' s'}^\dagger   f_{\RR ,\alpha' \eta' s'}\bigg\}
\eaa 
We then introduce the following mean-fields
\baa  
&V_3 =  \sum_{\RR, |\kk| <\Lambda_c}
\sum_{\alpha\eta, s}\frac{e^{i\kk \cdot \RR  }\delta_{1, \eta (-1)^{\alpha+1}}}{{N_M}\sqrt{N_M}} \langle \Psi| f_{\RR,\alpha \eta s}^\dag c_{\kk,\alpha+2\eta s} |\Psi\rangle 
\nonumber  \\ 
&V_4 =  \sum_{\RR, |\kk| <\Lambda_c}
\sum_{\alpha\eta, s}\frac{e^{i\kk \cdot \RR  }\delta_{-1, \eta (-1)^{\alpha+1}}}{{N_M}\sqrt{N_M}} \langle \Psi| \eta f_{\RR,\alpha \eta s}^\dag c_{\kk,\alpha+2\eta s} |\Psi\rangle  
\label{eq:sc_eq_2_v} 
\eaa  
and assume the ground state is translational invariant such that 
\baa  
& \sum_{ |\kk| <\Lambda_c}
\sum_{\alpha\eta, s}\frac{e^{i\kk \cdot \RR  }\delta_{1, \eta (-1)^{\alpha+1}}}{{}\sqrt{N_M}} \langle \Psi| f_{\RR,\alpha \eta s}^\dag c_{\kk,\alpha+2\eta s} |\Psi\rangle = \frac{1}{N_M}\sum_\RR \sum_{ |\kk| <\Lambda_c}
\sum_{\alpha\eta, s}\frac{e^{i\kk \cdot \RR  }\delta_{1, \eta (-1)^{\alpha+1}}}{{}\sqrt{N_M}} \langle \Psi| f_{\RR,\alpha \eta s}^\dag c_{\kk,\alpha+2\eta s} |\Psi\rangle =V_3
\nonumber  \\ 
& \sum_{ |\kk| <\Lambda_c}
\sum_{\alpha\eta, s}\frac{e^{i\kk \cdot \RR  }\delta_{-1, \eta (-1)^{\alpha+1}}}{{}\sqrt{N_M}} \langle \Psi| f_{\RR,\alpha \eta s}^\dag c_{\kk,\alpha+2\eta s} |\Psi\rangle = \frac{1}{N_M}\sum_\RR \sum_{ |\kk| <\Lambda_c}
\sum_{\alpha\eta, s}\frac{e^{i\kk \cdot \RR  }\delta_{-1, \eta (-1)^{\alpha+1}}}{{}\sqrt{N_M}} \langle \Psi| f_{\RR,\alpha \eta s}^\dag c_{\kk,\alpha+2\eta s} |\Psi\rangle =V_4
\eaa 
Then the Fock term can be written as 
\baa  
\text{F.T.} =&  - JN_M [ |V_3|^2 +|V_4|^2 ] \nonumber \\
&  +J    
    \sum_{\RR,|\kk|<\Lambda_c,\alpha \eta s}      \bigg\{  \frac{e^{i(-\kk')\cdot \RR}}{\sqrt{N_M}} \bigg[ \delta_{1,\eta (-1)^{\alpha+1}} f_{\RR , \alpha \eta s}^\dag  c_{ \kk', \alpha+2, \eta s} V_3^* + 
 \delta_{-1,\eta (-1)^{\alpha+1}} f_{\RR , \alpha \eta s}^\dag  c_{ \kk', \alpha+2, \eta s} V_4^* + 
\bigg] +\text{h.c.}\bigg\} 
 \label{eq:hj_fock}
\eaa 
\subsubsection{Hartree term}
The Hartree term takes the form of 
\baa  
\text{H.T.}=&
- \frac{J}{2N_M} \sum_{\RR } \sum_{\alpha\alpha'\eta\eta',ss'} \sum_{|\kk|,|\kk'|<\Lambda_c }   
     e^{i( \kk- \kk' )\cdot\RR } 
     ( \eta\eta' + (-1)^{\alpha+\alpha'} )
     \bigg\{   
     -\langle  :f_{\RR, \alpha \eta s}^\dag  f_{\RR ,\alpha' \eta' s'} :\rangle \langle : c_{\kk', \alpha'+2, \eta' s'}^\dagger  c_{ \kk, \alpha+2, \eta s}: \rangle   \nonumber \\
     &
     +: f_{\RR ,\alpha \eta s}^\dag  f_{\RR, \alpha' \eta' s'} : \langle : c_{\kk', \alpha'+2, \eta' s'}^\dagger  c_{ \kk, \alpha+2, \eta s}: \rangle 
     + \langle  :f_{\RR ,\alpha \eta s}^\dag  f_{\RR, \alpha' \eta' s'} :\rangle : c_{\kk', \alpha'+2, \eta' s'}^\dagger  c_{ \kk, \alpha+2, \eta s}: 
\bigg\} 
\eaa 
We introduce the following density matric which has also been used in Ref.~\cite{HF_MATBLG}
\baa  
& O^{c''}_{a \eta s,a'\eta's'} = \frac{1}{N_M}\sum_{|\kk|<\Lambda_c} \langle \Psi| :c_{\kk,a+2 \eta s}^\dag c_{\kk, a'+2 \eta' s'} : |\Psi\rangle,\quad a,a' \in \{1,2\} \, .
\label{eq:sc_eq_2} 
\eaa 
Using Eq.~\ref{eq:sc_eq_1} and Eq.~\ref{eq:sc_eq_2}, the Hartree term becomes 
\baa  
\text{H.T.}=&
- \frac{JN_M}{2} \sum_{\alpha\alpha'\eta\eta' ss'} 
     ( \eta\eta' + (-1)^{\alpha+\alpha'} )  O^f_{\alpha \eta s,\alpha'\eta's'}O^{c''}_{\alpha'\eta's',\alpha \eta s} \nonumber \\ 
& +\frac{J}{2}\sum_{\RR_c,\alpha \alpha'\eta \eta'ss'} 
 : f_{\RR ,\alpha \eta s}^\dag  f_{\RR, \alpha' \eta' s'} : O^{c''}_{\alpha'\eta's',\alpha\eta s} 
 +\frac{J}{2} \sum_{|\kk|<\Lambda_c,\alpha \alpha'\eta \eta'ss'} 
 O^f_{\alpha \eta s,\alpha'\eta's'} :c_{\kk', \alpha'+2, \eta' s'}^\dagger  c_{ \kk, \alpha+2, \eta s}:
 \label{eq:hj_hartree}
\eaa 
\subsubsection{Fock and Hartree terms}
Combing Hartree and Fock terms (Eq.~\ref{eq:hj_fock} and Eq.~\ref{eq:hj_hartree}), we have 
\baa  
\hH_J \approx &\hH_J^{MF} \nonumber \\ 
=& - JN_M \sum_{\xi = \pm } |V_3|^2 +|V_4|^2 \nonumber \\
&  +J    
    \sum_{\RR,|\kk|<\Lambda_c,\alpha \eta s}      \bigg\{  \frac{e^{i(-\kk')\cdot \RR}}{\sqrt{N_M}} \bigg[ \delta_{1,\eta (-1)^{\alpha+1}} f_{\RR , \alpha \eta s}^\dag  c_{ \kk', \alpha+2, \eta s} V_3^* + 
 \delta_{-1,\eta (-1)^{\alpha+1}} f_{\RR , \alpha \eta s}^\dag  c_{ \kk', \alpha+2, \eta s} V_4^*  
\bigg] +\text{h.c.}\bigg\} \nonumber \\ 
&- \frac{JN_M}{2} \sum_{\alpha\alpha'\eta\eta' ss'} 
     ( \eta\eta' + (-1)^{\alpha+\alpha'} )  O^f_{\alpha \eta s,\alpha'\eta's'}O^{c''}_{\alpha'\eta's',\alpha \eta s} \nonumber \\ 
& +\frac{J}{2}\sum_{\RR_c,\alpha \alpha'\eta \eta'ss'} 
 : f_{\RR ,\alpha \eta s}^\dag  f_{\RR, \alpha' \eta' s'} : O^{c''}_{\alpha'\eta's',\alpha\eta s} 
 +\frac{J}{2} \sum_{|\kk|<\Lambda_c,\alpha \alpha'\eta \eta'ss'} 
 O^f_{\alpha \eta s,\alpha'\eta's'} :c_{\kk', \alpha'+2, \eta' s'}^\dagger  c_{ \kk, \alpha+2, \eta s}:
 \label{eq:mf_kondo_hj}
\eaa 
$V_3,V_4$ describes the Fock contribution that characterize the hybridization between $f$- and $\Gamma_1\oplus\Gamma_2$ $c$-electrons. $O^f,O^{c''}$ are the mean fields taking the form of $\langle f^\dag f\rangle, \langle c^\dag c\rangle$ which represent the Fock contribution and have also been used in Ref.~\cite{HF_MATBLG}.


\subsection{Filling constraints and mean-field equations}
\label{sec:kondo_mf_filling}
We note that in the Kondo model the filling of $f$ electrons is fixed to be $\nu_f$ at each site. To simplify the calculation, we take a common approximation that only requires the average filling of $f$-electron to be $\nu_f$~\cite{coleman2015introduction,read1983solution}. In other words, we only require $\frac{1}{N_M}\sum_{\RR,\alpha \eta s} \langle \Psi|:f_{\RR,\alpha\eta s}^\dag f_{\RR,\alpha \eta s} :|\Psi\rangle =\nu_f$.
We then add the following term to the Hamiltonian 
\baa  
\hH_{\lambda_f}  = \sum_{\RR, \alpha \eta s }\lambda_f\bigg( :f_{\RR,\alpha \eta s} ^\dag f_{\RR,\alpha \eta s}: - \nu_f \bigg) 
\label{eq:ham_lamf}
\eaa 
and determine the Langrangian multiplier $\lambda_f$ from the following equation
\baa  
\frac{1}{N_M} \sum_{\RR, \alpha \eta s }\langle \Psi| :f_{\RR,\alpha \eta s} ^\dag f_{\RR,\alpha \eta s}:  |\Psi\rangle = \nu_f  
\label{eq:sc_eq_3}
\eaa

In practice, we perform calculations at fixed total filling $\nu= \nu_f+\nu_c$, where $\nu_f$ and $\nu_c$ are the average fillings of $f$ and $c$ electrons respectively. Since $\nu_f$ is also fixed in the Kondo model, we will self-consistently determine the chemical potential $\mu_c$ (in Eq.~\ref{eq:hmuc}) by requiring 
\baa  
\frac{1}{N_M}\sum_{|\kk|<\Lambda_c, a\eta s}\langle \Psi|: c_{\kk,a\eta s}^\dag c_{\kk,a \eta s }:|\Psi\rangle  = \nu_c = \nu-\nu_f 
\label{eq:sc_eq_4}
\eaa

Finally, our mean-field Hamiltonian takes the form of 
\baa  
\hH^{MF} = \hH_c + \hH_{cc} +\hH_K^{MF} +\hH_J^{MF} +\hH_{\lambda_f} +\hH_{\mu_c} 
 \eaa  
and we determine $V_1,V_2,V_3,V_4,O^f, O^{c',1},O^{c',2},O^{c''},\lambda_f,\mu_c$ 
from the self-consistent equations (Eq.~\ref{eq:sc_eq_1_v}, Eq.~\ref{eq:sc_eq_1}, Eq.~\ref{eq:sc_eq_2_v}, Eq.~\ref{eq:sc_eq_2},  Eq.~\ref{eq:sc_eq_3}, Eq.~\ref{eq:sc_eq_4}). During the self-consistent solution, at each step, we will adjust $\lambda_f, \mu_c$ according to the current filling of $f$- and $c$-electrons. We use $\nu_f^i$ and $\nu_c^i$ denote the filling of $f$ and $c$ at $i$-th step. For the $i+1$-th step, we will update $\lambda_f,\mu_c$ as $\lambda_f \rightarrow \lambda_f + r (\nu_f^i - \nu_f) , \mu_c \rightarrow \mu_c - r (\nu_c^i - \nu_c) $, where $r(>0)$ will be manually adjusted to improve the convergence (in practice, we take $r\sim 0.001$).

\subsection{Mean-field equations of the symmetric Kondo state}
\label{sec:symmetric_kondo}
We focus on the symmetric Kondo phase without any symmetry breaking. Therefore, we require our density matrix of $f$- and $c$- electrons (Eq.~\ref{eq:sc_eq_1}, Eq.~\ref{eq:sc_eq_2}) to be symmetric. We can then utilize symmetry to simplify the self-consistent equations (Eq.~\ref{eq:sc_eq_1}, Eq.~\ref{eq:sc_eq_2}). 
We first consider the $U(1)_v$ symmetry. From Eq.~\ref{eq:cont_sym}, a $U(1)_v$ symmetric solution satisfies
\baa  
O^{f}_{\alpha \eta s, \alpha ' \eta' s'} = O^{f}_{\alpha \eta s, \alpha '\eta's'}e^{-i\theta_\nu(\eta-\eta')} \Rightarrow O^{f}_{\alpha \eta s, \alpha  -\eta s'}  =0 \nonumber \\
O^{c',1}_{a \eta s, a' \eta' s'} = O^{c',1}_{a \eta s, a' \eta' s'}e^{-i\theta_\nu(\eta-\eta')} \Rightarrow O^{c',1}_{a \eta s, a' -\eta' s'}  =0 \nonumber \\
O^{c',2}_{a \eta s, \alpha' \eta' s'} = O^{c',2}_{a \eta s, \alpha' \eta' s'}e^{-i\theta_\nu(\eta-\eta')} \Rightarrow O^{c',2}_{a \eta s, \alpha' -\eta' s'}  =0 \nonumber \\
O^{c''}_{a \eta s, a' \eta' s'} = O^{c''}_{a \eta s, a' \eta' s'}e^{-i\theta_\nu(\eta-\eta')} \Rightarrow O^{c''}_{a \eta s, a' -\eta s'}  =0 \nonumber \\
\label{eq:valley_diag}
\eaa  
and $V_1,V_2,V_3,V_4$ are invariant under $U(1)_v$ transformation. 
From Eq.~\ref{eq:valley_diag}, $O^{f},O^{c',1},O^{c',2},O^{c''}$ are block diagonalized in valley index. We next consider a $SU(2)_\eta$ transformation acting on the valley $\eta$. We find 
\baa  
&\sum_{s,s'} [e^{i\sum_\mu \theta^\eta_\mu \sigma^\mu}]_{s_2,s}O^{f}_{\alpha \eta s, \alpha' \eta s'} [e^{i\sum_\mu \theta^\eta_\mu \sigma^\mu}]_{s',s_2'} = O^{f}_{\alpha \eta s_2, \alpha' \eta s_2'}\Rightarrow 
O^{f}_{a \eta s, a' \eta s'} \propto \mathbb{I}_{s,s'} \nonumber \\ 
&\sum_{s,s'} [e^{i\sum_\mu \theta^\eta_\mu \sigma^\mu}]_{s_2,s}O^{c',1}_{a \eta s, a' \eta s'} [e^{i\sum_\mu \theta^\eta_\mu \sigma^\mu}]_{s',s_2'} = O^{c',1}_{a \eta s, a' \eta s'}\Rightarrow 
O^{c',1}_{a \eta s, a' \eta s'} \propto \mathbb{I}_{s,s'} \nonumber \\ 
&\sum_{s,s'} [e^{i\sum_\mu \theta^\eta_\mu \sigma^\mu}]_{s_2,s}O^{c',2}_{a \eta s, a' \eta s'} [e^{i\sum_\mu \theta^\eta_\mu \sigma^\mu}]_{s',s_2'} = O^{c',2}_{a \eta s, a' \eta s'}\Rightarrow 
O^{c',2}_{a \eta s, a' \eta s'} \propto \mathbb{I}_{s,s'} \nonumber \\
&\sum_{s,s'} [e^{i\sum_\mu \theta^\eta_\mu \sigma^\mu}]_{s_2,s}O^{c''}_{a \eta s, a' \eta s'} [e^{i\sum_\mu \theta^\eta_\mu \sigma^\mu}]_{s',s_2'} = O^{c''}_{a \eta s, a' \eta s'}\Rightarrow 
O^{c''}_{a \eta s, a' \eta s'} \propto \mathbb{I}_{s,s'}
\label{eq:spin_diag} 
\eaa  
where $\mathbb{I}$ is an $2\times 2$ identical matrix. In addition, $V_1,V_2,V_3,V_4$ are invariant under $SU(2)_\eta$ transformation.

From Eq.~\ref{eq:valley_diag} and Eq.~\ref{eq:spin_diag}, the density matrices $O^{f}, O^{c',1},O^{c''}$ are diagonalized in valley and spin incdies. We then introduce $2\times 2$ matrices, $o^{f,\eta},o^{c',1,\eta}, o^{c',2,\eta}, o^{c'',\eta}$, such that 
\baa  
&O^{f}_{\alpha \eta s,\alpha'\eta's'} = o^{f,\eta}_{\alpha,\alpha'}\delta_{\eta,\eta'}\delta_{s,s'} ,\quad 
O^{c',1}_{a \eta s,a'\eta's'} = o^{c',1,\eta}_{a,a'}\delta_{\eta,\eta'}\delta_{s,s'} 
,\quad \nonumber \\
&O^{c',2}_{a \eta s,a'\eta's'} = o^{c',2,\eta}_{a,a'}\delta_{\eta,\eta'}\delta_{s,s'} 
,\quad 
O^{c''}_{a \eta s,a'\eta's'} = o^{c'',\eta}_{a,a'}\delta_{\eta,\eta'}\delta_{s,s'} 
\label{eq:small_o}
\eaa  

We now consider the effect of discrete symmetries in Eq.~\ref{eq:desc_sym_op}. Using Eq.~\ref{eq:desc_sym} and Eq.~\ref{eq:small_o}, we find 
\baa  
T:&\quad (o^{f,\eta})^* = o^{f,-\eta} ,
\quad (o^{c',1,\eta})^* = o^{c',1,-\eta} ,
\quad (o^{c',2,\eta})^* = o^{c',2,-\eta} ,\quad 
(o^{c'',\eta})^* = o^{c'',-\eta} \nonumber \\
C_{3z}:&\quad e^{i\frac{2\pi \eta}{3}  \sigma_z } o^{f,\eta}e^{-i\frac{2\pi \eta}{3} \sigma_z} = o^{f,\eta} ,\quad 
e^{i\frac{2\pi \eta}{3}  \sigma_z } o^{c',1,\eta}e^{-i\frac{2\pi\eta }{3}\sigma_z} = o^{c',1,\eta}, 
\quad 
e^{i\frac{\eta 2\pi}{3}}o^{c',2,\eta} e^{-i\frac{\eta 2\pi}{2}\sigma^z } = o^{c',2,\eta} ,\quad 
o^{c'',\eta} = o^{c'',\eta} \nonumber \\ 
C_{2x}:&\quad \sigma_x o^{f,\eta}\sigma_x = o^{f,\eta} ,\quad 
\sigma_x o^{c',1,\eta}\sigma_x = o^{c',1,\eta},\quad 
\sigma_x o^{c',2,\eta}\sigma_x = o^{c',2,\eta},\quad 
\sigma_x o^{c'',\eta}\sigma_x = o^{c'',\eta} \nonumber \\ 
C_{2z}T:&\quad (\sigma_x o^{f,\eta}\sigma_x)^* = o^{f,\eta} ,\quad 
(\sigma_x o^{c',1,\eta}\sigma_x)^* = o^{c',1,\eta},
\quad (\sigma_x o^{c',2,\eta} \sigma_x )^* = -o^{c',2,-\eta} ,
\quad (\sigma_x o^{c'',\eta}\sigma_x)^* =o^{c'',\eta} 
\label{eq:desc_sym_o}
\eaa 
From the definition (Eq.~\ref{eq:sc_eq_1}), $O^f,O^{c',1},O^{c''}$ are Hermitian matrices and then $o^{f},o^{c',1}, o^{c''}$ are also Hermitian matrices. 
Combining Eq.~\ref{eq:desc_sym_o} and the Hermitian properties, we can introduce real numbers $\chi^{f}_0,\chi^{c',1}_0,\chi^{c''}_0,\chi^{c''}_1$ and then $o^{f},o^{c'},o^{c''}$ take the following structure
\baa  
o^{f,\eta} = o^{f,-\eta }= \chi^f_0 \sigma_0 
,\quad 
o^{c',1,\eta} = o^{c',-\eta} =\chi^{c',1}_0 \sigma_0 
,\quad 
o^{c',2,\eta} =0
,\quad 
o^{c'',\eta} = \chi^{c''}_0\sigma_0 + \chi^{c''}_1\sigma_x 
\label{eq:o_in_alpha}
\eaa  
where $\sigma_0,\sigma_{x,y,z}$ are identity and Pauli matrices respectively with row and column indices $\alpha = 1,2$. 
Since the filling of $f$- and $\Gamma_1\oplus \Gamma_2$ $c$- electrons are $\nu_f = \text{Tr}[O^f],\nu_{c''} = \text{Tr}[O^{c''}]$ respectively, we find $\chi^f_0 = \nu_f/8, \chi^{c''} _0 = \nu_{c''}/8$. Using Eq.~\ref{eq:o_in_alpha} and Eq.~\ref{eq:small_o}, for the symmetric solution, we finally have 

\baa  
&O^f_{\alpha \eta s, \alpha '\eta 's'} = \delta_{\alpha ,\alpha'}\delta_{\eta,\eta'} \delta_{s,s'} \nu_f/8 \nonumber \\
&O^{c',1}_{a\eta s, a'\eta 's'} = \delta_{a,a'}\delta_{\eta,\eta'} \delta_{s,s'} \chi_0^{c',1} ,\quad O^{c',2}_{a\eta s, a'\eta 's'} = 0 ,\quad a,a' \in \{1,2\}\nonumber \\
&O^{c''}_{a\eta s, a'\eta 's'} = \delta_{\eta,\eta'} \delta_{s,s'} (\delta_{a,a'}\nu_{c''}/8  +\delta_{a,3-a'} \chi^{c''}_x), \quad a,a' \in \{1.2\}
\label{eq:ocf_from_alpha_kondo}
\eaa
As for the hybridization fields, we find discrete symmetries will not impose constraints on $V_1,V_2$. As for $V_3,V_4$, we have 
\baa  
T&: V_3 = V_4^* ;\quad \quad 
C_{3z}: V_3 = e^{i2\pi/3} V_3 ,\quad  V_4 = e^{i2\pi/3} V_4  \nonumber \\
C_{2x}&: V_3 = V_4 ;\quad\quad 
C_{2z}T : V_3 = V_4^* 
\label{eq:descrete_v}
\eaa  
therefore 
\baa  
V_3=V_4=0
\label{eq:v3v4_zero}
\eaa

In summary, instead of solving self-consistent equations of $O^{f},O^{c',1},O^{c''},V_3,V_4$ (Eq.~\ref{eq:sc_eq_1}, Eq.~\ref{eq:sc_eq_2_v}, Eq.~\ref{eq:sc_eq_2}), we can use
\baa  
&\nu_f = \frac{1}{N_M} \sum_{\RR, \alpha \eta s }\langle \Psi| :f_{\RR,\alpha \eta s} ^\dag f_{\RR,\alpha \eta s}:  |\Psi\rangle  \nonumber \\ 
&\chi^{c',1}_0=\frac{1}{8N_M} \sum_{|\kk|<\Lambda_c, a=1,2, \eta s }\langle \Psi| e^{-|\kk|^2\lambda^2}:c_{\kk,a \eta s} ^\dag c_{\kk,a \eta s}:  |\Psi\rangle  \nonumber \\ 
&\nu_{c''}=\frac{1}{N_M} \sum_{|\kk|<\Lambda_c, a=3,4, \eta s }\langle \Psi| :c_{\kk,a \eta s} ^\dag c_{\kk,a \eta s}:  |\Psi\rangle    \nonumber \\
&\chi_1^{c''}=\frac{1}{8N_M} \sum_{|\kk|<\Lambda_c, \eta s }\langle \Psi| c_{\kk,3 \eta s} ^\dag c_{\kk,4 \eta s}+c_{\kk, 4 \eta s} ^\dag c_{\kk,3 \eta s} |\Psi\rangle \nonumber \\
&V_3=V_4=0
\label{eq:sc_sym_kondo}
\eaa  
and obtain $O^{f},O^{c},O^{c''}$ via Eq.~\ref{eq:ocf_from_alpha_kondo}. We note that the first equation in Eq.~\ref{eq:sc_sym_kondo} is equivalent to Eq.~\ref{eq:sc_eq_3}. 
In summary, combining Eq.~\ref{eq:sc_eq_1_v}, Eq.~\ref{eq:sc_eq_4} and Eq.~\ref{eq:sc_sym_kondo}, we have a complete set of mean-field self-consistent equations for the symmetric Kondo state. We comment that Eq.~\ref{eq:sc_sym_kondo} are the same mean-field equations as we derived in Sec.~\ref{sec:kondo_mf_hk}, Sec.~\ref{sec:kondo_mf_hj} and Sec.~\ref{sec:kondo_mf_filling}, but with additional symmetry requirement, that is the ground states satisfy all symmetries.

We mention that, at $\nu=\nu_f=\nu_c=0$, we have $O^f_{\alpha \eta s,\alpha'\eta's'}=0, O^{c'}_{\alpha \eta s,\alpha'\eta' s'}=0$ and the Hartree term in Eq.~\ref{eq:mf_kondo_int} vanishes. We now prove the Hartree term in Eq.~\ref{eq:mf_kondo_hj} also vanishes. We note the only non-zero components of $O^{c''}$ are $O^{c''}_{1 \eta s, 2 \eta s},O^{c''}_{2 \eta s , 1 \eta s}$. From Eq.~\ref{eq:hj_hartree}, the Hartree term takes the form of (with $O^f=0$)
\baa  
&- \frac{J}{2N_M} \sum_{\RR s_1 s_2} \sum_{\alpha\alpha'\eta\eta'} \sum_{|\kk_1|,|\kk_2|<\Lambda_c }   
     e^{i( \kk_1- \kk_2 )\cdot\RR } 
     ( \eta\eta' + (-1)^{\alpha+\alpha'} ) 
     \bigg[ 
      :f_{\RR \alpha \eta s_1}^\dagger f_{\RR \alpha' \eta' s_2}:O^{c''}_{\alpha'\eta' s_2 ,\alpha \eta s_1} 
      \bigg]  \nonumber \\
=&- \frac{J}{2N_M} \sum_{\RR s } \sum_{\alpha \eta} \sum_{|\kk_1|,|\kk_2|<\Lambda_c }   
     e^{i( \kk_1- \kk_2 )\cdot\RR } 
     ( \eta\eta + (-1)^{\alpha+3-\alpha} ) 
     \bigg[ 
      :f_{\RR \alpha \eta s}^\dagger f_{\RR \alpha' \eta s}:O^{c''}_{3-\alpha\eta s ,\alpha \eta s} 
      \bigg]  \nonumber \\ 
    =&- \frac{J}{2N_M} \sum_{\RR s } \sum_{\alpha \eta} \sum_{|\kk_1|,|\kk_2|<\Lambda_c }   
     e^{i( \kk_1- \kk_2 )\cdot\RR } 
     ( 0 ) 
     \bigg[ 
      :f_{\RR \alpha \eta s}^\dagger f_{\RR \alpha' \eta s}:O^{c''}_{3-\alpha\eta s ,\alpha \eta s} 
      \bigg]  \nonumber \\ 
      =&0
\eaa  
and hence vanishes. In summary, at $\nu=0$, we only need to consider $V_1,V_2$, and other mean fields vanish.

\subsection{Properties of the symmetric Kondo state} 
\label{sec:prop_symm_kondo}
We solve the self-consistent equations Eq.~\ref{eq:sc_eq_1_v}, Eq.~\ref{eq:sc_eq_2_v}, Eq.~\ref{eq:sc_eq_3}, Eq.~\ref{eq:sc_eq_4} and Eq.~\ref{eq:sc_sym_kondo} at integer filling $\nu = 0,-1,-2$ with $\nu_f=\nu,\nu_c=0$. We identify the symmetric Kondo (SK) states at $\nu=0,-1,-2$ 
which are characterized by 
$V_1 \ne 0 $ ($|\gamma^2 V_1/D_{\nu_f,\nu_c}|=95\mathrm{meV},111\mathrm{meV},209\mathrm{meV}$ at $\nu=0,-1,-2$ respectively), 
$V_2\ne 0$ ($|v_\star^\prime \gamma V_2/D_{\nu_f,\nu_c}|=80\mathrm{meV},97\mathrm{meV},197\mathrm{meV}$ at $\nu=0,-1,-2$ respectively) and $V_{3}=0,V_4=0$. 
Even if we allow non-zero $V_3,V_4$ and initialize the mean-field calculations with non-zero $V_3,V_4$, $V_3,V_4$, we still find $V_3=V_4=0$ after self-consistent calculations (amplitudes smaller than $10^{-5}$). This is because $\hH_J$ describes ferromagnetic interactions and disfavors the development of non-zero $V_3,V_4$.
We also comment that the non-zero $V_1,V_2$, introduce an effective $f$-$c$ hybridization (Eq.~\ref{eq:hk_fock}) and characterize the Kondo physics.

We next discuss the topological feature of the bands. Since the SK states preserve all the symmetries, it is sufficient to only consider the bands in valley $+$ and spin $\up$. We find at $\nu=0,-1,-2$, the representations formed by flat bands at $\Gamma,K,M$ are $\Gamma_1 \oplus \Gamma_2, K_2K_3, M_1\oplus M_2$ respectively. We note that the representations formed by flat bands here are equivalent to that of the non-interacting THF model. Thus, the flat bands form a fragile topology at $\nu=-1,-2$ and a stable topology at $\nu=0$ due to the additional particle-hole symmetry at $\nu=0$~\cite{SON19,HF_MATBLG}.

In addition, we also calculate the Wilson loop of the flat bands (valley $+$ spin $\up$). In the calculation of Wilson loop, we let $k_1 \in \{\frac{i}{N}\}_{i=1,...,N-1}, k_2 = \{\frac{j}{N}\}_{j=1,...,N-1}$ and $k_1 \in \{\frac{i}{N}\}_{i=1,...,N-1}, k_2 = \{\frac{j}{N}\}_{j=1,...,N-1}$ and $\kk = k_1 \textbf{b}_{M,1} +k_2 \textbf{b}_{M,2}$, where $\textbf{b}_{M,1} = \frac{4\pi}{3a_M}(\sqrt{3},0),\textbf{b}_{M,2}=\frac{4\pi}{3a_M}(\frac{\sqrt{3}}{2}, \frac{3}{2})$ are two moir\'e reciprocal lattice vectors and $a_M$ is the moir\'e lattice constant. We then let $|u_{n,\kk}\rangle$ denote the $n$-th eigenvectors of the single-particle Hamiltonian $H(\kk)$ (of valley $+$ spin $\up$). We focus on the subset of the bands, which we denote with band indices $n=1,..,n_{band}$. Here, we take the flat bands as the subset of the bands that we are interested in. We then define the matrix $U_{\kk}$ as a matrix formed by the eigenvectors of the flat bands $ U_\kk= [|u_{1,\kk}\rangle, |u_{2,\kk}\rangle,...,|u_{n_{band},\kk}\rangle ]$. The Wilson loop~\cite{SON19} along the $k_2$ direction is defined as 
\baa  
W(k_1) =U_{k_1,k_2=0}^\dag \prod_{j=1}^{N-1} \bigg(U_{k_1, k_2 =\frac{2\pi j}{N} } U_{k_1, k_2 =\frac{2\pi (j+1)}{N}  }^\dag \bigg) V^{(k_1=0,k_2=2\pi)}U_{k_1,k_2=2\pi}\, . 
\eaa 
where $V^{\bm{G}}$ is defined as $H(\kk+\bm{G}) = V^{\bm{G}}H(\kk)V^{\bm{G},\dag}$, $\bm{G} = n\textbf{b}_{M,1} +m\textbf{b}_{M,2}$, $n,m\in\mathbb{Z}$. We mention that $c$-electrons are defined in the momentum space that can be larger than the first MBZ (depending on the momentum cutoff $\Lambda_c$). Thus we introduce $V^\GG$ that maps $c_\kk$ to $c_{\kk+\GG}$ to restore the periodic condition $H(\kk+\bm{G}) = V^{\bm{G}}H(\kk)V^{\bm{G},\dag}$. 
The corresponding Wilson loop Hamiltonian~\cite{SON19} is 
\baa  
\mathcal{H}(k_1) =-i \ln(W(k_1))
\eaa  
We plot the Wilson loop spectrum (eigenvalues of $\mathcal{H}(k_1)$) in Fig.~\ref{fig:wilson_loop}, where we observe the Wilson loop has winding number $1$. As shown in Ref.~\cite{SON19}, in the presence of additional particle-hole symmetry at $\nu=0$~\cite{SON19, HF_MATBLG}, $(-1)^{n}$ with $n$ the winding number of Wilson loop is a stable topological index. We conclude that at $\nu=0$, the symmetric Kondo state has a stable topology that is characterized by the odd winding number of the Wilson loop.

From Fig.~\ref{fig:wilson_loop}, we observe the behaviors of the Wilson loop are similar at different fillings. We check the overlapping of the flat-band wavefunctions between different bands. We let $\{|u_{i,\kk}^{\nu}\rangle\}_{i=1,...,n_{band}} $ denote the wavefunction of flat bands at filling $\nu$. We define the overlapping between wavefunctions at fillings $\nu$ and $\nu'$ as 
\baa  
Overlap(\nu,\nu') = \frac{1}{N}\sum_{\kk} 
\sum_{i,j \in \{1,...,n_{band}\}}
\langle u_{i,\kk}^{\nu} | u_{j,\kk}^{\nu'}\rangle \langle u_{j,\kk}^{\nu'} | u_{i,\kk}^{\nu}\rangle 
\eaa  
We find $Overlap(0,-1) = 99.1\%, Overlap(-1,-2) = 91.6\%$. The large overlapping of wavefunctions between different fillings indicates similar behaviors of the Wilson loop at different fillings as we showed in Fig.~\ref{fig:wilson_loop}.

\begin{figure}
    \centering
    \includegraphics[width=0.8\textwidth]{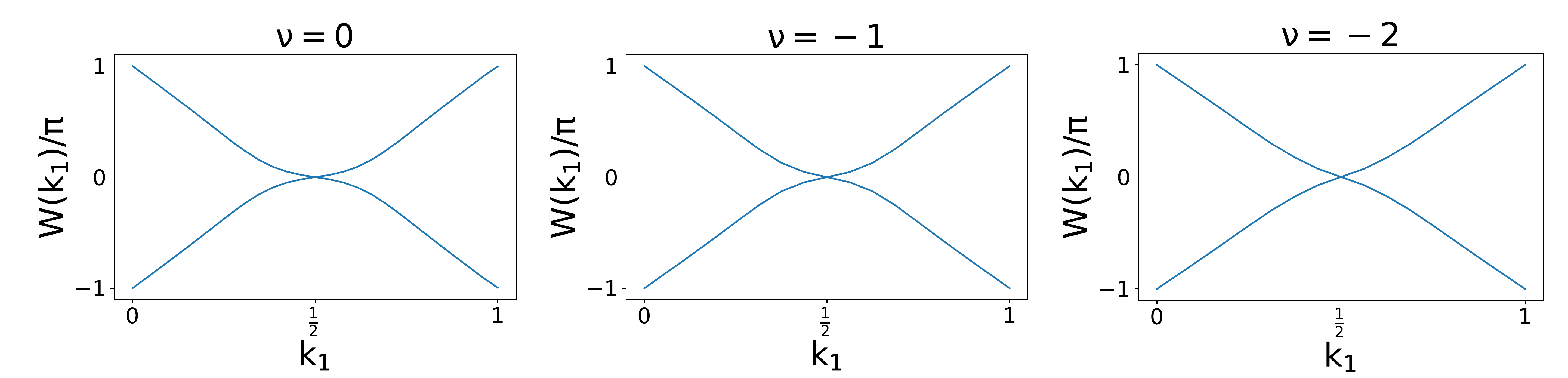}
    \caption{Wilson loop spectrum of flat bands of SK states at $\nu=0,-1,-2$.}
    \label{fig:wilson_loop}
\end{figure}

Finally, we analyze the mean-field Hamiltonian of the symmetric Kondo state. The mean-field single-particle Hamiltonian of valley $\eta$ and spin $\up$ (spin $\up$ and spin $\dn$ are equivalent) of the Kondo symmetric state can be approximately written as 
\baa  
&\tilde{H}^{(\eta)}(\kk) = \begin{bmatrix}
    \tilde{H}^{(f,\eta)}(\kk) & \tilde{H}^{(fc,\eta)}(\kk) \\ 
  \tilde{H}^{(fc,\eta),\dag}(\kk) & \tilde{H}^{(c,\eta)}(\kk) 
\end{bmatrix} \\ 
&\tilde{H}^{(f,\eta)} (\kk)= E_f \mathbb{I}_{ 2\times 2 } \nonumber \\ 
&\tilde{H}^{(c,\eta)}(\kk) = 
\begin{bmatrix}
    E_{c}\mathbb{I}_{2\times 2} & v_\star(\eta k_x \sigma_0 +ik_y\sigma_z) \nonumber \\ 
    v_\star(\eta k_x \sigma_0 -ik_y\sigma_z) &  E_{c''}\mathbb{I}_{2\times 2}
\end{bmatrix} \nonumber \\ 
&\tilde{H}^{(fc,\eta)}(\kk) = 
\begin{bmatrix}
   \tilde{\gamma} \sigma_0 + \tilde{v_\star^\prime}(\eta k_x \sigma_x + k_y\sigma_y) & 0_{2\times 2 }
\end{bmatrix} 
\label{eq:single_particle_ham_kondo}
\eaa  
where $\tilde{H}^{(f,\eta)},\tilde{H}^{(c,\eta)},\tilde{H}^{(fc,\eta)}$ denote the single-particle Hamiltonian of the $f$-block, $c$-block and $fc$-block respectively. $E_f,E_c,E_{c''}$ denote the energy shifting induced by the Hartree term, one-body scattering term $\hH_{cc}$ and chemical potential. $E_f,E_c$ can be $\kk$ dependent and we only keep its $\kk$-independent part which makes dominant contributions. $E_{c''}$ comes from the Hartree contribution of $\hH_J$ (Eq.~\ref{eq:hj_hartree}, which is relatively small and we set $E_{c''}=0$. We also set $M=0$, since it is small compared to the other parameters. 
$\tilde{\gamma}, \tilde{v_\star}^\prime$ denote the renormalized $f$-$c$ hybridization emerged from Kondo interactions (Eq.~\ref{eq:hk_fock}). We also drop the damping factor $e^{-|\kk|^2\lambda^2/2}$ to simplify the analysis. In practice, we find $|\tilde{\gamma} |= \frac{1}{D_{\nu_c,\nu_f}}|\gamma^2 V_1^* + \gamma v_\star^\prime V_2^*|  \approx 175\text{meV}, 209\text{meV},406\text{meV}$. 
In the chiral limit $v_\star^\prime =0 $, we also have $\tilde{v_\star^\prime} = 0 $ ($\tilde{v_\star^\prime} = v_\star^\prime = \gamma v_\star^\prime V_1^*/D_{\nu_c,\nu_f} $). We note that $|\tilde{v}_\star^\prime| |\kk|$ can reach a similar amplitude as the $k$-independent hybridization $|\tilde{\gamma}|$. However, we expect in most regions of MBZ, $|\tilde{\gamma}|$ makes the dominant contribution. We, therefore, drop the set $\tilde{v_\star^\prime}=0$ or equivalently $v_\star^\prime =0$ as an approximation. By setting $v_\star^\prime =0$, we can further separate $\tilde{H}^{(\eta)}(\kk)$ into two blocks. The first block corresponds to the row and column indices $1,3,5$ with electron operators,$f_{\kk, 1\eta s}, c_{\kk,1\eta s}, c_{\kk,3 \eta s}$. The second block corresponds to the row and column indices $2,4,6$ with electron operators,$f_{\kk, 2\eta s}, c_{\kk,2\eta s}, c_{\kk,4 \eta s}$. We focus on the first block whose single-particle Hamiltonian is 
\baa  
&\tilde{h}^{(\eta)}(\kk) = \begin{bmatrix}
    E_f & \tilde{\gamma} & 0  \\ 
  \tilde{\gamma}  & E_{c} & v_\star(\eta k_x +i  k_y) \\ 
  0& v_\star(\eta k_x -i  k_y) & 0
\end{bmatrix} 
\label{eq:single_particle_ham_kondo_chiral}
\eaa  
We next analyze the eigensystems of $\tilde{h}^{(\eta)}(\kk)$. We note that $\tilde{\gamma}$ provides the largest energy scales near $\Gamma_M$ point and will gap out $f$- and $\Gamma_3$ $c$-electrons. To observe this, we first consider the first $2\times 2$ block of $\tilde{h}^{(\eta)}(\kk)$ which describes the single-particle Hamiltonian of  $f$- and $\Gamma_3$ $c$-electrons 
\baa  
\begin{bmatrix}
    E_f & \tilde{\gamma}   \\ 
  \tilde{\gamma}  & E_{c} 
\end{bmatrix} 
\eaa  
The eigenvalues and eigenvectors are 
\baa  
E_{1} &= \frac{E_c+E_f}{2} -\sqrt{\tilde{\gamma}^2+\frac{(E_c-E_f)^2}{4}},\quad 
E_{2}= \frac{E_c+E_f}{2} +\sqrt{\tilde{\gamma}^2+\frac{(E_c-E_f)^2}{4}}\nonumber \\ 
v_1 &= \frac{1}{\sqrt{2E_{fc}(E_{fc}-E_3)}}
\begin{bmatrix}
    E_3 -E_{fc} & \tilde{\gamma} 
\end{bmatrix}^T ,\quad  
v_2 = \frac{1}{\sqrt{2E_{fc}(E_{fc}+E_3)}}
\begin{bmatrix}
    E_3 +E_{fc} &\tilde{\gamma}
\end{bmatrix}^T \nonumber \\ 
\eaa 
where $E_3 = \frac{E_f-E_c}{2}, E_{fc}= \sqrt{\tilde{\gamma}^2 + E_3^2}$. 
Since $\tilde{\gamma}$ is larger than $E_c,E_f$, $\Gamma_3$ $c$-electrons and $f$-electrons are gapped out by the hybridization. Consequently, the flat bands are mostly formed by $\Gamma_1\oplus\Gamma_2$ $c$-electrons. Numerically, we indeed find the orbital weights of $\Gamma_1\oplus\Gamma_2$ $c$-electrons are large ($71\%,77\%,89\%$ at $\nu=0,-1,-2$ respectively). 

We next treat $v_\star$ perturbatively. We find the dispersion of the flat band becomes
\baa  
E_{\kk}^{flat} \approx  \frac{-E_f |\kk|^2 (v_\star )^2 }{E_1E_2}  =   \frac{E_f |\kk|^2 (v_\star )^2 }{\tilde{\gamma}^2 -E_cE_f }  
\eaa  
At $\nu=0$ with particle-hole symmetry, $E_f=E_c=0$ and $E_{\kk}^{flat} \approx 0$. However, at $\nu=-1,-2$, where $E_f \ne 0, E_c \ne 0$, flat bands become dispersive. 
We observe that $|E_c|$ is much smaller than $|\tilde{\gamma}|$ at $\nu=-1,-2$. $E_f$ increases as we change from $\nu=0$ to $\nu=-2$, because we are doping more holes to the $f$-orbitals. At $\nu=-2$, $E_f$ can reach $\sim 0.5|\tilde{\gamma}|$, but at $\nu=-1$, $E_f\sim 0.1|\tilde{\gamma}|$. Approximately, the dispersion of the flat band is $E_\kk^{flat} \approx (v_\star)^2 E_f/\tilde{\gamma}^2$. At $\nu=-1$, we have $E_\kk \approx 13\mathrm{meV}\cdot \mathrm{\mathring{A}^2} |\kk|^2 $ and, at $\nu=-2$, we have $E_\kk \approx 45\mathrm{meV}\cdot \mathrm{\mathring{A}^2} |\kk|^2$. This indicates a larger dispersion at $\nu=-2$, which is consistent with our numerical result shown in the main text Fig.1.

We next analyze the wavefunctions of the flat bands. The corresponding electron operator of the flat band $d^\dag_{flat,\kk}$ is 
\baa  
d^\dag_{flat,\kk } \approx \frac{1}{A_\kk} \bigg[ 
c^\dag_{\kk, 3\eta s} + \frac{ v_\star (\eta k_x -ik_y) }{E_cE_f-\tilde{\gamma}^2} \bigg(-E_f c^\dag_{\kk,1\eta s} + \tilde{\gamma} f^\dag_{\kk,1\eta s} \bigg) \bigg] 
\label{eq:flat_band_d_op}
\eaa 
where the normalization factor 
\baa  
A_\kk = \sqrt{ 1 + \frac{ |v_\star|^2 |\kk|^2 (E_f^2 +\tilde{\gamma}^2) }{(E_cE_f-\tilde{\gamma}^2)^2} }
\eaa  
We observe that in the large $|\tilde{\gamma}|$ limit, the flat bands are mostly formed by $\Gamma_1\oplus \Gamma_2$ $c$-electrons ($c_{\kk,3\eta s}^\dag$). 
We also provide the Berry curvature derived from the wavefunction in Eq.~\ref{eq:flat_band_d_op}
\baa  
\Omega(\kk) = \frac{-2(E_cE_f-\tilde{\gamma}^2)^2(E_f^2+\tilde{\gamma}^2)v_\star^2 }{\bigg[ 
(E_cE_f-\tilde{\gamma}^2)^2 + (E_f^2+\tilde{\gamma}^2)v_\star^2|\kk|^2
\bigg]^2} 
\eaa  

We next calculate the Wilson loop from the wavefunction in Eq.~\ref{eq:flat_band_d_op}. The wavefunctions of $d_{flat,\kk}^\dag$ is 
\baa  
u(\kk) = \frac{1}{A_\kk} 
\begin{bmatrix}
    1 & \frac{v_\star \eta k_x -i k_y }{E_cE_f-\tilde{\gamma}^2} (-E_f) & \frac{v_\star \eta k_x -i k_y }{E_cE_f-\tilde{\gamma}^2} \tilde{\gamma}  
\end{bmatrix}^T
\label{eq:eigenvector_flat_band}
\eaa  
where the first, second and third rows denote $c^\dag_{\kk,3 \eta s}, f^\dag_{\kk,1 \eta s},c^\dag_{\kk, 1\eta s}$ respectively. 
We then parametrize the momentum as 
\baa  
&\kk = x_1 a_M\textbf{b}_{M,1} +x_2 a_M \textbf{b}_{M,2}, \quad \textbf{b}_{M,1} = \frac{4\pi}{3a_M}(\sqrt{3},0),\quad \textbf{b}_{M,2}=\frac{4\pi}{3a_M}(\frac{\sqrt{3}}{2}, \frac{3}{2}) \\
& 
x_1,x_2 \in [-\frac{1}{2},\frac{1}{2}]\frac{1}{a_M}
\eaa   
and define $|u(x_1,x_2)\rangle $ as $|u(\kk)\rangle$ with $\kk= x_1 a_M\textbf{b}_{M,1} +x_2 a_M \textbf{b}_{M,2}$.
The Wilson loop can be written as 
\baa  
W(x_1) = \prod_{j=0}^{N-1} \langle u( x_1, x_2 = x_{2,i}) |u(x_1,x_2 = x_{2,j+1} \rangle  \langle u(x_1,x_{2,N} ) | u(x_1,x_{2,0})\rangle,\quad x_{2,i} =-\frac{1}{2a_M} + \frac{1}{a_M} \frac{i}{N}
\eaa  
The spectrum of the Wilson loop is  
\baa  
N(x_1) = -i\ln(W(x_1)) =-i \int_{-\frac{1}{2a_M}}^{\frac{1}{2a_M}} 
\langle u(x_1,x_2) | \partial_{x_2} | u(x_1,x_2)\rangle 
dx_2  -i\ln(\langle u(x_1,1/(2a_M) ) | u(x_1,-1/(2a_M)\rangle)
\eaa  
In the continuous limit with $a_M\rightarrow 0$, we find
\baa 
N(x_1) = -i\int_{-\infty }^{\infty } 
\langle u(x_1,x_2) | \partial_{x_2} | u(x_1,x_2)\rangle 
dx_2  -i\ln(\langle u(x_1,\infty ) | u(x_1,-\infty \rangle)
\label{eq:wilson_loop_cont}
\eaa  
Combining Eq.~\ref{eq:eigenvector_flat_band} and Eq.~\ref{eq:wilson_loop_cont}, we find 
\baa  
N(x_1) =& \pi \bigg( 1+ 
\frac{v_\star^2 x_1 }{ \sqrt{x_1^2 v_\star^2  + \frac{(\tilde{\gamma}^2 -E_cE_f)^2}{E_f^2+\tilde{\gamma}^2} } }\bigg) 
\eaa  
Even though Eq.~\ref{eq:wilson_loop_cont} is calculated from 
the perturbative wavefunction in Eq.~\ref{eq:eigenvector_flat_band}, it qualitatively captures the behaviors of the Wilson loop shown in Fig.~\ref{fig:wilson_loop}. We observe that $N(-\infty) = 0$ and $N(\infty) = 2\pi$, which indicates a $2\pi$ winding at $\nu=0,-1,-2$. We also mention that current calculations correspond to one of the two flat bands for each valley and each spin, because we only pick one block of the single-particle Hamiltonian as we discussed near Eq.~\ref{eq:single_particle_ham_kondo_chiral}. The other flat band can be derived in the same manner and has similar behaviors, since it has a similar single-particle Hamiltonian.

\section{Mean-field solutions of the topological heavy-fermion model }
We now discuss the mean-field equations of topological heavy-fermion mode in Eq.~\ref{eq:thf_ham}. We use a similar Hartree-Fock approximation as introduced in Ref.~\cite{HF_MATBLG}. However, we decouple $\hH_J$ via Eq.~\ref{eq:mf_kondo_hj}.
The mean-field expectation values we considered are
\baa  
&O^f_{\alpha \eta s,\alpha'\eta' s'} = \frac{1}{N_M}\sum_\RR \langle \Psi| :f_{\RR,\alpha \eta s}^\dag f_{\RR,\alpha' \eta' s'} : |\Psi\rangle 
\nonumber \\ 
&O^{c'}_{a \eta s,a'\eta's'} = \frac{1}{N_M}\sum_{|\kk|<\Lambda_c} \langle \Psi| :c_{\kk,a \eta s}^\dag c_{\kk, a' \eta' s'} : |\Psi\rangle, \quad a\in \{1,2\}
\nonumber \\ 
& O^{c''}_{a \eta s,a'\eta's'} = \frac{1}{N_M}\sum_{|\kk|<\Lambda_c} \langle \Psi| :c_{\kk,a+2 \eta s}^\dag c_{\kk, a'+2 \eta' s'} : |\Psi\rangle,\quad a,a' \in \{1,2\} \nonumber \\
&O^{c'f}_{a\eta s, \alpha'\eta's'} =  \frac{1}{\sqrt{N}N}\sum_{|\kk|<\Lambda_c, \RR }e^{-i\kk\cdot \RR}
\langle \Psi| c_{\kk,a\eta s}^\dag f_{\RR,\alpha '\eta' s'} |\Psi\rangle ,\quad a\in\{1,2\}\nonumber \\
&V_3 =  \sum_{\RR, |\kk| <\Lambda_c}
\sum_{\alpha\eta, s}\frac{e^{i\kk \cdot \RR  }\delta_{1, \eta (-1)^{\alpha+1}}}{{N_M}\sqrt{N_M}} \langle \Psi| f_{\RR,\alpha \eta s}^\dag c_{\kk,\alpha+2\eta s} |\Psi\rangle 
\nonumber  \\ 
&V_4 =  \sum_{\RR, |\kk| <\Lambda_c}
\sum_{\alpha\eta, s}\frac{e^{i\kk \cdot \RR  }\delta_{-1, \eta (-1)^{\alpha+1}}}{{N_M}\sqrt{N_M}} \langle \Psi| \eta f_{\RR,\alpha \eta s}^\dag c_{\kk,\alpha+2\eta s} |\Psi\rangle  
\label{eq:thf_mf}
\eaa  
where $O^{f},O^{c''},V_3,V_4$ have also been used in the Kondo lattice mean-field calculations (Eq.~\ref{eq:sc_eq_1}, Eq.~\ref{eq:sc_eq_2} and Eq.~\ref{eq:sc_eq_2_v}). 

In addition, THF model also has a chemical potential term $\hH_\mu$ (Eq.~\ref{eq:thf_mu}) 
and we determine $\mu$ by requiring the total filling of $f$- and $c$-electrons to be $\nu$:
\baa 
\nu = \text{Tr}[O^f] +\text{Tr}[O^{c'}]+\text{Tr}[O^{c''}]
\label{eq:fill_const_thf}
\eaa 
where we note that the filling of $f$-, $\Gamma_3$ $c$- and $\Gamma_1\oplus\Gamma_2$ $c$-electrons are 
\baa 
\nu_f = \text{Tr}[O^f] ,\quad \nu_{c'} = \text{Tr}[O^{c'}] ,\quad \nu_{c''} = \text{Tr}[O^{c''}]\, ,
\eaa 
respectively.

We discuss the difference and similarities between the mean-field equations of the KL model and that of the THF model. For the THF model, we introduce mean fields $O^f,O^{c'},O^{c''},O^{c'f}, V_3,V_4$ (Eq.~\ref{eq:thf_mf}) (for a generic state without enforcing any symmetries). As for KL model, we introduce mean fields $V_1,V_2,O^f, O^{c',1},O^{c',2}, O^{c''}, V_3,V_4$ (Eq.~\ref{eq:sc_eq_1}, Eq.~\ref{eq:sc_eq_1_v}, Eq.~\ref{eq:sc_eq_2}, Eq.~\ref{eq:sc_eq_2_v}) (for a generic state without enforcing any symmetry). 
\begin{itemize}
    \item For both models, $O^{c''},O^{f}, V_3,V_4$ are part of mean fields and contribute the mean-field decoupling of $\hH_J$ (Eq.~\ref{eq:mf_kondo_hj}).
    \item In the THF model, we introduce a chemical potential term $\mu$ that couples to both the $f$-electron density operators and $c$-electron density operator (Eq.~\ref{eq:thf_mu}. We enforce the total filling of $f$- and $c$-electrons to be $\nu$ by tuning $\mu$. In the KL model, we introduce a Lagrangian multiplier $\lambda_f$ (Eq.~\ref{eq:ham_lamf}) that couples to $f$-electron density operators, and a chemical potential $\mu_c$ (Eq.~\ref{eq:hmuc}) that couples to the $c$-electron density operators. We enforce the fillings of $f$-electrons and $c$-electrons to be $\nu_f$ and $\nu_c$ respectively by tuning $\lambda_f$ and $\mu_c$ in the KL model. 
    \item We also mention that $O^{c'}$ in THF model (Eq.~\ref{eq:thf_mf}) and $O^{c',1}$ in the KL model (Eq.~\ref{eq:sc_eq_1}) are different, where the latter one has included an additional damping factor $e^{-|\kk|^2\lambda^2}$.
    \item In the THF model, we do not need hybridization fields $V_1,V_2$ (Eq.~\ref{eq:sc_eq_1_v}), since both come from the decoupling of Kondo interactions that only appear in the KL model.
\end{itemize}

\subsection{Mean-field equations of fully symmetric state}
We next discuss the solution of the symmetric state The fully symmetric state is characterized by density matrices $O^f, O^{c'},O^{c''},O^{c'f}$ and hybridization fields $V_3,V_4$ that satisfy all symmetries. The structures of $O^{f},O^{c''}$ in the fully symmetric state are given in Eq.~\ref{eq:ocf_from_alpha_kondo}. We also prove that $V_3=V_4=0$ in a fully symmetric state (near Eq.~\ref{eq:v3v4_zero}).
We now discuss the symmetry properties of $O^{c'}, O^{c'f}$.
 From Eq.~\ref{eq:cont_sym}, a $U(1)_v$ symmetric solution satisfies
\baa  
O^{c'}_{a \eta s, \alpha ' \eta' s'} = O^{c'}_{a \eta s, \alpha '\eta's'}e^{-i\theta_\nu(\eta-\eta')} \Rightarrow O^{c'}_{a \eta s, \alpha  -\eta s'}  =0 \nonumber \\
O^{c'f}_{a \eta s, \alpha ' \eta' s'} = O^{c'f}_{a \eta s, \alpha '\eta's'}e^{-i\theta_\nu(\eta-\eta')} \Rightarrow O^{c'f}_{a \eta s, \alpha  -\eta s'}  =0 
\label{eq:ofc_valley_diag}
\eaa  
Then, $O^{c'},O^{c'f}$ is block diagonalized in valley indices. We next consider a $SU(2)_\eta$ transformation acting on the valley $\eta$. It indicates
\baa  
\sum_{s,s'} [e^{i\sum_\mu \theta^\eta_\mu \sigma^\mu}]_{s_2,s}O^{c'}_{\alpha \eta s, \alpha' \eta s'} [e^{i\sum_\mu \theta^\eta_\mu \sigma^\mu}]_{s',s_2'} = O^{c'}_{\alpha \eta s_2, \alpha' \eta s_2'} \Rightarrow O^{c'}_{\alpha \eta s, \alpha' \eta s'}  \nonumber \\ 
\sum_{s,s'} [e^{i\sum_\mu \theta^\eta_\mu \sigma^\mu}]_{s_2,s}O^{c'f}_{\alpha \eta s, \alpha' \eta s'} [e^{i\sum_\mu \theta^\eta_\mu \sigma^\mu}]_{s',s_2'} = O^{c'f}_{\alpha \eta s_2, \alpha' \eta s_2'} \Rightarrow O^{c'f}_{\alpha \eta s, \alpha' \eta s'} 
\label{eq:ofc_spin_diag}
\eaa  
Combining Eq.~\ref{eq:ofc_valley_diag} and Eq.~\ref{eq:ofc_spin_diag}, we can introduce $2\times 2$ matrices $o^{c',\eta}, o^{c'f,\eta}$, such that
\baa  
&O^{c'}_{\alpha \eta s, \alpha'\eta's'} = o^{c'}_{\alpha,\alpha'}\delta_{s,s'}\delta_{\eta,\eta'},\quad 
O^{c'f}_{\alpha \eta s, \alpha'\eta's'} = o^{c'f}_{\alpha ,\alpha'}\delta_{s,s'}\delta_{\eta,\eta'} 
\label{eq:ofc_small_o}
\eaa   
We now consider the effect of discrete symmetries in Eq.~\ref{eq:desc_sym_op}. Using Eq.~\ref{eq:desc_sym} and Eq.~\ref{eq:ofc_small_o}, we find 
\baa  
T:&\quad  (o^{c',\eta})^* = o^{c',-\eta},\quad (o^{c'f,\eta})^* = o^{c'f,-\eta} 
\nonumber \\
C_{3z}:&\quad 
e^{i2\pi/3 \eta  \sigma_z } o^{c',\eta}e^{-i2\pi\eta /3\sigma_z} = o^{c',\eta} ,\quad 
e^{i2\pi/3 \eta  \sigma_z } o^{c'f,\eta}e^{-i2\pi\eta /3\sigma_z} = o^{c'f,\eta}  \nonumber \\ 
C_{2x}:&\quad 
 \sigma_x o^{c',\eta}\sigma_x = o^{c',\eta} ,\quad 
\sigma_x o^{c'f,\eta}\sigma_x = o^{c'f,\eta} 
\nonumber \\ 
C_{2z}T:&\quad
\sigma_x (o^{c',\eta})^*\sigma_x = o^{c',\eta} ,\quad
\sigma_x (o^{c'f,\eta})^*\sigma_x = o^{c'f,\eta} 
\label{eq:desc_sym_ofc}
\eaa 
Then we can introduce a single real number $\chi_0^{c'}, \chi_0^{c'f}$ to characterize the density matrices
\baa  
O^{c'}_{\alpha \eta s,\alpha'\eta's'} = \chi_0^{c'}\delta_{\alpha,\alpha'}\delta_{\eta,\eta'},\quad 
O^{c'f}_{\alpha \eta s,\alpha'\eta's'} = \chi_0^{c'f}\delta_{\alpha,\alpha'}\delta_{\eta,\eta'}\delta_{s,s'} \, .
\label{eq:ocf_from_alpha}
\eaa  
Since the filling of $\Gamma_3$ $c$-electrons is $\nu_{c'} = \text{Tr}[O^{c'}]=8\chi_0^{c'}$, we let $\chi_0^{c'} = \nu_{c'}/8$. 
Therefore, instead of calculating the original density matrices in Eq.~\ref{eq:thf_mf}, we can calculate the following quantities 
\baa  
&\nu_f = \frac{1}{N_M} \sum_{\RR, \alpha \eta s }\langle \Psi| :f_{\RR,\alpha \eta s} ^\dag f_{\RR,\alpha \eta s}:  |\Psi\rangle  \nonumber \\ 
&\nu_{c'}=\frac{1}{N_M} \sum_{|\kk|<\Lambda_c, a=1,2, \eta s }\langle \Psi| :c_{\kk,a \eta s} ^\dag c_{\kk,a \eta s}:  |\Psi\rangle  \nonumber \\ 
&\nu_{c''}=\frac{1}{N_M} \sum_{|\kk|<\Lambda_c, a=3,4, \eta s }\langle \Psi| :c_{\kk,a \eta s} ^\dag c_{\kk,a \eta s}:  |\Psi\rangle    \nonumber \\
& \chi^{c'f} = \frac{1}{8N_M\sqrt{N_M}}\sum_{|\kk|<\Lambda_c,\RR, \alpha \eta s}e^{-i\kk\cdot\RR}\langle \Psi |c^\dag_{\kk, \alpha \eta s} f_{\RR,\alpha \eta s}|\Psi \rangle 
\label{eq:sc_sym_thf}
\eaa  
and construct density matrices via Eq.~\ref{eq:ocf_from_alpha_kondo} and Eq.~\ref{eq:ocf_from_alpha}. In addition, the filling constraints in Eq.~\ref{eq:fill_const_thf} becomes 
\baa  
\nu = \nu_f + \nu_{c'} +\nu_{c''}
\label{eq:fill_const_thf_sym}
\eaa  
Combining Eq.~\ref{eq:sc_sym_thf} and Eq.~\ref{eq:fill_const_thf_sym}, we have a complete set of the mean-field self-consistent equations of symmetric state.

Here we discuss the differences and similarities between the symmetric solution in the KL model (Sec.~\ref{sec:prop_symm_kondo}) and the symmetric solution in the THF model as introduced in this section.
\begin{itemize}
    \item Both Kondo symmetric (KS) state in KL model and the symmetric state in the THF model preserve all the symmetries. 
    \item The KS state is adiabatically connected to the symmetric state in the THF model. 
    \item The mean-field solutions are exact at $N=\infty$.
    \item To obtain a more precise description of the Kondo state, we need to introduce a Gutzwiller projector to our symmetric-state wavefunction in the THF model. The Gutzwiller projector will suppress the charge fluctuations of $f$-electrons and is expected to further lower the energy of the symmetric state. 
    \item We also comment that, in the THF model, the flat bands are mainly formed by $f$-electrons with $f$-electron orbital weights $80\%$, $85\%$ and $87\%$ at $\nu=0,-1,-2$ respectively. However, in the KL model, the flat bands are mainly formed by $\Gamma_1\oplus\Gamma_2$ electrons as discussed in Sec.~\ref{sec:prop_symm_kondo}. This is because, in the KL model, we observed an enhanced $f$-$c$ hybridization driven by Kondo interactions which is absent in the symmetric state solution of THF model. We expect the enhanced hybridization will be recovered after introducing the Gutzwiller projector. 
    \item We find the spectrum of the Wilson loop in the symmetric state of the THF model has winding number one and three crossing points(Fig.~\ref{fig:wilsonloop_hf}).  However, in the SK state of the KL model, the spectrum of the Wilson loop spectrum has winding number one and one crossing point (Fig.~\ref{fig:wilson_loop}). However, at $\nu=0$, for both the THF model and the KL model, the symmetric state has a stable topology with an odd Winding number of Wilson loop spectrum~\cite{SON19}.  We also mention the difference in the Wilson loop spectrum comes from the absence of enhanced $f$-$c$ hybridization in the THF model.
\end{itemize}

\begin{figure}
    \centering
    \includegraphics[width=1.0\textwidth]{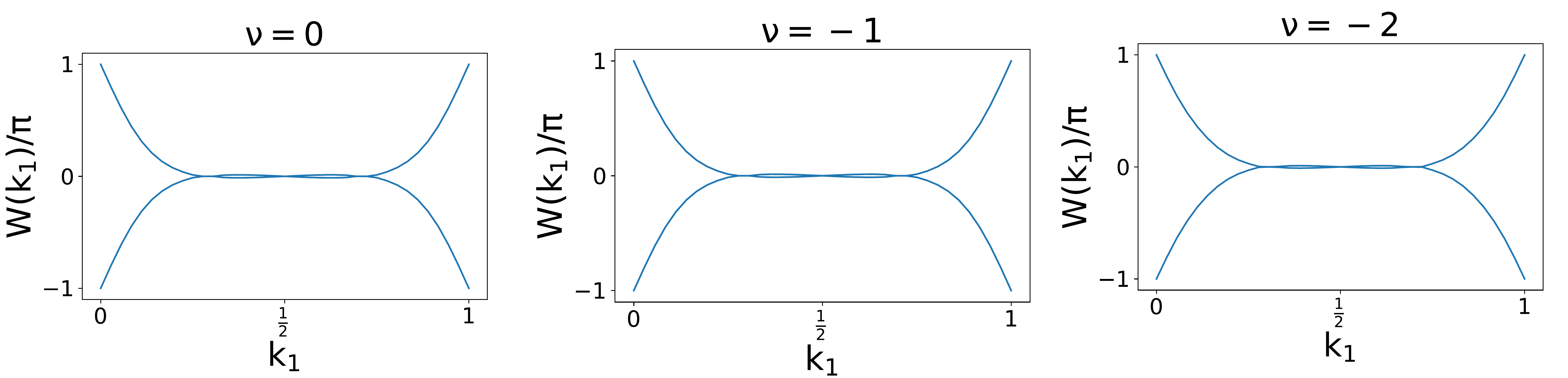}
    \caption{Wilson loop spectrum of symmetric state in the THF model at $\nu=0,-1,-2$}
    \label{fig:wilsonloop_hf}
\end{figure}
\subsection{Mean-field equations of the symmetric state in the presence of strain}
We now discuss the mean-field solution of the symmetric state in the presence of strain. We add the following term to the Hamiltonian (Eq.~\ref{eq:thf_ham})
\baa  
\hH_{strain} = \alpha \sum_{\RR,\eta s}(f_{\RR,1\eta s}^\dag f_{\RR,2\eta s} +\text{h.c.}) 
\eaa 
We note that $\hH_{strain}$ only breaks $C_{3z}$ symmetry. To show this, we rewrite the $\hH_{strain}$ as 
\baa  
&\hH_{strain} = \sum_{\RR,\alpha \eta s,\alpha'\eta's'} \alpha f_{\RR,\alpha \eta s}^\dag  [h_{strain}]_{\alpha\eta s,\alpha'\eta' s'} f_{\RR,\alpha'\eta' s'} \nonumber \\ 
&
h_{strain} =\sigma_x\tau_0\varsigma_0 \label{eq:hstrain}
\eaa  
where the matrix structure of $\hH_{strain}$ is denoted by $h_{strain} \sigma_x \tau_0 \varsigma_0$. We now show it commutes with $D^f(T),D^f(C_{2x}),D^f(C_{2z}T),D^{f}(g_{SU(2)_\eta}(\theta_\eta^\mu) ), D^f( g_{U(1)_v}((\theta_v) ),D^f( g_{U(1)_c}((\theta_c) )$ (Eq.~\ref{eq:desc_sym}, Eq.~\ref{eq:cont_sym}) and hence $\hH_{strain}$ preserves all symmetries except for $C_{3z}$
\baa  
&[h_{strain}, D^f(T)] = [\sigma_x\tau_0\varsigma_0,\sigma_0\tau_x\varepsilon_0] =0 \nonumber \\
&[h_{strain}, D^f(C_{2x})] = [\sigma_x\tau_0\varsigma_0,\sigma_x\tau_0 \varsigma_0] =0 \nonumber \\
&[h_{strain}, D^f(C_{2z}T)] = [\sigma_x\tau_0\varsigma_0,\sigma_x\tau_0 \varsigma_0] =0 \nonumber \\
&[h_{strain}, D^f( g_{U(1)_c}((\theta_c) )] = [\sigma_x\tau_0\varsigma_0,e^{-i\theta_c }\sigma_0\tau_0\varsigma_0] =0 \nonumber \\
&[h_{strain}, D^f( g_{U(1)_v}((\theta_v) )] = [\sigma_x\tau_0\varsigma_0, \sigma_0e^{-i\theta_v \tau_z  }\varsigma_0 ] =0 \nonumber \\
&[h_{strain},D^{f}(g_{SU(2)_\eta}(\theta_\eta^\mu) )] = [\sigma_x\tau_0\varsigma_0, \sigma_0 e^{ -i \sum_\mu \theta_\mu^\eta  
\frac{\tau_0+\eta \tau_z}{4} \varsigma_\mu }] =0  \nonumber 
\eaa  

In the presence of strain, the symmetric state is defined as the state that preserves all the symmetries except for $C_{3z}$ which is broken by the strain.
In the presence of $C_{3z}$-breaking strain, Eq.~\ref{eq:small_o} and Eq.~\ref{eq:ofc_small_o} still hold, because the system still has $U(1)_c\times U(1)_v \times SU(2)_+ \times SU(2)_-$ symmetry. As for Eq.~\ref{eq:desc_sym_o} and Eq.~\ref{eq:desc_sym_ofc}, we only need to consider $T,C_{2x},C_{2z}T$ and we find 
\baa  
&O^f_{\alpha\eta s,\alpha'\eta's'} =\bigg[ \chi_0^f \sigma_0 + \chi_1^f\sigma_x\bigg]_{\alpha ,\alpha'} \delta_{\eta ,\eta'}\delta_{s,s'} \nonumber \\ 
&O^{c'}_{\alpha\eta s,\alpha'\eta's'} =\bigg[ \chi_0^{c'} \sigma_0 + \chi_1^{c'}\sigma_x\bigg]_{\alpha ,\alpha'} \delta_{\eta ,\eta'}\delta_{s,s'} ,\quad 
O^{c''}_{\alpha\eta s,\alpha'\eta's'} =\bigg[ \chi_0^{c''} \sigma_0 + \chi_1^{c''}\sigma_x\bigg]_{\alpha ,\alpha'} \delta_{\eta ,\eta'}\delta_{s,s'} ,\nonumber \\
&
O^{c'f}_{\alpha\eta s,\alpha'\eta's'} =\bigg[ \chi_0^{c'f} \sigma_0 + \chi_1^{c'f}\sigma_x\bigg]_{\alpha ,\alpha'} \delta_{\eta ,\eta'}\delta_{s,s'} 
\label{eq:o_from_chi_strain}
\eaa  
where $\chi_0^f,\chi_1^f,\chi_0^{c'},\chi_1^{c'},\chi_0^{c''},\chi_1^{c''},\chi_0^{c'f},\chi_1^{c'f}$ are real numbers tha characterize the density matrices. 
Combining Eq.~\ref{eq:sc_eq_1}, Eq.~\ref{eq:sc_eq_2} and Eq.~\ref{eq:o_from_chi_strain}, we find 
\baa  
&\chi_0^f = \frac{1}{8N_M} \sum_{ \RR,\alpha 
eta s}\langle \Psi|:f_{\RR,\alpha \eta s}^\dag f_{\RR,\alpha \eta s}: |\Psi\rangle ,\quad \chi_1^f = \frac{1}{8N_M} \sum_{\RR,\eta s}\langle \Psi|f_{\RR,1\eta s}^\dag f_{\RR,2\eta s} + 
f_{\RR,2\eta s}^\dag f_{\RR,1\eta s} |\Psi\rangle \nonumber \\
&\chi_0^{c'} = \frac{1}{8N_M} \sum_{ |\kk|<\Lambda_c, =1,2,
\eta s}\langle \Psi|:c_{\kk,a\eta s}^\dag c_{\kk,a \eta s}^\dag: |\Psi\rangle ,\quad \chi_1^{c'} = \frac{1}{8N_M} \sum_{|\kk|<\Lambda_c,\eta s}\langle \Psi|c_{\kk,1\eta s}^\dag c_{\kk,2\eta s} + 
c_{\kk,2\eta s}^\dag c_{\kk,1\eta s} |\Psi\rangle \nonumber \\
&\chi_0^{c''} = \frac{1}{8N_M} \sum_{ |\kk|<\Lambda_c, =3,4,
\eta s}\langle \Psi|:c_{\kk,a\eta s}^\dag c_{\kk,a \eta s}^\dag: |\Psi\rangle ,\quad \chi_1^{c''} = \frac{1}{8N_M} \sum_{|\kk|<\Lambda_c,\eta s}\langle \Psi|c_{\kk,3\eta s}^\dag c_{\kk,4\eta s} + 
c_{\kk,4\eta s}^\dag c_{\kk,3\eta s} |\Psi\rangle \nonumber \\
&\chi_0^{c'f} = \frac{1}{8N_M\sqrt{N_M}}\sum_{|\kk|<\Lambda_c,\RR, \alpha \eta s}e^{-i\kk\cdot\RR}\langle \Psi |c^\dag_{\kk, \alpha \eta s} f_{\RR,\alpha \eta s}|\Psi \rangle  ,\quad \nonumber \\
&\chi_1^{c'f} = \frac{1}{8N_M\sqrt{N_M}}\sum_{|\kk|<\Lambda_c,\RR, \eta s}e^{-i\kk\cdot\RR}\langle \Psi |c^\dag_{\kk, 1 \eta s} f_{\RR,2 \eta s}+c^\dag_{\kk, 2 \eta s} f_{\RR,1 \eta s}|\Psi \rangle  \nonumber \\ 
&\chi_0^{c''f} = \frac{1}{8N_M\sqrt{N_M}}\sum_{|\kk|<\Lambda_c,\RR, \alpha \eta s}e^{-i\kk\cdot\RR}\langle \Psi |c^\dag_{\kk, \alpha+2 \eta s} f_{\RR,\alpha \eta s}|\Psi \rangle  ,\quad \nonumber \\
&\chi_1^{c''f} = \frac{1}{8N_M\sqrt{N_M}}\sum_{|\kk|<\Lambda_c,\RR, \eta s}e^{-i\kk\cdot\RR}\langle \Psi |c^\dag_{\kk, 3 \eta s} f_{\RR,2 \eta s}+c^\dag_{\kk, 4 \eta s} f_{\RR,3 \eta s}|\Psi \rangle 
\label{eq:sc_sym_strain_thf} 
\eaa  
where we also have $\chi_0^f = \nu_f/8, \chi_0^{c'}=\nu_{c'}/8, \chi_0^{c''}=\nu_{c''}/8$. 
As for $V_3,V_4$, we only consider the $T,C_{2x},C_{z}T$ of Eq.~\ref{eq:descrete_v}, which indicates 
\baa  
V_3 =V_4 = V_3^* =V_4^*
\label{eq:descrete_v_c3z}
\eaa  
Combining Eq.~\ref{eq:sc_eq_2_v}, Eq.~\ref{eq:fill_const_thf_sym}, Eq.~\ref{eq:sc_sym_strain_thf} and Eq.~\ref{eq:descrete_v_c3z} with filling constraints $\nu=\nu_f+\nu_{c'}+\nu_{c''}$, we have a complete set of the mean-field self-consistent equations of symmetric state in the presence of strain. We perform calculations with non-zero strain at $\nu=0,-1,-2,-3$. We initialize the calculations with the fully symmetric solutions derived at zero strain, and the procedure converges within 500 iterations. 
The results are illustrated and discussed in Sec.~\ref{sec:strain}.

\subsection{Effect of doping}
We now discuss the effect of doping at zero strain. For hole doping at $\nu=0,-1,-2,-3$ and electron doping at $\nu=0$, we mainly dope electrons to the light bands that are mostly formed by $c$-electrons (Fig.~\ref{fig:disp_heavy_light}). Consequently, the energy difference between the symmetric state and the ordered state decreases since we have more conduction $c$-electrons near the Fermi energy, and the system favors the symmetric state. 

 We now point out the complexity of electron dopings at $\nu=-1,-2$.
Doping electrons at $\nu=-1,-2$ is equivalent to dope electrons to the heavy bands that are mostly formed by $f$-electrons (Fig.~\ref{fig:disp_heavy_light}). The heavy (flat) bands become closer to the Fermi energy, and hence, the energy cost of putting $f$-electrons into flat bands will be small. Then we can fill the heavy (flat) bands with a small energy cost. By filling the heavy (flat) bands, the type of orders formed by $f$-electrons can change a lot. 
To observe the change of the ordered states, we consider the following order parameters
\baa  
&O_{x} =\frac{1}{N_M} \sum_{\RR,\alpha \eta s,\alpha'\eta's'} f_{\RR,\alpha \eta s}^\dag [o_{x}]_{\alpha \eta s,\alpha'\eta' s'} f_{\RR,\alpha'\eta's'},\quad  x \in \{KIVC,S_z,V_z,V_y\}\nonumber \\
&o_{KIVC} = \sigma_y \tau_y\varsigma_0,\quad o_{S_z} =  \sigma_0 \tau_0 \varsigma_z ,
\quad 
o_{V_z} = \sigma_0 \tau_z \varsigma_0 
,\quad 
o_{V_y} = \sigma_0 \tau_y \varsigma_0
\label{eq:order_para}
\eaa 
We measure the expectation values of $O_{x=KIVC,S_z,V_z,V_y}$ with respect to the ordered states at each filling. In Fig.~\ref{fig:order_dop}, we show the evolution of $\langle O_x\rangle$ as a function of doping. We find for hole doping at $\nu=0,-1,-2$ and electron doping at $\nu=0$ (where carriers go to light bands in both cases), the system stays in the same ordered states (compared to the integer filling). 
However, for electron doping at $\nu=-1,-2$, we can observe the changes of the order parameters. This is because we are mainly dope $f$-electrons 
for electron doping at $\nu=-1,-2$. 
We thus conclude that electron doping at $\nu=-1,-2$ will introduce sizeable changes of the order parameters. Both the change of order parameters and the doping effect will affect the energy competition between the symmetric state and the ordered state. 

Finally, we comment on the $\nu=-3$ case. At $\nu=-3$, the actual ground state might be a CDW state which breaks the translational symmetry~\cite{xie2022phase} and is beyond our current consideration. In addition, at $\nu=-3$, even for the valley polarized state we currently considered, electron doping is equivalent to doping both heavy and light bands (Fig.~\ref{fig:disp_heavy_light}), which is different from $\nu=-1,-2$. We leave the detailed study of $\nu=-3$ for future study.

\begin{figure}
    \centering
    \includegraphics[width=1.0\textwidth]{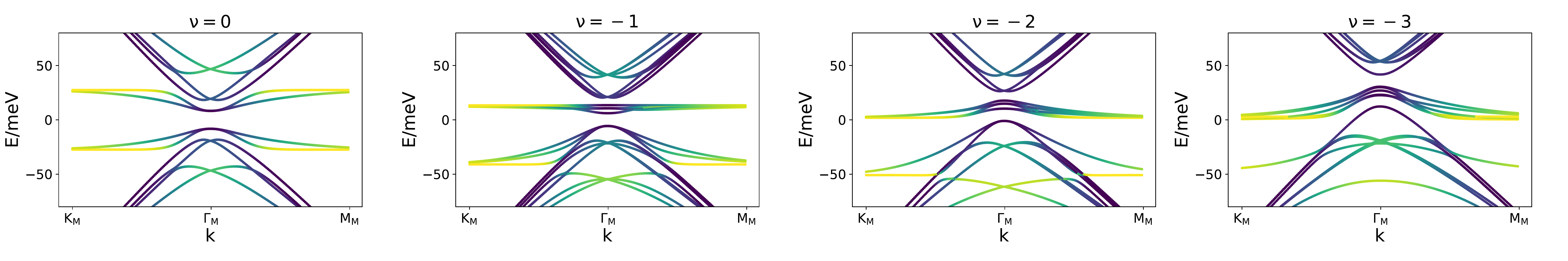}
    \caption{Dipsersions of KIVC state at $\nu=0$, KIVC+VP state at $\nu=-1$, KIVC state at $\nu=-2$ and VP state at $\nu=-3$. The color represents the weight of $f$-(yellow) and $c$-(blue) electrons.  }
    \label{fig:disp_heavy_light}
\end{figure}
\begin{figure}
    \centering
    \includegraphics[width=1.0\textwidth]{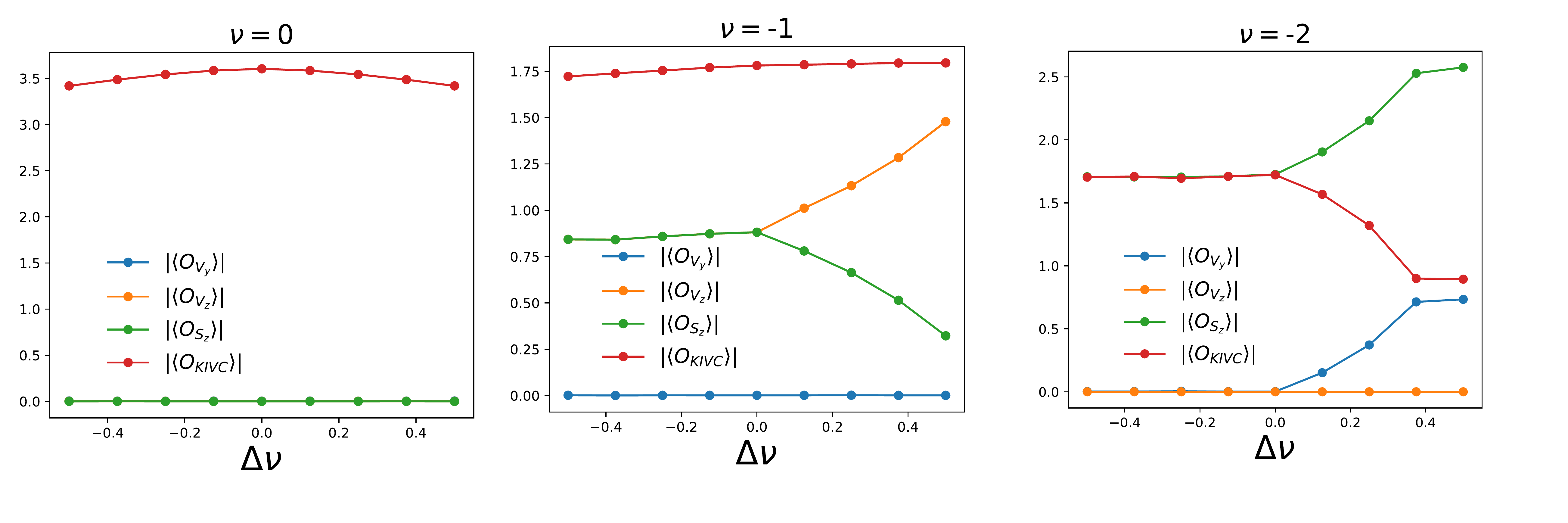}
    \caption{Evolution of order parameter as a function of doping at $\nu=0,-1,-2$}
    \label{fig:order_dop}
\end{figure}

\subsection{Effect of strain}
\label{sec:strain}
We next analyze the effect of the strain. 
We first note that as we increase $\alpha$ (Eq.~\ref{eq:hstrain}), the strain will gradually suppress the KIVC order ($O^{KIVC}$, Eq.~\ref{eq:order_para}). This can be observed from 
\baa  
\{ o_{KIVC} ,h_{strain}\} = \{\sigma_y \tau_y\varsigma_0, \sigma_x \tau_0\varsigma_0 \} = 0
\eaa  
Heuristically, the anti-commuting nature indicates the competition between $o_{KIVC}$ and $h_{strain}$. Thus, as we increase $h_{strain}$, $o_{KIVC}$ will be suppressed. We also find the spin-polarization $O_{S_z}$ and valley polarization $O_{V_z}$ commute with $h_{strain}$
\baa  
[o_{S_z}, h_{strain}] = [\sigma_0\tau_0\varsigma_z, \sigma_x \tau_0\varsigma_0] = 0 ,
\quad 
[o_{V_z}, h_{strain}] = [\sigma_0\tau_z\varsigma_0, \sigma_x \tau_0\varsigma_0] = 0 
\eaa  
Heuristically, this indicates the valley and spin polarization do not directly compete with $h_{strain}$. However, as we will show in this section, a sufficiently large strain could still destroy the valley and spin polarization in the THF model. 

For future convenience, we also introduce the eigenstates of the strain Hamiltonian $\hH_{strain}$ (Eq.~\ref{eq:hstrain})
\baa  
d_{\RR,1\eta s }^\dag = \frac{1}{\sqrt{2}}(f_{\RR,1\eta s}^\dag -f_{\RR,2\eta s}^\dag )
,\quad 
d_{\RR,2\eta s }^\dag = \frac{1}{\sqrt{2}}(f_{\RR,1\eta s}^\dag +f_{\RR,2\eta s}^\dag )
\eaa  
We will call $d_{\RR,1\eta s }^\dag$ and $d_{\RR,2\eta s }^\dag $ as $d_1$ and $d_2$ electrons (orbitals), respectively, for short. We mention that $d_1,d_2$ are $f$-electrons. 
The strain Hamiltonian can then be written as
\baa  
\hH_{strain} = \alpha \sum_{\RR,\alpha \eta s}( -d_{\RR,1\eta s}^\dag d_{\RR,1\eta s} +d_{\RR,2\eta s}^\dag d_{\RR,2\eta s})
\label{eq:def_eig_strain}
\eaa  
Thus, for a positive strain amplitude $\alpha >0$, the energy of $d_1$ electrons will be lowered and the energy of $d_{2}$ electrons will be raised. 
We introduce $\langle h_{strain}\rangle$ to characterize the population imbalance between $d_1$ and $d_2$ electrons
\baa  
\langle h_{strain} \rangle  = \langle \frac{1}{N_M} \sum_{\RR,\alpha \eta s,\alpha'\eta's'} f_{\RR,\alpha \eta s}^\dag [h_{strain}]_{\alpha \eta s,\alpha'\eta' s'} f_{\RR,\alpha'\eta's'}
\rangle 
= \frac{1}{N_M}\sum_{\RR,\alpha \eta s}
\langle d_{\RR,1\eta s}^\dag d_{\RR,1\eta s} -d_{\RR,2\eta s}^\dag d_{\RR,2\eta s} \rangle
\label{eq:def_hstrain_exp}\, 
\eaa  

In Fig.~\ref{fig:order}, we plot the evolution of various order parameters and also $|\langle h_{strain}\rangle|$ 
where the expectation value is taken with respect to the ordered state solution. 
In all cases, $|\langle h_{strain} \rangle |$ increases as we increase $\alpha$, since $\alpha$ linearly coupled to $h_{strain}$ term. The KIVC order will be suppressed and fully destroyed at sufficient strong strain at $\nu=0,-1,-2$. At $\nu=0$, after the destruction of the KIVC order, self-consistent calculation produces a symmetric ground state that only breaks $C_{3z}$ symmetry, even though we initialize the mean-field calculation with an ordered state.

However, at $\nu=-1,-2$, after the destruction of KIVC order, the spin polarization and valley polarization still exist. By further increasing the strain, the ordered states will finally become unstable (Fig.~\ref{fig:order}), which means the mean-field calculations that are initialized with ordered solutions converge to a symmetric state. 

We next analyze the transition from an ordered state to a symmetric state at a large strain at $\nu=-1,-2$. 

\subsubsection{$\nu=-1$}
We first consider the $\nu=-1$ with $4\text{meV}\lesssim\alpha \lesssim 18\text{meV} $. In this parameter region, the KIVC order is destroyed but valley and spin polarization persist (Fig.~\ref{fig:order}). In Fig.~\ref{fig:disp} (a) (b), we plot the band structures in this parameter region. We note that flat bands that are mostly formed by $f$-electrons (marked by red circles, Fig.~\ref{fig:disp}) move towards the Fermi level, as we increase strain. Near the transition point to the symmetric state, the flat bands are very close to the Fermi level. This signals an instability of the ordered states since we can fill the flat band without any energy cost. 
By diagonalizing the mean-field Hamiltonian, we find the flat bands (marked by red circles, Fig.~\ref{fig:disp}) correspond to $d_{1}$ electrons (Eq.~\ref{eq:def_eig_strain}). By filling the flat bands, we have more populations in $d_{1}$ orbitals, which increase $|\langle h_{strain}\rangle |$ (Eq.~\ref{eq:def_hstrain_exp}) and drive the system to a symmetric state.

We now estimate the critical value of strain $\alpha_c$ at which a transition from an ordered state to a symmetric state happens. At $\alpha_c$, the flat bands (marked by red circles, Fig.~\ref{fig:disp}) are very close to the Fermi energy and induce the transition. To estimate $\alpha_c$, we calculate the excitation gap of the flat bands (marked by red circles, Fig.~\ref{fig:disp}): $\Delta E_{flat}$. Then we have 
\baa  
\Delta E_{flat}\bigg|_{\alpha \approx \alpha_c} = 0
\eaa  

We estimate $\Delta E_{flat}$ using the zero-hybridization limit~\cite{Spin_MATBLG} of the model, where $\gamma =0 ,v_\star^\prime =0$ (Eq.~\ref{eq:def_fc_mat}). In addition, we also set $J=0$ to simplify the calculation (Eq.~\ref{eq:hj_def}). 
The zero-hybridization model with non-zero strains are 
\baa  
\hH_{\text{zero-hyb}} = \hH_U +\hH_W +\hH_V + \hH_{stain} +\hH_c +\hH_{\mu}
\label{eq:zero_hyb_mod}
\eaa  
where $\hH_U,\hH_W,\hH_V,\hH_{strain},\hH_c,\hH_{\mu}$ are defined in Eq.~\ref{eq:hu_def}, Eq.~\ref{eq:hw_def}, Eq.~\ref{eq:hv_def}, Eq.~\ref{eq:hstrain}, Eq.~\ref{eq:hc_def} and Eq.~\ref{eq:thf_ham} respectively, and $\hH_V$ are treated with mean-field methods. In the zero-hybridization model, the filling of $f$-electrons $\nu_f$ and $c$-electrons-$\nu_c$ are good quantum numbers. 
We solve the zero-hybridization model at fixed total filling $\nu$ with the assumption that the ground state does not break translational symmetry (fillings of $f$-electrons are uniform)~\cite{Spin_MATBLG}. To estimate the excitation gap of the flat bands, we calculate the energy cost of adding one $d_{\RR,1\eta s}$ electron. 
We mention that, in our mean-field calculations with finite $f$-$c$ hybridization, the relevant flat bands (marked by red circles in Fig.~\ref{fig:disp}) correspond to $d_{1}$ electrons. We let $|\Psi_{\text{zero-hyb}}\rangle $ denote the ground state of the zero-hybridization model. The state with one-more $d_{\RR,1\eta s}$ is
\baa  
|\Psi^{exct}_{\text{zero-hyb}}\rangle = d_{\RR,1\eta s}^\dag |\Psi_{\text{zero-hyb}} \rangle \, .
\eaa  
We next calculate 
\baa  
\Delta E_{flat} =\langle \Psi^{exct}_{\text{zero-hyb}} |\hH_{\text{zero-hyb}} |\Psi^{exct}_{\text{zero-hyb}}\rangle - 
\langle \Psi_{\text{zero-hyb}} |\hH_{\text{zero-hyb}} |\Psi_{\text{zero-hyb}}\rangle 
\eaa  
The energy loss from Hubbard interaction term is  
\baa  
\Delta E_U  =\langle \Psi^{exct}_{\text{zero-hyb}} |\hH_{U} |\Psi^{exct}_{U}\rangle - 
\langle \Psi_{\text{zero-hyb}} |\hH_{U} |\Psi_{\text{zero-hyb}}\rangle =  \frac{U}{2} (\nu_f + 1)^2 - \frac{U}{2}\nu_f^2  = U(\nu_f+\frac{1}{2}) \nonumber 
\eaa  
The energy loss from $\hH_W$ term is 
\baa  
\Delta E_W =&\langle \Psi^{exct}_{\text{zero-hyb}} |\hH_{W} |\Psi^{exct}_{\text{zero-hyb}}\rangle - 
\langle \Psi_{\text{zero-hyb}} |\hH_{W} |\Psi_{\text{zero-hyb}}\rangle \nonumber \\
=& \sum_{a=1,2,3,4} W_a \nu_{c,a} (\nu_f+1) - \sum_{a=1,2,3,4} W_a \nu_{c,a} \nu_f  =  \sum_{a=1,2,3,4} W_a \nu_{c,a} 
\eaa  
where $\nu_{c,a}$ denotes the filling of $c$-electrons in orbital $a$. 
The energy change from $\hH_V$ is 
\baa  
\Delta E_V =\langle \Psi^{exct}_{\text{zero-hyb}} |\hH_{V} |\Psi^{exct}_{\text{zero-hyb}}\rangle - 
\langle \Psi_{\text{zero-hyb}} |\hH_{V} |\Psi_{\text{zero-hyb}}\rangle =0
\eaa  
The energy change from $\hH_{strain}$ is 
\baa  
\Delta E_{strain} = \langle \Psi^{exct}_{\text{zero-hyb}} |\hH_{strain} |\Psi^{exct}_{\text{zero-hyb}}\rangle - 
\langle \Psi_{\text{zero-hyb}} |\hH_{strain} |\Psi_{\text{zero-hyb}}\rangle  = -\alpha 
\eaa  
The energy change from chemical potential $\hH_\mu$ is 
\baa  
\Delta E_\mu =\langle \Psi^{exct}_{\text{zero-hyb}} |\hH_{\mu} |\Psi^{exct}_{\text{zero-hyb}}\rangle - 
\langle \Psi_{\text{zero-hyb}} |\hH_{\mu} |\Psi_{\text{zero-hyb}}\rangle  -\mu 
\eaa 
Then the excitation energy of adding one $d_{\RR,1\eta s}$ electron is 
\baa  
\Delta E_{flat} = \Delta E_U + \Delta E_W +\Delta_{strain}+\Delta E_\mu
= \frac{U}{2}(\nu_f+1/2) + \sum_a W_a \nu_{c,a} -\mu -\alpha 
\eaa 
We further take the following approximation: $W_{1,2,3,4} =W =47$meV (the difference between $W_{1,2,3,4}$ is about 15$\%$). Then 
\baa  
\Delta E_{flat}\approx  \frac{U}{2}(\nu_f+1/2) +W \nu_{c} -\mu -\alpha 
\label{eq:estimate_exct_gap_comp}
\eaa  

At $\nu=-1$ and $0\text{meV} \le  \alpha\le 43 $meV, the ground state of the zero-hybridization model has $\nu_f=-1,\nu_c=\nu-\nu_f=0$ (Fig.~\ref{fig:zero_hyb}). Then 
\baa  
\Delta E_{flat}= -\frac{U}{2}  -\mu -\alpha 
\label{eq:estimate_exct_gap}
\eaa  
We next determine chemical potential $\mu$. Chemical potential
$\mu$ is determined by requiring the $c$-electrons filling to be $\nu_c=0$. The single-particle Hamiltonian of $c$-electron in the zero-hybridization limit takes the form of 
\baa  
\hH_{c,\text{zero-hyb}} = \hH_c + \sum_{\kk,a\eta s} (W\nu_f +\frac{V(0)}{\Omega_0} \nu_c -\mu ) c_{\kk,a\eta s}^\dag c_{\kk,a\eta s}
\label{eq:hc_zero_hyb}
\eaa  
where we have set $W_{1,2,3,4}=W$. We note that when 
\baa  
W\nu_f -\frac{V(0)}{\Omega_0}\nu_c - \mu = 0
\eaa  
$\hH_{c,\text{zero-hyb}}=\hH_c$ and we have $\nu_c=0$. Therefore,
\baa  
\mu = W\nu_f +\frac{V(0)}{\Omega_0}\nu_c =-W 
\label{eq:mu_at_nu_1}
\eaa  
where we take $\nu_c=0,\nu_f=-1$ (Fig.~\ref{fig:zero_hyb}). Using Eq.~\ref{eq:estimate_exct_gap} and Eq.~\ref{eq:mu_at_nu_1}, we find 
\baa 
\Delta E_{flat} = W - \frac{U}{2} -\alpha 
\label{eq:del_e_nu_1}
\eaa  

Then the flat bands reach Fermi energy when $\Delta E_{flat} =0$, which leads to
\baa  
\Delta E_{flat}=0 \Rightarrow \alpha_c = W-\frac{U}{2} = 18\text{meV}
\label{eq:estimation_transition_nu_1}
\eaa 
which is close to the value (also around $\alpha=18$meV as shown in Fig.~\ref{fig:order} (b)) from self-consistent calculations of the finite-hybridization model. Here, the finite-hybridization model refers to the original THF model with finite $\gamma,v_\star^\prime$. Therefore, we conclude the transition from an ordered state to a symmetric state happens at $\alpha =\alpha_c \approx 18$meV at $\nu=-1$.

We also discuss the solutions of the zero-hybridization model here.
In Fig.~\ref{fig:zero_hyb} (a), we show the ground state properties of the zero-hybridization model at various strains and $\nu=-1$, where
\baa  
\nu^f_1 =\frac{1}{N_M}\sum_{\RR,  \eta s}:d_{\RR,1\eta s}^\dag d_{\RR,1\eta s}:,\quad \nu^f_2 =\frac{1}{N_M}\sum_{\RR,  \eta s}:d_{\RR,2\eta s}^\dag d_{\RR,2\eta s}:
\eaa  
denotes the filling of $d_1$ and $d_{2}$ electrons respectively with $\nu^f =\nu^f_1 +\nu^f_2$. We find a transition happens at $\alpha \approx 25$meV. 
We note that this transition is described by filling one more $d_{\RR,1\eta s}$ electrons at each site.
After the transition, there will be 4 $f$-electrons filling $d_{1}$ orbitals, and zero $f$-electrons filling the $d_{1}$ orbitals. Thus for $d_{1}$ orbitals, all the valleys and spins are filled, but for $d_{2}$ orbitals all the valleys and spins are empty. Therefore, there is no room to develop order and the ground state is a symmetric state. We note that the transition in the zero-hybridization limit and the transition in the finite-hybridization model (at $\nu=-1,\alpha \approx 16 $meV, Fig.~\ref{fig:order}) share the same origin. They are both driven by filling electrons in $d_{1}$ orbitals (in the finite-hybridization model, we fill the flat bands) and, after the transition, both ground states are symmetric. Thus, the results between zero-hybridization and finite-hybridization models are consistent. However, the critical values $\alpha_c$ for the two models are different, since we have finite $f$-$c$ hybridization in the finite-hybridization model.

\subsubsection{$\nu=-2$}
We next discuss the transition from an ordered state to a symmetric state at $\nu=-2$. We focus on the parameter region $10\text{meV}\lesssim \alpha \lesssim 45$meV, where the KIVC order is destroyed but valley and spin polarization exist (Fig.~\ref{fig:order}). In Fig.~\ref{fig:disp} (c) (d), we plot the band structures in this parameter region. As we increase strain, we note that flat bands (marked with red circles, Fig.~\ref{fig:disp}), move towards the Fermi level. Similar to the $\nu=-1$ case, when the flat bands reach the Fermi energy, a transition to the symmetric state happens. However, at $\nu=-2$, we need a much larger strain to destroy the ordered state as shown in Fig.~\ref{fig:order} (d). To understand this, we start from the zero-hybridization limit of the model (Eq.~\ref{eq:zero_hyb_mod}). As shown in Eq.~\ref{eq:estimate_exct_gap_comp}, the excitation energy of the relevant flat bands (marked by red circles in Fig.~\ref{fig:disp})
\baa 
\Delta E_{flat}= \frac{U}{2}(\nu_f+1/2) +W \nu_{c} -\mu -\alpha
\label{eq:delta_e_complete}
\eaa 
By solving the zero-hybridization model, we find the ground states have $\nu_f=-1$ and $\nu_c=-1$ in the parameter region we focused $4\text{meV}\lesssim\alpha \lesssim 18\text{meV} $, as shown in Fig.~\ref{fig:zero_hyb}. Then
\baa  
\Delta E_{flat}= -\frac{U}{2} - W -\mu -\alpha 
\label{eq:de_nu_1}
\eaa 

We now calculate the chemical potential. $\mu$ is determined by requiring the filling of $c$-electrons to be $\nu_c=-1$. The single-particle Hamiltonian of $c$-electron in the zero-hybridization limit (Eq.~\ref{eq:hc_zero_hyb}) takes the form of 
\baa  
\hH_{c,\text{zero-hyb}} = \hH_c + \sum_{\kk,a\eta s} (W\nu_f +\frac{V(0)}{\Omega_0} \nu_c -\mu ) c_{\kk,a\eta s}^\dag c_{\kk,a\eta s}
\eaa  
where we have set $W_{1,2,3,4}=W$, and take the mean-field treatment of $\hH_V$(Eq.~\ref{eq:mf_hV}). At $M=0$ limit ($M=3.697$meV, which is relatively small), the dispersion of $c$-electrons are $E_\kk = \pm v_\star |\kk| -E_c$, where we define 
\baa  
E_c = W\nu_f +\frac{V(0)}{\Omega_0} \nu_c -\mu 
\label{eq:Ec_def}
\eaa  
Then all the $c$-states with energy smaller than $0$ will be filled. 
The corresponding Fermi momentum $k_F$ is 
\baa  
|v_\star k_F| = E_c \Rightarrow k_F = \frac{1}{|v_\star|} E_c
\eaa 
Then the filling of $c$-electrons is 
\baa  
\nu_c = - \frac{8}{A_{MBZ}} \int_{|\kk|<k_F} dk_x dk_y  = -\frac{8\pi k_F^2}{A_{MBZ}} = -\frac{8\pi E_c^2 }{A_{MBZ}|v_\star|^2}
\eaa   
where the prefactor $8$ comes from the 8-fold degeneracy of the bands (2 spins, 2 valleys, and 2 orbitals) and $A_{MBZ}$ is the area of MBZ. 
We require $\nu_c=-1$ which indicates
\baa 
\nu_c = -\frac{8\pi E_c^2 }{A_{MBZ}|v_\star|^2} = -1 \Rightarrow E_c = \sqrt{ \frac{A_{MBZ}|v_\star|^2}{8\pi} } 
\label{eq:nu_c_eq_1}
\eaa 
Using Eq.~\ref{eq:Ec_def} and Eq.~\ref{eq:nu_c_eq_1}, we find 
\baa  
\mu = -W -\frac{V(0)}{\Omega_0} -\sqrt{ \frac{A_{MBZ}|v_\star|^2}{8\pi} } 
\label{eq:mu_at_nu_2}
\eaa  
where we take $\nu_f=-1,\nu_c=-1$. Combining Eq.~\ref{eq:de_nu_1} and Eq.~\ref{eq:nu_c_eq_1}, we find 
\baa  
\Delta E_{flat} = -\frac{U}{2} +\frac{V(0)}{\Omega_0} + \sqrt{ \frac{A_{MBZ}|v_\star|^2}{8\pi} } -\alpha 
\label{eq:del_e_nu_2}
\eaa  
And the transition happens at
\baa  
\Delta E_{flat} = 0 \Rightarrow 
\alpha_c =- \frac{U}{2} +\frac{V(0)}{\Omega_0} +\sqrt{ \frac{A_{MBZ}|v_\star|^2}{8\pi} } =  62\text{meV}
\label{eq:single_par_E}
\eaa  
We note that our estimation of transition in Eq.~\ref{eq:single_par_E} is based on the single-particle picture and has not included the effect of hybridization. Thus the estimated value is larger than the transition value $\alpha \approx 45$meV from mean-field self-consistent calculations with finite $f$-$c$ hybridization.

We also discuss the solution obtained from the zero-hybridization model at $\nu=-2$.
In Fig.~\ref{fig:zero_hyb} (b), we show the ground state properties of the zero-hybridization model at various strains and $\nu=-2$. We observe two transitions, both characterize by the increments of $\nu^f_1$. After the first transition $\alpha \sim 15$meV, we have three $d_{1}$ electrons at each site. Then not all valleys and spins of $d_{1}$ orbitals are filled. This gives the possibility of developing spin and valley polarized order after we turn on the $\hH_J$ and $f$-$c$ hybridization. Indeed, from Fig.~\ref{fig:order}, we can observe the non-vanishing order at 
$15\text{meV}\lesssim \alpha \lesssim 40$meV. By further increasing strain, at $\alpha \approx 50$meV, we observe a second transition in the zero-hybridization model. After the second transition, we have $\nu_1^f=2$, $\nu^f_2=-2$ and there is no room for $f$-electrons to develop order. Correspondingly, for the finite-hybridization model, we also observe a transition to the symmetric stat at $\alpha \approx 45$meV. Thus we conclude the consistency between zero-hybridization and finite-hybridization models. We also point out that, the transition value $\alpha_c \approx 62$meV suggested from single-particle excitation in Eq.~\ref{eq:single_par_E} is also larger than the transition value $\alpha_c \approx 50$meV obtained by solving many-body wavefunction in the zero-hybridization model. This points out the limitation of the single-particle picture.

Finally, we comment on the difference between $\nu=-1$ and $\nu=-2$. Eq.~\ref{eq:single_par_E} and Eq.~\ref{eq:estimation_transition_nu_1} indicate that a relatively large strain is required to destroy the ordered state at $\nu=-2$ compared to $\nu=-1$. This is because, near the transition point with $\alpha<\alpha_c$, $c$-electrons have filling $\nu_c=0$ at $\nu=-1$, but have filling $\nu_c=-1$ at $\nu=-2$ (Fig.~\ref{fig:zero_hyb}). In other words, we dope more holes to the $c$-bands at $\nu=-2$ compared to $\nu=-1$. In order to dope more holes to $c$-bands, the chemical potential (or Fermi energy) needs to be increased. The increment of chemical potential also leads to a larger Fermi surface, which has been observed in the model with finite hybridization (Fig.~\ref{fig:disp}), where the Fermi surface at $\nu=-2$ is larger than the Fermi surface at $\nu=-1$. In the zero-hybridization model, the change of chemical potential can be observed analytically from Eq.~\ref{eq:mu_at_nu_1} and Eq.~\ref{eq:mu_at_nu_2}. We note that the excitation gap of the flat band ($\Delta E_{flat}$) is measured with respect to the Fermi energy (or chemical potential). Then at fixed $\alpha$, we find the value of $\Delta E_{flat}$ at $\nu=-2$ is much larger than its value at $\nu=-1$ (Eq.~\ref{eq:del_e_nu_1} and Eq.~\ref{eq:del_e_nu_2}) due to the larger chemical potential at $\nu=-2$.  
Thus, at $\nu=-2$, a much larger strain is needed to make $\Delta E_{flat}=0$ and destroy the ordering.

\subsubsection{Discussions about $\nu=-3$}
We now discuss $\nu=-3$. We first comment that, at $\nu=-3$, other low energy states that break translational symmetry exists even at zero strain~\cite{xie2022phase}. Furthermore, even in the zero-hybridization limit at zero strain, $\nu=-3$ is close to the transition point between $\nu_f=-3$ and $\nu_f=-2$ states~\cite{HF_MATBLG}. These all point out the complexity of $\nu=-3$ even in the absence of strain. Thus, we leave a more systematical analysis at $\nu=-3$ for future study.

\begin{figure}
    \centering
    \includegraphics[width = 1.0\textwidth]{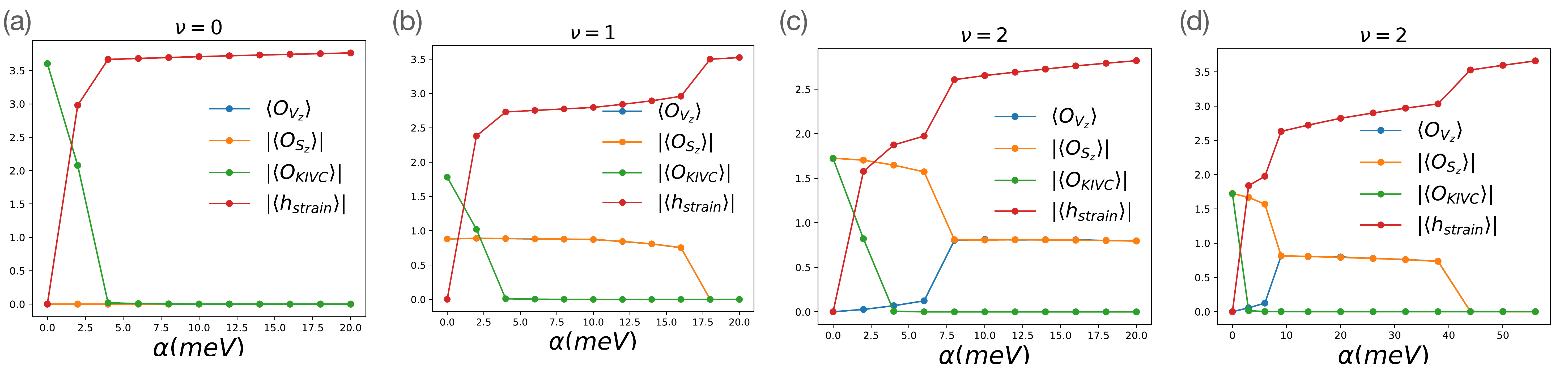}
    \caption{(a), (b), (c) Evolution of order parameters as a function of strain amplitude $\alpha$ at $\nu=0,-1,-2$. (d) Evolution of order parameters at $\nu=-2$ with an extended parameter region $0\text{meV} \le  \alpha\le 55 $meV  }
    \label{fig:order}
\end{figure}

\begin{figure}
    \centering
    \includegraphics[width = 1.0\textwidth]{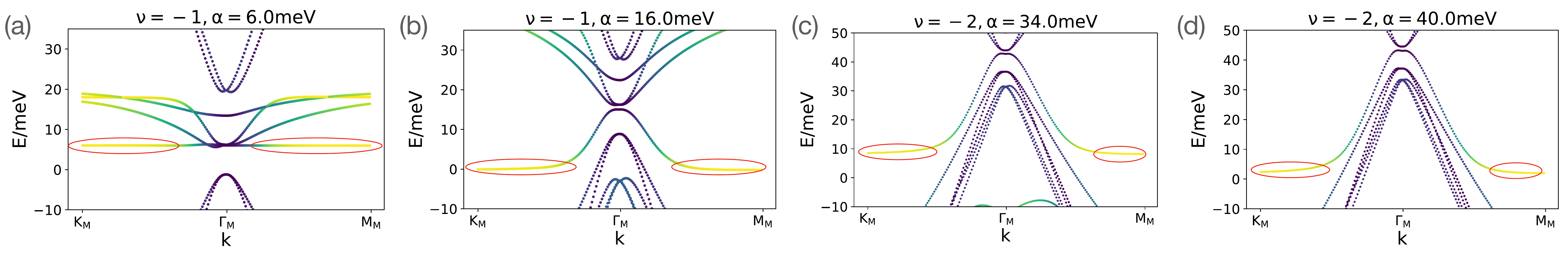}
    \caption{Band structure at $\nu=-1$ (a), (b) and $\nu=-2$ (c), (d). Red circles mark the relevant flat bands. }
    \label{fig:disp}
\end{figure} 

\begin{figure}
    \centering
    \includegraphics[width=0.9\textwidth]{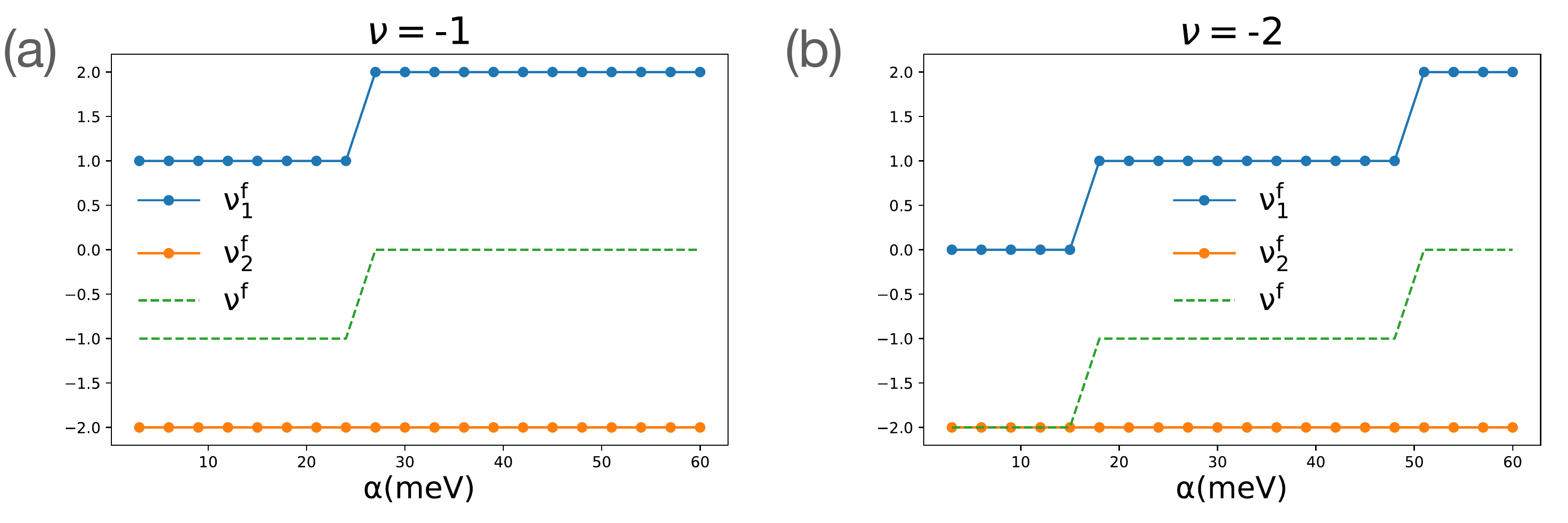}
    \caption{Evolution of filling as a function of the strain in the zero-hybridization model at $\nu=-1,-2$.}
    \label{fig:zero_hyb}
\end{figure}

\subsection{Strain}

We now discuss the derivation of strain term in the Eq.~\ref{eq:hstrain}. Here we will take a minimal model to capture the effect of strain, and leave a detailed analysis (based on Ref.~\cite{vafek2022continuum}) for future study. 
We take the single-particle Hamiltonian of twisted bilayer graphene at valley $\eta$ and with non-zero strain~\cite{koshino_22}
\baa 
H^\eta(\kk)  = \begin{bmatrix}
    H^\eta_{1}(\kk) & U \\ 
    U^\dag & H^\eta_2(\kk) 
\end{bmatrix}
\eaa  
where $H_{l}(\kk) $ is a $2\times 2$ matrix that characterizes the Hamiltonian of $l$-layer graphene. $U$ is the interlayer coupling matrix. In real space, $U$ takes the form of
\baa  
U = \sum_{j=1}^3 U_je^{i\eta \bm{q}_j\cdot \rr} ,\quad 
U_j = \begin{bmatrix}
    w_0 & w_1 e^{-i\eta 2(j-1)\pi/3} \nonumber \\
    w_1e^{i\eta 2(j-1)\pi/3} & w_0 
\end{bmatrix}
\eaa  
with $w_0,w_1$ are two constant parameters that characterize the interlayer couplings.
$H_l(\kk)$ is given by 
\baa 
H^\eta_{l=1,2}(\kk) = -\hbar v_F \bigg[ \bigg(
R(\mp \theta) +\mathcal{E}^l\bigg)^{-1}(\kk + \frac{e}{\hbar}\bm{A}^{l,\eta})
\bigg ]\cdot \bm{\sigma}^\eta \nonumber \\
\eaa 
where $\bm{\sigma}^\eta = (\eta \sigma_x, \sigma_y ) $ , $R(\theta)$ is a rotation matrix, $\theta$ is the twist angle and~\cite{koshino_22}
\baa  
&\mathcal{E}^l =
\begin{bmatrix}
    \epsilon_{xx}^{(l)} &  -\Omega^{(l)} +\epsilon_{xy}^{(l)} \\
     -\Omega^{(l)} +\epsilon_{xy}^{(l)}  &  \epsilon_{yy}^{(l)}
\end{bmatrix}
\nonumber \\ 
&\bm{A}^{l,\eta} =\eta \frac{3\beta \gamma_0}{2ev_F}
\begin{bmatrix}
    \epsilon_-^{(l)}  \\   \epsilon_{xy}^{(l)}
\end{bmatrix} \nonumber
,\quad \gamma_0=2.7\text{eV},\quad \beta \approx 3.14
\\ 
& \epsilon_{\pm}^{(l)} =\frac{1}{2} (\epsilon_{xx}^{(l)} \pm \epsilon_{yy}^{(l)}) \,.
\eaa  
 $\epsilon_{xx}^{(l)},\epsilon_{yy}^{(l)}$ are normal strains along $x$ and $y$ directions respectively, $\epsilon_{xy}^{(l)}$ is the sheer strain. $\Omega^{(l)}$ denotes a rotation from the twist angles~\cite{koshino_22}. Since the flat band is sensitive to $\epsilon_{-}= (\epsilon_{-}^{(1)}-\epsilon_{-}^{(2)})/2$ and $\epsilon_{xy}= (\epsilon_{xy}^{(1)}-\epsilon_{xy}^{(2)})/2$~\cite{koshino_22}, we consider the heterostrain with $\Omega^{(l)}=0,\epsilon_+^{(l)}=0, \epsilon_{-}^{(l)} =(-1)^{l+1} \epsilon_{-},  \epsilon_{xy}^{(l)} =(-1)^{l+1} \epsilon_{xy}$. We further focus on the sheer strain $\epsilon_{xy}$ and set the normal anisotropic strain to be zero $\epsilon_- =0$. For small twist angle, we approximately have 
\baa  
H^\eta_{l=1,2}(\kk)
 \approx & -\hbar v_F \bigg[ \bigg(R(\pm \theta) -\mathcal{E}^l \bigg) (\kk + \frac{e}{\hbar}\bm{A}^{l,\eta})
\bigg ]\cdot \bm{\sigma}^\eta \nonumber \\
 \approx & -\hbar v_F \bigg[R(\pm \theta) \kk -\mathcal{E}^l \kk  + \frac{e}{\hbar}\bm{A}^{l,\eta}
\bigg ]\cdot \bm{\sigma}^\eta
\eaa 
The effect of strain is then captured by the following single particle Hamiltonian of valley $\eta$
\baa  
&H_{strain}^\eta(\kk) = H_{strain,1}^\eta + H_{strain,2}^\eta   \nonumber \\ 
&H_{strain,1}^\eta =-\frac{3\beta \gamma_0 \eta }{2}\epsilon_{xy} \begin{bmatrix}
   \sigma_y &  \\ 
  & - \sigma_y 
\end{bmatrix} ,\quad \quad 
H_{strain,2}^\eta = \hbar v_F \epsilon_{xy}
\begin{bmatrix}
  \eta k_y \sigma_x + k_x \sigma_y \\ 
  & -\eta k_y \sigma_x - k_x \sigma_y 
\end{bmatrix}
\label{eq:single_particle_strain}
\eaa  

We next project the $H_{strain}^\eta(\kk)$ to the Wannier basis of $f$-electrons. 
We take the following wavefunction of $f$ electrons at $\RR$ with orbital $\alpha$, valley $\eta $ and spin $s$ ~\cite{HF_MATBLG}
\baa  
&|W_{\RR,\alpha=1,\eta, s}\rangle  
= \sqrt{\frac{2\pi \lambda_0^2}{\Omega_{tot}}}
\sum_{l = \pm } \sum_{\kk} \sum_{\QQ \in \mathcal{Q}_{l\eta}} e^{i\frac{\pi}{4}l\eta-i\kk\cdot \RR -\frac{1}{2}\lambda_0^2 (\kk-\QQ)^2 } |\kk,\QQ,1,\eta ,s\rangle \nonumber \\
&|W_{\RR,\alpha=2,\eta, s}\rangle  
= \sqrt{\frac{2\pi \lambda_0^2}{\Omega_{tot}}}
\sum_{l = \pm } \sum_{\kk} \sum_{\QQ \in \mathcal{Q}_{l\eta}} e^{-i\frac{\pi}{4}l\eta-i\kk\cdot \RR -\frac{1}{2}\lambda_0^2 (\kk-\QQ)^2 } |\kk,\QQ,2,\eta ,s\rangle 
\label{eq:wannier_basis}
\eaa  
where $\lambda_0 = 0.1a_M$, $\Omega_{tot}$ is the total area of the sample and $\mathcal{Q}_{l\eta} = \{ \pm \bm{q}_{1} + n_1 \bm{b}_{M,1} + n_2 \bm{b}_{M,2}|n_1,n_2 \in \mathbb{Z} \}$, $\bm{q}_1 = k_\theta (0,-1)$~\cite{HF_MATBLG}. From Eq.~\ref{eq:single_particle_strain} and Eq.~\ref{eq:wannier_basis}, 
we can calculate
\baa  
h^{strain}_{\alpha\eta s,\alpha_2\eta_2s_2}=\frac{1}{N_M}\sum_\RR \langle W_{\RR,\alpha, \eta ,s } | \sum_{\kk,\alpha'l',\alpha_2'l_2',\eta',s'} \psi^\dag_{\kk, l,\alpha',\eta' ,s'} [H_{strain}^{\eta'}(\kk)]_{l'\alpha',l_2'\alpha_2'}\psi_{\kk,l_2'\alpha',\eta',s'}
|W_{\RR,\alpha_2,\eta_2,s_2} \rangle 
\label{eq:hstrain_f_basis}
\eaa  
where $\psi_{\kk,l,\alpha , \eta ,s}$ is the electron operator of the original electron basis with momentum $\kk$, layer $l$, sublattice $\alpha$, valley $\eta$ and spin $s$. Here, we take the average over all the moir\'e unit cells (sum over $\RR$), and thus we only keep the momentum-independent (in $f$-electron basis) contributions. We leave the momentum-dependent term for future studies. In the $f$-basis, the strain can be characterized by
\baa  
\hH_{strain} =\sum_{\RR,\alpha\eta s,\alpha_2\eta_2s_2} h^{strain}_{\alpha\eta s,\alpha_2\eta_2s_2}f_{\RR,\alpha \eta s}^\dag f_{\RR,\alpha_2\eta_2s_2} 
\eaa  

We next evaluate $h^{strain}$ from Eq.~\ref{eq:hstrain_f_basis}.  We separate $h^{strain}$ into two parts (Eq.~\ref{eq:single_particle_strain})
\baa  
h^{strain,1}_{\alpha\eta s,\alpha_2\eta_2s_2}=\frac{1}{N_M}\sum_\RR \langle W_{\RR,\alpha, \eta ,s } | \sum_{\kk,\alpha'l',\alpha_2'l_2',\eta',s'} \psi^\dag_{\kk, l,\alpha',\eta' ,s'} [H_{strain,1}^{\eta'}(\kk)]_{l'\alpha',l_2'\alpha_2'}\psi_{\kk,l_2'\alpha',\eta',s'}
|W_{\RR,\alpha_2,\eta_2,s_2} \rangle 
\nonumber \\
h^{strain,2}_{\alpha\eta s,\alpha_2\eta_2s_2}=\frac{1}{N_M}\sum_\RR \langle W_{\RR,\alpha, \eta ,s } | \sum_{\kk,\alpha'l',\alpha_2'l_2',\eta',s'} \psi^\dag_{\kk, l,\alpha',\eta' ,s'} [H_{strain,2}^{\eta'}(\kk)]_{l'\alpha',l_2'\alpha_2'}\psi_{\kk,l_2'\alpha',\eta',s'}
|W_{\RR,\alpha_2,\eta_2,s_2} \rangle 
\label{eq:hstrain_sep}
\eaa  

For $h^{strain,1}$, using Eq.~\ref{eq:single_particle_strain}, Eq.~\ref{eq:wannier_basis} and Eq.~\ref{eq:hstrain_sep}, we find
\baa  
&h^{strain,1}_{\alpha\eta s,\alpha_2\eta_2s_2}\nonumber \\
=& \delta_{\eta,\eta_2}\delta_{s,s_2}\delta_{\alpha,3-\alpha_2} \frac{2\pi\lambda_0^2}{\Omega_{tot}} \frac{- 3\beta \gamma_0}{2}\epsilon_{xy} 
\sum_{l=\pm }\sum_{\QQ \in \mathcal{Q}_{l\eta}  }  \sum_{\kk}
i (-1)^{\alpha }l\eta 
e^{-\lambda_0^2(\kk-\QQ)^2}e^{-i\frac{\pi \eta l}{2}(-1)^{\alpha+1} }  \nonumber \\
=&  \delta_{\eta,\eta_2}\delta_{s,s_2}\delta_{\alpha,3-\alpha_2} \frac{3\pi \lambda_0^2 \beta \gamma_0}{\Omega_{tot}} \epsilon_{xy} \sum_\kk 
\bigg[  \sum_{\QQ \in \mathcal{Q}_{+}  }e^{-\lambda_0^2(\kk-\QQ)^2} +\sum_{\QQ \in \mathcal{Q}_{-}  }e^{-\lambda_0^2(\kk-\QQ)^2} 
\bigg] 
\eaa 
In a more compact form, we find 
\baa  
&h^{strain,1} = \alpha \sigma_x \tau_0\varsigma_0 \nonumber \\
&\alpha = \frac{3\pi \lambda_0^2 \beta \gamma_0}{\Omega_{tot}} \epsilon_{xy} \sum_\kk 
\bigg[  \sum_{\QQ \in \mathcal{Q}_{+}  }e^{-\lambda_0^2(\kk-\QQ)^2} +\sum_{\QQ \in \mathcal{Q}_{-}  }e^{-\lambda_0^2(\kk-\QQ)^2} 
\bigg] \approx \bigg(1.3 \times 10^4 \epsilon_{xy} \bigg) \text{meV} \, .
\label{eq:hstrain_1}
\eaa  

We next calculate $h^{strain,2}$, using Eq.~\ref{eq:single_particle_strain}, Eq.~\ref{eq:wannier_basis} and Eq.~\ref{eq:hstrain_sep},
we find 
\baa  
&h^{strain,2}_{\alpha \eta s,\alpha_2\eta_2 s_2} \nonumber \\ 
=&\delta_{\eta,\eta_2}\delta_{s,s_2} \delta_{\alpha,3-\alpha_2}\frac{2\pi \lambda_0^2 \hbar \epsilon_{xy}}{\Omega_{tot}} 
\sum_{l=\pm }\sum_{Q \in \mathcal{Q}_{l\eta} }\sum_{\kk} l \bigg( 
\eta(k_y-Q_y) +i(-1)^\alpha(k_x-Q_x)
\bigg) e^{i\frac{\pi}{2}l\eta(-1)^\alpha }
 e^{-\lambda_0^2 (\kk-\QQ)^2} \nonumber \\
=&\delta_{\eta,\eta_2}\delta_{s,s_2} \delta_{\alpha,3-\alpha_2}\frac{2\pi \lambda_0^2 \hbar \epsilon_{xy}}{\Omega_{tot}} \sum_{\kk}
\bigg[ \sum_{Q \in \mathcal{Q}_{\eta} }\bigg( 
\eta(k_y-Q_y) +i(-1)^\alpha(k_x-Q_x)
\bigg) i\eta(-1)^\alpha
 e^{-\lambda_0^2 (\kk-\QQ)^2} \nonumber \\
 &+\sum_{Q \in \mathcal{Q}_{-\eta} }\bigg( 
\eta(k_y-Q_y) +i(-1)^\alpha(k_x-Q_x)
\bigg) i\eta(-1)^\alpha
 e^{-\lambda_0^2 (\kk-\QQ)^2}  \bigg] \nonumber \\ 
 =&\delta_{\eta,\eta_2}\delta_{s,s_2} \delta_{\alpha,3-\alpha_2}\frac{2\pi \lambda_0^2 \hbar \epsilon_{xy}}{\Omega_{tot}} \sum_{\kk}\sum_{\QQ \in \mathcal{Q}_0 } \bigg( 
\eta(k_y-Q_y-\eta q_{1,y}) +i(-1)^\alpha(k_x-Q_x-\eta q_{1,x})
\bigg) i\eta(-1)^\alpha
 e^{-\lambda_0^2 (\kk-\QQ-\eta \qq_{1})^2} \nonumber \\
 &+\bigg( 
\eta(k_y-Q_y+\eta q_{1,y}) +i(-1)^\alpha(k_x-Q_x+\eta q_{1,x})
\bigg) i\eta(-1)^\alpha
 e^{-\lambda_0^2 (\kk-\QQ+\eta \qq_{1})^2}  \bigg] 
 \eaa  
 where we let $\mathcal{Q}_{0} = \{ n_1 \bm{b}_{M,1} + n_2 \bm{b}_{M,2}|n_1,n_2 \in \mathbb{Z} \}$. Since $\qq_{1}= k_\theta(0,1)$, $q_{1,x}=0,q_{1,y}=k_\theta$, we find 
 \baa  
 &h^{strain,2}_{\alpha \eta s,\alpha_2\eta_2 s_2} \nonumber \\ 
 =&\delta_{\eta,\eta_2}\delta_{s,s_2} \delta_{\alpha,3-\alpha_2}\frac{2\pi \lambda_0^2 \hbar \epsilon_{xy}}{\Omega_{tot}} \sum_{\kk}\sum_{\QQ \in \mathcal{Q}_0 }\bigg[ i(-1)^\alpha (k_y-Q_y)f_+(\kk-\QQ) 
 -i q_{1,y}(-1)^\alpha f_{-}(\kk-\QQ)
 \nonumber \\ 
 &- (k_x-Q_x)\eta  f_+(\kk-\QQ) 
 \eaa 
 where we define 
 \baa  
 f_\pm (\kk-\QQ) = \pm  e^{-\lambda_0^2 (\kk-\QQ+ \qq_{1})^2} + e^{-\lambda_0^2 (\kk-\QQ- \qq_{1})^2} 
 \eaa  
Since $\qq_1 = k_\theta(0,1)$, $f_+(\kk-\QQ)$ is an even function of both $(\kk-\QQ)_x$ and $(\kk-\QQ)_y$ and $f_-(\kk-\QQ)$ is an even function of $(\kk-\QQ)_x$ and an odd function of $(\kk-\QQ)_y$. Therefore, 
\baa 
&\sum_\kk \sum_{\QQ \in \mathcal{Q}_0} (k_y-Q_y)f_+(\kk-\QQ) = 0,\quad\quad 
\sum_\kk \sum_{\QQ \in \mathcal{Q}_0} f_-(\kk-\QQ) = 0 ,\quad\quad 
&\sum_\kk \sum_{\QQ \in \mathcal{Q}_0}(k_x-Q_x) f_+(\kk-\QQ) = 0 
\eaa 
Thus,
\baa  
h^{strain,2}_{\alpha \eta s,\alpha_2\eta_2 s_2} = 0 
\label{eq:hstrain_2}
\eaa 
However, we mention that, even though $h^{strain,2}=0$, $H_{strain,2}^\eta$ in Eq.~\ref{eq:single_particle_strain} can contribute a $\kk$-denpendent term in the THF model which has been omitted here.

 Combining Eq.~\ref{eq:hstrain_1} and Eq.~\ref{eq:hstrain_2}, we have
\baa  
\hH_{strain} =& \sum_{\RR,\alpha \eta s,\alpha_2\eta_2s_2} 
\alpha_{\alpha\eta s,\alpha_2\eta_2s_2} f_{\RR,\alpha \eta s}^\dag f_{\RR,\alpha_2\eta_2s_2} 
=&\alpha \sum_{\RR,\alpha \eta s,\alpha_2\eta_2s_2} 
[h_{strain}]_{\alpha\eta s,\alpha_2\eta_2s_2} f_{\RR,\alpha \eta s}^\dag f_{\RR,\alpha_2\eta_2s_2} 
\eaa  
which is equivalent to the Eq.~\ref{eq:hstrain}. $\alpha$ is connected to $\epsilon_{xy}$ via
\baa  
\alpha = \frac{3\pi \lambda_0^2 \beta \gamma_0}{\Omega_{tot}} \epsilon_{xy} \sum_\kk 
\bigg[  \sum_{\QQ \in \mathcal{Q}_{+}  }e^{-\lambda_0^2(\kk-\QQ)^2} +\sum_{\QQ \in \mathcal{Q}_{-}  }e^{-\lambda_0^2(\kk-\QQ)^2} 
\bigg] \approx \bigg(1.3 \times 10^4 \epsilon_{xy} \bigg) \text{meV} \, .
\eaa 
For a typical strain at the order of $\epsilon_{xy} \sim 10^{-3}$~\cite{koshino_22}, $\alpha\sim 13$meV. 
We mention that the strain could also induce addition terms to the conduction $c$-electron blocks ($c^\dag c$) and $f$-$c$ hybridization blocks ($c^\dag f,f^\dag c$), which are expected to be small. Here, we only consider a minimal model, that is zeroth order in $\kk$, to capture the essential effect of the strain. We leave a comprehensive construction of strain term (based on Ref.~\cite{vafek2022continuum}) for future study.

\section{Dynamical mean field theory: implementation}
We study dynamical effects of the $f$-$f$ density-density term using DMFT. The $c$-$f$ and $c$-$c$ density terms are included in this model on the Hartree level. We neglect the $f$-$c$ exchange interaction $\hat{H}_J$. We treat a Hamiltonian of the form $\hat{H}_c + \hat{H}_{fc} + \hat{H}_U + \hat{H}^{MF}_W + \hat{H}^{MF}_V$ where the first two terms compose the non-interacting THF model and the remaining terms are given by
\begin{align}
    \hat{H}_U =& \frac{U}{2}\sum_{(\alpha,\eta,\sigma)\neq (\alpha', \eta', \sigma')} f^\dagger_{\alpha,\eta,\sigma}f_{\alpha,\eta,\sigma}f^\dagger_{\alpha',\eta',\sigma'}f_{\alpha',\eta',\sigma'}\nonumber  \\&- 3.5 U \sum_{(\alpha,\eta,\sigma)} f^\dagger_{\alpha,\eta,\sigma}f_{\alpha,\eta,\sigma}\\
    \hat{H}^{MF}_W =& W\nu_f\sum_{(a,\eta,\sigma)} c^\dagger_{a,\eta,\sigma} c_{a,\eta,\sigma}  + W\nu_c \sum_{(\alpha,\eta,\sigma)} f_{\alpha,\eta,\sigma}^\dagger f_{\alpha,\eta,\sigma},\\
    \hat{H}^{MF}_V =& V\nu_c\sum_{(a,\eta,\sigma)} c^\dagger_{a,\eta,\sigma} c_{a,\eta,\sigma}.
\end{align}
  Here, the last two terms are mean-field decoupled and therefore composed of single particle terms, and we have neglected all constant terms as an overall shift does not affect the observables we are interested in. The two particle term $\hat{H}_U$ is restricted to the f-~subspace. We proceed with an LDA+DMFT-style calculation with a correlated subspace made up of f-~electrons embedded in a larger space of f- and c-~electrons. By allowing an arbitrary shift of the chemical potential $\mu$ and fixing the total filling $\nu = \nu_f + \nu_c$, we can consider an effective interacting Hamiltonian restricted to the the f- subspace:
\begin{align}
    \hat{H}^{eff}_I =& \hat{H}_{ff} + \hat{H}_{DC}\\
    \hat{H}_{ff} =& \frac{U}{2}\sum_{(\alpha,\eta,\sigma)\neq (\alpha', \eta', \sigma')}f^\dagger_{\alpha,\eta,\sigma}f_{\alpha,\eta,\sigma}f^\dagger_{\alpha',\eta',\sigma'}f_{\alpha',\eta',\sigma'} \\ 
    \hat{H}_{DC}=&
    \underbrace{\left(-3.5U + (\nu-2\nu_f)W -V(\nu-\nu_f)\right)}_{\mu_{DC}(\nu_f)}\sum_{(\alpha,\eta,\sigma)} f^\dagger_{\alpha,\eta,\sigma}f_{\alpha,\eta,\sigma}.
\end{align}
The $f$-electron occupation $\nu_f$ is self-consistently calculated along with the the $f$- subspace self-energy $\Sigma_f$ in at each step of the DMFT loop for a set of different total fillings ($\nu$) and temperatures (T). Written in this way, the single particle terms can be treated as an iteration-dependent double counting potential, $\mu^{(i)}_{DC} = \mu_{DC}\left(\nu_{f}^{(i)}\right)$, where the superscript labels the iteration number.

In the $i$th iteration of the self-consistency loop, we perform the following sequence of steps:
\begin{enumerate}
    \item Embed the old impurity self-energy $\Sigma_f^{(i-1)}$ into the lattice Green's function associated to the Hamiltonian $\hat{H}_c + \hat{H}_{fc} + \hat{H}_{DC}^{(i-1)}$.
    \item Fix the total filling to $\nu$.
    \item Calculate the new orbital-resolved fillings $\nu_f^{(i)}$ and $\nu_c^{(i)}$.
    \item Determine the new double counting term $\mu_{DC}^{(i)}$.
    \item Use a CT-QMC solver to obtain the new self energy $\Sigma_f^{(i)}$.
\end{enumerate}

We performed our calculations in parallel using the w2dynamics suite \cite{w2dyn1,w2dyn2}\footnote{https://github.com/w2dynamics/w2dynamics} as well as the TRIQS family of packages \cite{parcollet_triqs_2015, seth_triqscthyb_2016, aichhorn_triqsdfttools_2016}.

In Fig.~3(a) of the main text, the zero frequency spectral weight and the scattering rate are computed by extrapolating the green function and the self-energy respectively on the Matsubara axis. We approximate these by using the value at the first Matsubara frequency so that
\begin{align}
\Gamma &= -\text{Im} \Sigma(\omega=0) \approx -\text{Im} \Sigma(\omega_0),\\A(\omega=0) &= -\frac{1}{\pi}\left(\sum_k \text{Tr} G(k,\omega = 0)\right) 
\approx-\frac{1}{\pi}\left(\sum_k \text{Tr}G(k,\omega_0)\right).
\end{align}

\end{document}